\documentclass[aps,prd,twocolumn,showpacs,10pt,superscriptaddress, footinbib]{revtex4-1}

\usepackage{amsmath, amsthm, amssymb,bm,relsize}
\usepackage{amsmath,autobreak}
\usepackage{graphicx}  
\usepackage{amssymb}   
\usepackage{amsmath}
\usepackage{hyperref}
\usepackage{color}
\usepackage{comment}
\usepackage{ulem}
\usepackage{hyperref}
\usepackage{cleveref}
\usepackage{todonotes}
\usepackage{comment}

\usepackage{dsfont}
\newcommand\BbbGamma{\reflectbox{\rotatebox[origin=c]{180}{$\mathds L$}}}

\usepackage{caption}
\usepackage{booktabs,siunitx}
\allowdisplaybreaks
\usepackage{tabularx}
\newcolumntype{P}[1]{>{\centering\arraybackslash}p{#1}}
\usepackage{multirow, caption, booktabs}
\usepackage{slashed}
\usepackage{array}
\allowdisplaybreaks
\usepackage{endnotes}
\let\footnote=\endnote

\usepackage{multirow, caption, booktabs} 

\usepackage{graphicx}
\usepackage{psfrag}
\usepackage[caption=false]{subfig}
\definecolor{urlblue}{rgb}{0.2,0.4,0.7}
\definecolor{citegreen}{rgb}{0,0.6,0.2}
\definecolor{linkred}{rgb}{0.9,0.2,0.1}
\usepackage{hyperref}
\hypersetup{colorlinks=true, citecolor=citegreen, linkcolor=blue,urlcolor=urlblue}

\usepackage{slashed}
\usepackage{array}

\allowdisplaybreaks

\usepackage{amsmath,bm,amssymb}
\usepackage{amsbsy}
\usepackage{autobreak}
\usepackage{endnotes}
\let\footnote=\endnote

\def\la{\leftarrow}

\allowdisplaybreaks

\begin{document}


\title{NNLO QCD corrections to unpolarized and polarized SIDIS}

\author{Saurav Goyal}
\email{sauravg@imsc.res.in}
\affiliation{The Institute of Mathematical Sciences,  Taramani, 600113 Chennai, India}
\affiliation{Homi Bhabha National Institute, Training School Complex, Anushakti Nagar, Mumbai 400094, India}
\author{Roman N. Lee}
\email{r.n.lee@inp.nsk.su}
\affiliation{Budker Institute of Nuclear Physics, 630090, Novosibirsk, Russia}
\author{Sven-Olaf Moch}
\email{sven-olaf.moch@desy.de}
\affiliation{II. Institute for Theoretical Physics, Hamburg University, D-22761 Hamburg, Germany} 
\author{Vaibhav Pathak}
\email{vaibhavp@imsc.res.in}
\affiliation{The Institute of Mathematical Sciences, Taramani, 600113 Chennai, India}
\affiliation{Homi Bhabha National Institute, Training School Complex, Anushakti Nagar, Mumbai 400094, India}
\author{Narayan Rana}
\email{narayan.rana@niser.ac.in}
\affiliation{School of Physical Sciences, National Institute of Science Education and Research, 752050 Jatni, India}
\affiliation{Homi Bhabha National Institute, Training School Complex, Anushakti Nagar, Mumbai 400094, India}
\author{V. Ravindran}
\email{ravindra@imsc.res.in}
\affiliation{The Institute of Mathematical Sciences, Taramani, 600113 Chennai, India}
\affiliation{Homi Bhabha National Institute, Training School Complex, Anushakti Nagar, Mumbai 400094, India}

\date{\today}

\begin{abstract}
The semi-inclusive deep-inelastic scattering (SIDIS) process requires the presence of an identified hadron H$'$
in the final state, which arises from the scattering of a lepton with an initial hadron P. 
By employing factorization in quantum chromodynamics (QCD), SIDIS provides essential knowledge on the hadron structure, enabling the exploration of parton distribution functions (PDFs) and fragmentation functions (FFs). 
The coefficient functions for SIDIS can be calculated in perturbative QCD and are currently known to the next-to-next-to-leading order (NNLO) for the cases, where the incoming lepton and the hadron P are either both polarized or unpolarized. 
We present a detailed description of these NNLO computations, 
including a thorough discussion of all the partonic channels, the calculation of the amplitudes and master integrals for the phase-space integration as well as the renormalization of ultraviolet divergences and mass factorization of infrared divergences in dimensional regularization through NNLO. 
We provide an extensive phenomenological analysis of the effects of NNLO corrections on SIDIS cross sections for different PDFs and FFs 
and various kinematics, including those of the future Electron-Ion Collider (EIC). 
We find that these corrections are not only significant but also crucial for reducing the dependence on the renormalization and factorization  scales $\mu_R$ and $\mu_F$ to obtain stable predictions.
\end{abstract}

\maketitle

\section{Introduction}
The exploration of hadron structure and the underlying dynamics, has been a primary focus in high-energy physics. 
Historically, deep inelastic scattering (DIS) of a lepton on an initial hadron P has proven to be especially useful for investigating short-distance phenomena and understanding hadron structure.
The colliding leptons, e.g. electrons or muons interact with hadrons through either electromagnetic or 
weak interactions, depending on the energy transferred from leptons to hadrons.
Inclusive DIS~\cite{Blumlein:2012bf} depends on the momenta $P$ of the incoming hadron and 
the momentum transfer $q$ from the incoming lepton to the hadron through $Q^2 = -q^2 > 0$ and 
the Bjorken variable $x=Q^2/(2 P\cdot q)$.
The DIS cross sections, written in terms of the structure functions (SFs) which depend on $x$ and $Q^2$, 
can be described within the framework of quantum chromodynamics (QCD), the gauge theory of strong interaction thanks to the QCD factorization. Factorization property of QCD allows for the separation of long- and short-distance phenomena in the so-called Bjorken limit, 
where $x$ is kept fixed and $Q^2\rightarrow \infty, P\cdot q \rightarrow \infty$.
The SFs can then be written as a convolution of parton distribution functions (PDFs) which parametrize the dynamics of partons such as quarks, antiquarks and gluons in the hadron, and coefficient functions (CFs). 
The PDFs are typically extracted from a comparison of experimental data to theoretical predictions. 
Inclusive DIS data~\cite{H1:2015ubc} serve as one of the primary sources of information in that endeavor, as documented by its use in all global fits of unpolarized PDFs~\cite{Alekhin:2017kpj,Hou:2019efy,Bailey:2020ooq,NNPDF:2021njg,Alekhin:2024bhs}.
The CFs on the other hand, are calculable in perturbative QCD and currently known to next-to-next-to-next-to-leading order (N$^3$LO)~\cite{Moch:2004xu,Vermaseren:2005qc,Moch:2008fj,Blumlein:2022gpp}, 
both, in the case of unpolarized and polarized initial leptons and hadrons.

Semi-inclusive DIS (SIDIS) experiments~\cite{Aschenauer:2019kzf} observe one specific
hadron in the final state, in addition to the scattered lepton, 
which adds sensitivity to dynamics that governs the fragmentation of partons into the final state hadron. 
Through QCD factorization, the hadronic cross sections for SIDIS are expressed in terms of a set of SFs, 
which share an additional dependence on the scaling variable $z=P_H\cdot P/P\cdot q$, where $P_H$ is the momentum of final state hadron and the long distance dynamics of the fragmentation process is parametrized by fragmentation functions (FFs)~\cite{Metz:2016swz}. Like PDFs, FFs are process independent, and not calculable in perturbative QCD. In contrast to PDFs, there is limited data available for their determination, cf.\ e.g.~\cite{ParticleDataGroup:2024cfk}, 
which makes SIDIS an interesting reaction for expanding this area of research.

SIDIS measurements were pioneered at DESY's HERA collider by the HERMES experiment~\cite{HERMES:2004zsh} in electron-nucleon scattering. 
Other experiments include, e.g.\ COMPASS~\cite{COMPASS:2010hwr} at CERN's SPS, which has studied hadron structure with high intensity muon and hadron beams. 
The Electron-Ion Collider (EIC) which is going to come up at the BNL, is a new initiative in the field of high-energy physics to further explore strong dynamics and the fundamental structure of hadrons using high-energy electron and ion beams~\cite{AbdulKhalek:2021gbh}. 
The SIDIS reaction will be one of the flagship measurements at the EIC, covering a large range of hadron kinematics in $(x, Q^2)$ and $(z, Q^2)$, 
for the simultaneous extraction of PDFs and FFs, in particular for the case of polarized beams, 
that will allow one to study spin-dependent PDFs in largely unexplored regions. 
In recent times, measurements from SIDIS experiments \cite{HERMES:2012uyd,COMPASS:2016crr,Aschenauer:2015rna} 
have already been used~\cite{de_Florian_2017,Anderle_2017,Leader:2015hna,Bertone:2018ecm,Abdul_Khalek_2021} to directly constrain FFs, 
and also in combined fits of FFs and spin-dependent PDFs~\cite{Ethier_2017,de_Florian_2019,Moffat_2021}, 
most recently at next-to-next-to-leading order (NNLO) accuracy~\cite{Bertone:2024taw,Borsa:2024mss}, 
improving earlier fits of polarized PDFs ~\cite{deFlorian:2009vb,deFlorian:2014yva}.

The CFs for SIDIS have been computed in perturbative QCD to next-to-leading order (NLO) QCD corrections long ago~\cite{Altarelli:1979kv}, see also
\cite{Nason:1993xx,Furmanski:1981cw,Graudenz:1994dq,deFlorian:2012wk,deFlorian:2012wk}.
Beyond NLO, only partial results have been available for along time, see \cite{Daleo:2003jf,Daleo:2003xg,Anderle:2012rq,Anderle:2016kwa}. 
In the soft and collinear limit ($x \to 1$ and $z \to 1$), the dominant threshold logarithms 
have been obtained recently up to N$^3$LO accuracy in QCD~\cite{Abele:2021nyo,Abele:2022wuy}, and have been resummed to all orders in perturbation theory to next-to-next-to-next-to-leading logarithmic (N$^{3}$LL) accuracy, 
extending earlier resummations for the SIDIS process, which had achieved only lower logarithmic accuracy~\cite{Cacciari:2001cw,Anderle:2012rq,Anderle:2013lka,Sterman:2006hu}, i.e.\ 
to next-to-leading logarithm (NLL) level.
This progress has exploited relations between SIDIS and the rapidity distribution of pair of leptons in the Drell-Yan (DY) process, relying on results for the DY rapidity distribution to the required accuracy~\cite{Anderle:2012rq,Westmark:2017uig,Banerjee:2018vvb}. 

Recently, the complete NNLO QCD corrections to the SIDIS CFs, for both polarized and unpolarized initial lepton and hadron beams, have been computed~\cite{Goyal:2023zdi,Bonino:2024qbh,Bonino:2024wgg,Goyal:2024tmo}.
The results have been obtained independently by two groups in full agreement with each other, and also agree with the so-called soft-plus-virtual corrections at NNLO, derived in \cite{Abele:2021nyo}.
However, details of the complete NNLO computation have not been documented thus far, which is a gap that the current article aims to close. 
The outline of the article is as follows. 
In Sec.~\ref{sec:theory} we set the stage with a short summary of the QCD theory description for the SIDIS process. 
Sec.~\ref{sec:partonlevel} gives a detailed discussion of all partonic channels and amplitudes entering in the computation of the CFs through NNLO as well as the corresponding master integrals for the phase-space and virtual loop integration. 
Sec.~\ref{sec:massfact} discusses the mass factorization, both for the incoming and the final state partons, for the extraction of the finite coefficient functions.
In Sec.~\ref{sec:pheno} we apply the NNLO result in a comprehensive phenomenological study.
We conclude in Sec.~\ref{sec:conclusion} with a summary and an outlook on further developments.
Appendixes \ref{Appendix:A}, \ref{Appendix:B} and \ref{Appendix:C}  
collect known results on splitting functions.

\section{Theoretical Framework}
\label{sec:theory}
We consider the SIDIS process given by
\begin{equation}
l(k_l) + \text{P}(P) \rightarrow l({k}'_l)+\text{H}'(P_H) + \text{X}\, ,
\end{equation}
where the lepton $l$ has incoming (outgoing) momentum $k_l$(${k}'_l$), 
and \text{P}(\text{H}$'$) is the incoming (outgoing) hadron with momentum $P$($P_H$). 
The remaining set of final state particles is collectively denoted by \text{X}.

We restrict ourselves to the differential observable 
$d^3 (\Delta) \sigma/dxdydz$ where
the spin-averaged cross section is defined by
\begin{equation}
d \sigma = \frac{1}{4}\sum_{s_l,S,s'_l,S_H} d\sigma_{s_l ,S}^{s'_l,S_H} \,,
\end{equation}
and the corresponding spin-dependent cross section is
\begin{equation}
d \Delta \sigma ={\frac{1}{2} } \sum_{s'_l,S_H}
  \left(d\sigma_{s_l=\frac{1}{2},S=\frac{1}{2}}^{s'_l,S_H}
-d\sigma_{s_l=\frac{1}{2},S=-\frac{1}{2}}^{s'_l,S_H} \right)
\, ,
\end{equation}
where $s_l$, $S$ are the spins of incoming lepton and hadron, 
and $s'_l$, $S_H$ are the spins of the outgoing ones, respectively.
The differential DIS cross section depends on 
the space-like momentum transfer $Q^2=-q^2$ where $q={k}_l-{k}'_l$,
the Bjorken variable $x=Q^2/(2 P\cdot q)$, 
the inelasticity $y={P\cdot q}/{P\cdot k_l}$, 
and the scaling variable $z={P\cdot P_H}/{P\cdot q}$ 
for the fraction of the initial energy transferred to the final-state hadron.

In the approximation of a single-photon exchange between the incoming lepton and hadron, the differential cross section factorizes into a leptonic tensor denoted by $(\Delta){\cal L}_{\mu\nu}$
and the hadronic tensor $(\Delta)W_{\mu \nu}$:
\begin{align}
d(\Delta) \sigma &= \frac{d^3 k'_l}{(2\pi)^3 2k'^{0}_l} \frac{1}{4 \sqrt{(k_l\cdot P)^2 -m^2M^2}}
(\Delta)  {\cal L}^{\mu \nu}(k_l,k'_l,q) 
\nonumber\\
& \times \left(\frac{e^4}{Q^4}\right)  (4 \pi M)  ~~(\Delta) W_{\mu\nu}(q,P,P_H)
\, ,
\label{lwmunu}
\end{align}
where $e$ denotes the electric charge and $M(m)$ is the mass of the incoming hadron (lepton) and we drop the latter mass throughout. Thus 
\begin{equation*}
 {\cal L}^{\mu\nu}=2{k}_l^{\mu}{k}_{l}^{\prime\nu}+2{k}_{l}^{\prime\mu}{k}_l^{\nu} -Q^2 g^{\mu\nu}, 
 \, \,
\Delta{\cal L}^{\mu \nu} = 2 i \epsilon^{\mu \nu \sigma\lambda} q_{\sigma} s_{l,\lambda}
\, ,
\end{equation*}
where $\epsilon^{\mu\nu\sigma\lambda}$ is the Levi-Civita tensor (with $\epsilon^{0123} = -\epsilon_{0123} = +1$).

The hadronic tensor is not calculable in perturbation theory due to the composite structure of hadrons.
In the single-photon exchange approximation, one can use electromagnetic current conservation, parity and time reversal symmetries to parametrize $(\Delta) W_{\mu\nu}$ in terms of tensors
such as $g_{\mu \nu},\epsilon_{\mu\nu \sigma\rho}$ and the vectors $q,P$ and $S$, 
weighted by two unpolarized SFs denoted by $F_i(q,P,P_H)$ 
and the polarized SFs denoted by $g_i(q,P,P_H)$ for $i =1,2$, 
see, e.g.~\cite{ParticleDataGroup:2024cfk}. 
These SFs can be expressed through the scaling variables $x$, $z$ and the photon virtuality $Q^2$, i.e. $F_i=F_i(x,z,Q^2)$ and $g_i=g_i(x,z,Q^2)$. 
The (un)polarized hadronic tensor is then expanded as,
\begin{align}
\label{wmunu}
W_{\mu\nu} &= \sum_{i=1,2} W_i(x,z,Q^2) T_{i,\mu \nu}(P,q),\\
\label{dwmunu}
\Delta W_{\mu \nu} &= \sum_{i=1,2} g_i(x,z,Q^2) S_{i,\mu \nu}(P,q,S)
\, ,
\end{align}
where the respective tensors are given by
\begin{eqnarray}
   T_{1,\mu \nu} &=&-g_{\mu\nu} + \frac{q_\mu q_\nu}{q^2}
   \nonumber \, ,\\
   T_{2,\mu \nu} &=&\frac{1}{M^2}\biggl(P_{\mu} - \frac{P.q}{q^2} q_\mu\biggr)\biggl(P_{\nu} - \frac{P.q}{q^2} q_\nu\biggr)\, ,
\end{eqnarray}
and
\begin{eqnarray}
S_{1,\mu\nu} &=& -\frac{i}{P.q} \epsilon_{\mu \nu \sigma \lambda} q^\sigma S^\lambda \nonumber \, ,\\
S_{2,\mu\nu}&=& -\frac{i}{P.q} \epsilon_{\mu \nu \sigma \lambda} q^\sigma  \left(S^\lambda - \frac{S\cdot q}{P\cdot q }P^\lambda\right)\, ,
\end{eqnarray}
with $S^2=1$ and $S\cdot P=0$.
Substituting $(\Delta) W_{\mu \nu}$ in eq.~(\ref{lwmunu})
and expressing the phase space of the leptonic tensor in terms of $x$ and $y$, we obtain
\begin{equation}
    \frac{d^3\sigma}{dxdydz}=\frac{4\pi  \alpha_e^{2}}{Q^2} \left[y F_1(x,z,Q^2) + \frac{(1-y)}{y} F_2(x,z,Q^2)\right]
    \, ,
\end{equation}
where $\alpha_e=e^2/(4\pi)$ is the fine structure constant.
The SFs $F_i$ are related to $W_i$ through
$W_1 = F_1/M$ and $W_2 = F_2/(y E)$ with 
the energy $E$ of the incoming lepton.
Similarly, the spin-dependent cross section is found to be
\begin{equation}
\frac{d^3\Delta \sigma}{dxdydz} = \frac{4\pi\alpha_e^2}{Q^2} \big(2-y\big)g_1 (x,z,Q^2)\, ,
\end{equation}
Note that $g_2$ does not contribute since we restrict ourselves to longitudinally polarized hadron in the initial state.

The SFs are subject to the standard factorization in the QCD improved parton model.
They are expressed as a convolution of PDFs, $(\Delta) f_{a/\text{P}}(x_1,\mu_F^2)$ and FFs, $D_{\text{H}'/b}(z_1,\mu_F^2)$ 
with CFs, $(\Delta ){\cal C}_{i,ab}(x/x_1,z/z_1,Q^2,\mu_F^2)$, the latter being calculable in perturbation theory. 
Here $x_1$ is the incoming parton's momentum fraction  with respect to the hadron P, i.e.\ $x_1= p_a/P$ 
and $z_1$ is the momentum fraction of the final state parton $b$  carried away by the outgoing hadron H$'$, i.e.\ $z_1 = P_H/p_b$. 
The SFs $g_{1}(x,z,Q^2)$ and $F_i(x,z,Q^2)$ factorize as
\begin{widetext}
\begin{align}
\label{eq:StrucCoeff}
\big(g_{1}\big) F_i (x,z,Q^2)
&= x^{i-1}\sum_{a,b = q,\overline{q},g}
\int_x^1 \frac{dx_1}{x_1} (\Delta )f_{a/\text{P}}\left( x_1,\mu_F^2\right)
\int_z^1 \frac{dz_1}{z_1}
  D_{\text{H}'/b}\left( z_1,\mu_F^2\right)
  (\Delta) {\cal C}_{i,ab}\left(\frac{x}{x_1},\frac{z}{z_1},Q^2,\mu_F^2\right)\, ,
\end{align}
\end{widetext}
where $f_{a/\text{P}}$ denotes the spin averaged PDF and
$\Delta f_{a/\text{P}} = f_{a(\uparrow)/\text{P}(\uparrow)}
-f_{a(\downarrow)/\text{P}(\uparrow)}$.

The CFs can be extracted from parton level sub-process cross sections $d (\Delta) \hat \sigma_{i,ab}/dx dy  dz$ 
through the relation $x^{(1-i)}d (\Delta) \hat \sigma_{i,ab}/dx dy  dz = (\Delta) \hat {\cal C}_{i,ab}$. 
After renormalization, these CFs $(\Delta)\hat {\cal C}_i$ are UV finite but still contain collinear divergences, which factorize through Altarelli-Parisi (AP) kernels $(\Delta)\Gamma$ and $\tilde \Gamma$. 
The mass factorization reads
\begin{eqnarray}
\label{eq:massfact}
{\lefteqn{
(\Delta) \hat  {\cal C}_{i,ab}(x,z)(\varepsilon) \,=\,}}
\nonumber \\
&& 
(\Delta) \Gamma_{c\leftarrow a}(x,\varepsilon) \otimes
(\Delta)  {\cal C}_{i,cd}(x,z,\varepsilon)
\, \tilde \otimes \,
\tilde\Gamma_{b\leftarrow d}(z,\varepsilon)\, ,
\qquad
\end{eqnarray}
where $\Gamma$ and $\Delta \Gamma$ are the space-like AP kernels, including spin dependence for the latter,
and $\tilde\Gamma$ are the corresponding time-like AP kernels. 
These kernels remove all collinear divergences, so that the CFs $(\Delta) {\cal C}_{i,cd}(x,z,\varepsilon)$ are finite at every order in perturbation theory. 
Details will be described below. 
The convolutions $\otimes$ and $\tilde \otimes$ between the various functions in eq.~(\ref{eq:massfact}) are defined by
\begin{eqnarray}
{\lefteqn{
A(x)\otimes {\cal C}(x,z)\, \tilde \otimes\, B(z) \,=\,}}\nonumber \\
&&
   \int_x^1 \frac{dx_1}{x_1}
   \int_z^1 \frac{dz_1}{z_1}
A(x_1)\, {\cal C}\left(\frac{x}{x_1},\frac{z}{z_1}\right) B(z_1)\, .
\end{eqnarray}

The partonic cross sections are defined as,
\begin{eqnarray}
\label{eq:parton-crs}
\frac{d^3{(\Delta) \hat \sigma}_{i,ab}}{dx dy dz} &=& \frac{  (\Delta) \mathcal{P}_{i}^{\mu\nu}}{4\pi}\int \text{dPS}_{X+b}\, {\Sigma}|{(\Delta) M}_{ab}|^{2}_{\mu\nu}\,
\nonumber \\
&&\times \delta\bigg(\frac{z}{z_1}-\frac{p_a \cdot p_b}{p_a\cdot q}\bigg)
\end{eqnarray}
where the  partonic projectors in $d$ space-time dimensions read,
\begin{eqnarray}
\mathcal{P}^{\mu\nu}_1&=&\frac{1}{(d-2)}\left(t_{1,\mu\nu}+\frac{Q^2}{p_a\cdot q}t_{2,\mu\nu}\right) \nonumber \\
\mathcal{P}^{\mu\nu}_2&=&\frac{Q^2}{(d-2)p_a.q}\left(t_{1, \mu\nu}+(d-1)\frac{Q^2}{p_a \cdot q}t_{2,\mu\nu}\right) \nonumber \\
\Delta \mathcal{P}^{\mu\nu}_{1}&=& \frac{i}{(d-2)(d-3)}
\epsilon^{\mu \nu \sigma \lambda} \,  \frac{q_\sigma p_{a,\lambda}}{p_a\cdot q}
\, .
\end{eqnarray}
where, $t_{1,\mu\nu} = T_{1,\mu\nu}$, $t_{2,\mu\nu}= x_1 T_{2,\mu\nu}$
 and $(\Delta) M_{ab} = M_{a(\uparrow)b}+(-)M_{a(\downarrow)b}$  is the spin-independent (dependent) amplitude for the process $\gamma^* + a(p_a,s_a)\rightarrow  b(p_b) + \text{X} $,
where the parton `$b$' fragments into hadron H$'$.  Here $s_a$ denotes the spin of the incoming parton $a$.  $\text{dPS}_{X+b}$ is the phase space for the final state particles consisting of X and $b$.
${\Sigma}$  denotes the summation over final state spin/polarization and their color quantum numbers in addition to the average over spin/polarization and color of incoming parton $a$.  
For spin-dependent cross sections, we take difference of the polarizations of the incoming partons instead of averaging over them.

\section{Partonic sub-processes}
\label{sec:partonlevel}
The computation of CFs requires parton level cross sections, the UV renormalization constants and the AP kernels to desired accuracy in the strong coupling constant ($a_s$) and $\varepsilon$.
Since the latter are all known, we only need to compute the partonic cross sections in eq.~(\ref{eq:parton-crs}) to second order in $a_s$. 
We use \texttt{QGRAF}~\cite{Nogueira:1991ex}
to generate Feynman diagrams for all the sub-processes.  With a set of in-house routines written in \texttt{FORM}  \cite{Kuipers:2012rf,Ruijl:2017dtg}, the output of
\texttt{QGRAF} is converted into a suitable format to apply Feynman rules and to perform Dirac algebra, Lorentz contractions and simplifications of color factors.
In the computation of partonic cross section beyond 
leading order (LO) in perturbation theory, we encounter both 
UV and IR divergences. 
The latter originate from soft and collinear partons due to massless gluons and quarks (anti-quarks).  
We work in $d=4+\varepsilon$ space-time dimensions to regulate them.

The spin dependent partonic amplitudes squared, $|\Delta M_{ab}|^2$ in eq.~(\ref{eq:parton-crs}), contain the Dirac matrix $\gamma_5$ or the Levi-Civita tensor from
spin dependent quarks (anti-quarks) wave functions or the polarization of gluons, respectively, see, e.g.~\cite{Zijlstra:1993sh}.
Since the $\gamma_5$ matrix and the Levi-Civita tensor are intrinsically four-dimensional objects, we need to choose a prescription to define them in $d=4+\varepsilon$ dimensions.
There are several schemes to do so, 
however none of them is known to preserve the chiral Ward identity.
In our work, we use Larin's scheme \cite{Larin:1993tq} to define $\gamma_5$ in $d=4+\varepsilon$, 
\begin{align}
\slash\!\!\!p_a\gamma_5 = -{\frac{i}{6}} ~\epsilon_{\mu \nu \sigma \lambda} p_a^\mu \gamma^\nu \gamma^\sigma \gamma^\lambda.
\end{align}
The product of two Levi-Civita tensors can be expressed in terms of a determinant of Kronecker deltas defined 
in $d=4 +\varepsilon$ dimensions, 
\begin{align}
  \label{eqn:LeviContract}
\epsilon_{\mu_1\nu_1\lambda_1\sigma_1}\,\epsilon^{\mu_2\nu_2\lambda_2\sigma_2}=
\large{\left |
  \begin{array}{cccc}
    \delta_{\mu_1}^{\mu_2} &\delta_{\mu_1}^{\nu_2}&\delta_{\mu_1}^{\lambda_2} & \delta_{\mu_1}^{\sigma_2}\\
    \delta_{\nu_1}^{\mu_2}&\delta_{\nu_1}^{\nu_2}&\delta_{\nu_1}^{\lambda_2}&\delta_{\nu_1}^{\sigma_2}\\
    \delta_{\lambda_1}^{\mu_2}&\delta_{\lambda_1}^{\nu_2}&\delta_{\lambda_1}^{\lambda_2}&\delta_{\lambda_1}^{\sigma_2}\\
    \delta_{\sigma_1}^{\mu_2}&\delta_{\sigma_1}^{\nu_2}&\delta_{\sigma_1}^{\lambda_2}&\delta_{\sigma_1}^{\sigma_2}
  \end{array}
\right |}
\, .
\end{align}
The results obtained in Larin's scheme are converted to the $\overline {\rm{MS}}$ scheme through an additional renormalization scheme transformation.  
The renormalization constant~\cite{Matiounine:1998re,Ravindran:2003gi,Moch:2014sna} used for this scheme transformation restores the Ward identity and also transforms the spin dependent PDFs in ${\overline {\rm{MS}}}$ scheme. 
Further details will be presented below.

\subsection{Amplitudes}
\begin{figure}[!ht]
\includegraphics[width=0.35\textwidth]{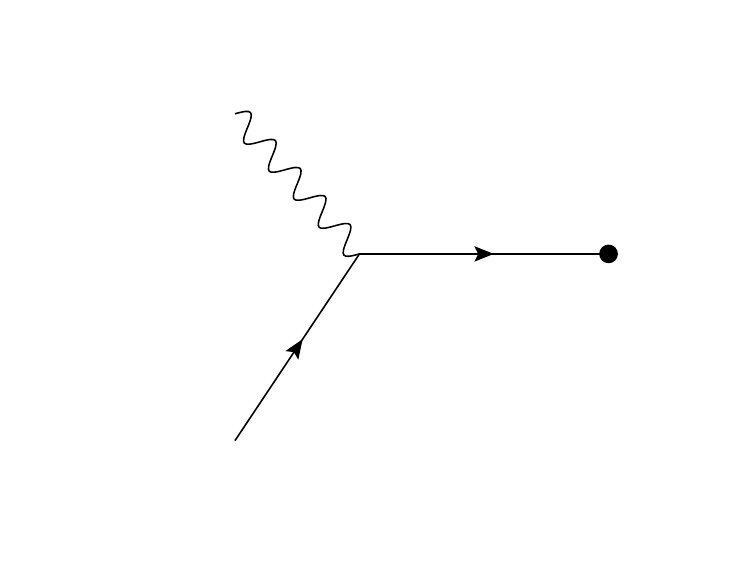}
\caption{Feynman diagram for the LO process.}
\label{fig:lo}
\end{figure}

In the following, we list the amplitudes corresponding
to sub-processes that contribute to 
the CFs, denoted by $(\Delta) {\cal C}_i$,
up to NNLO level.
We begin with the partonic sub-process at LO as shown in Fig.~\ref{fig:lo}:
\begin{align}
    q(\Bar{q}) + \gamma^* \rightarrow q(\Bar{q})\, .
\end{align}
Here, the quark (antiquark) in the final state will fragment into hadron H$'$.
%
At the NLO level, two types of processes contribute, namely the real parton emission sub-processes (R) and virtual gluon corrections (V) to the Born amplitude.
\begin{align}
    q(\Bar{q}) + \gamma^* &\rightarrow q(\Bar{q}) + {\text {one loop}}\, ,\\
    q(\Bar{q}) + \gamma^* &\rightarrow q(\Bar{q}) + g\, ,\\
    g + \gamma^* &\rightarrow q + \Bar{q}\, .
\end{align}
Any one of the final state partons in this list can fragment into the observed hadron H$'$. 
\begin{figure}[!ht]
\includegraphics[width=0.5\textwidth]{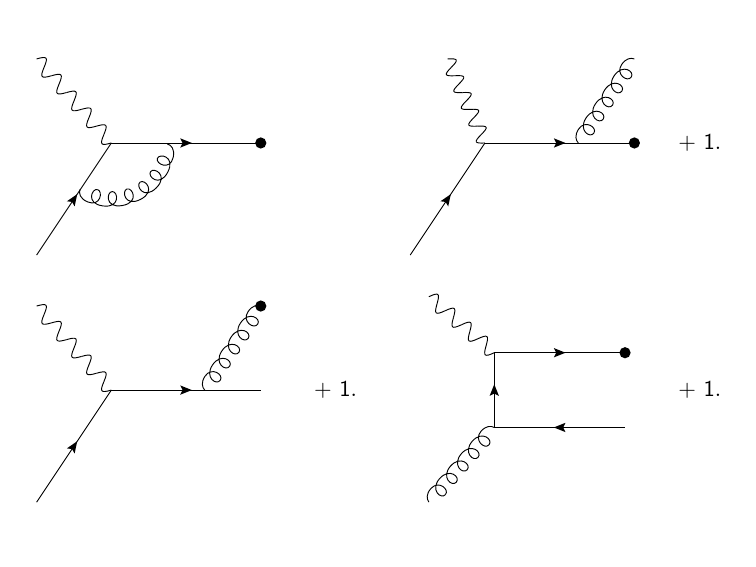}
\caption{Sub-processes at NLO.
The integers denote the numbers of additional diagrams to be considered.}
\label{fig:nlo}
\end{figure}
%
At NNLO, we encounter two-loop amplitudes, which we classify as VV.
The sub-process is given by
\begin{align}
    &q(\Bar{q}) + \gamma^* \rightarrow q(\Bar{q})+{\text{two loops}}\, ,
\end{align}
for which, the schematic Feynman diagrams are given as
\begin{figure}[!ht]
\includegraphics[width=0.32\textwidth, trim={1cm 2cm 1cm 1.9cm}, clip]{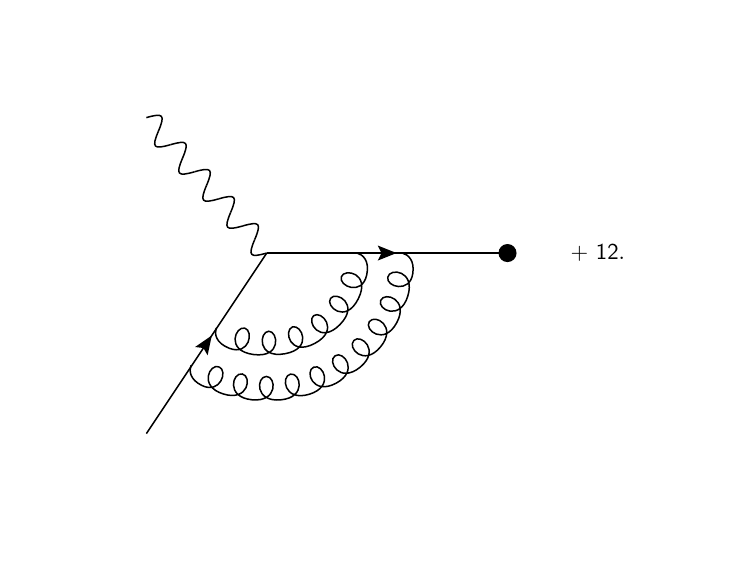}
\caption{Schematic representation of two-loop VV Feynman diagrams.}
\label{fig:nnlovv}
\end{figure}
%

\noindent
The class RV consists of single real emission amplitudes with one-loop virtual correction, as given by:
\begin{align}
    q(\Bar{q}) + \gamma^* &\rightarrow q(\Bar{q}) + g  +{\text {one loop}}\, ,\\
    g + \gamma^* &\rightarrow q + \Bar{q} 
    +{\text {one loop}}\, .
\end{align}
The Feynman diagrams can be drawn as:
\begin{figure}[!ht]
\includegraphics[width=0.45\textwidth, trim={0.1cm 0.2cm 0.1cm 0.2cm}, clip]{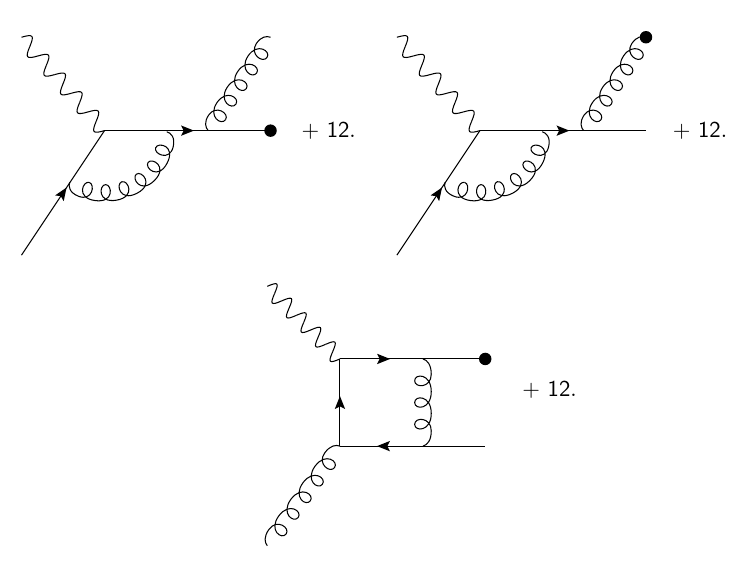}
\caption{RV diagrams (one-loop with one real emission) at NNLO.}
\label{fig:nnlorv}
\end{figure}
%

\noindent
The list of sub-processes with two real emissions reads: 
\begin{align}
    q(\Bar{q}) + \gamma^* &\rightarrow q(\Bar{q}) + g + g\, ,\\
    g + \gamma^* &\rightarrow  g + q + \Bar{q}\, ,\\
    q(\Bar{q}) + \gamma^* &\rightarrow q(\Bar{q}) + q' + \Bar{q}'\, ,\\
    q(\Bar{q}) + \gamma^* &\rightarrow q(\Bar{q}) + q + \Bar{q}\, .
\end{align}\\
Again, any one of the final state partons in the above list of sub-processes can fragment into hadron H$'$. 
\begin{figure}[!ht]
\includegraphics[width=0.48\textwidth, trim={0cm 3cm 0cm 2.2cm}, clip]{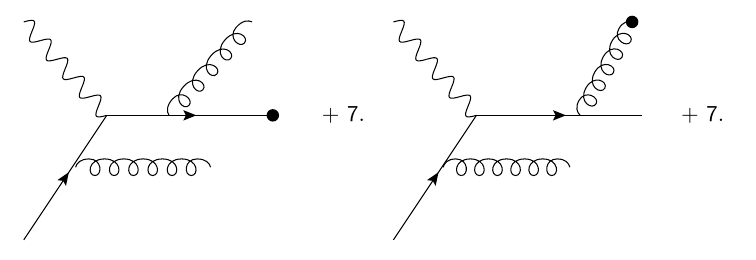}
\caption{RR diagrams: emissions of two gluons. The integer
indicates the number of additional diagrams.}
\label{fig:q2qgg}
\end{figure}
%
\begin{figure}[!ht]
\includegraphics[width=0.45\textwidth, trim={0cm 2.2cm 0cm 1.6cm}, clip]{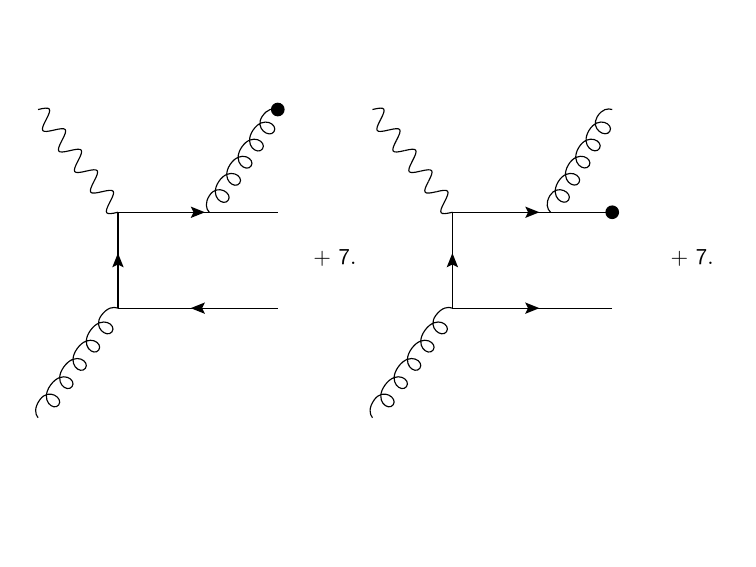}
\caption{RR diagrams: gluon initiated processes. The integer
indicates the number of additional diagrams.}
\label{fig:g2gqQ}
\end{figure}
\begin{figure}[!ht]
\includegraphics[width=0.45\textwidth, trim={0cm 2.4cm 0cm 2.9cm}, clip]{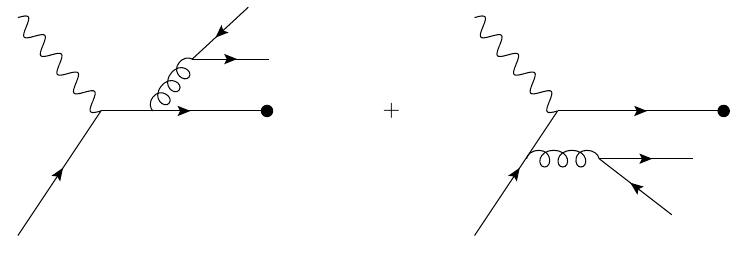}
\caption{RR diagrams: A type as defined in the text.}
\label{fig:nnlorrA}
\end{figure}
%
\begin{figure}[!ht]
\includegraphics[width=0.45\textwidth, trim={0cm 2.6cm 0cm 2.6cm}, clip]{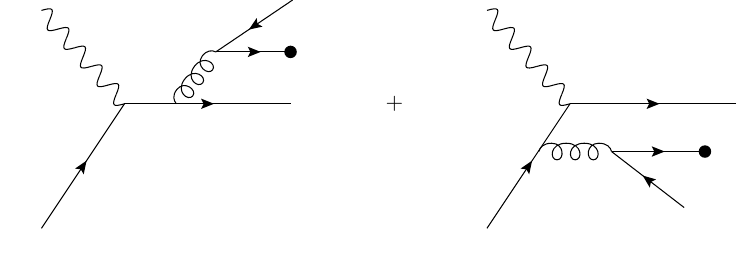}
\caption{RR diagrams: B type as defined in the text.}
\label{fig:nnlorrB}
\end{figure}
%
\begin{figure}[!ht]
\includegraphics[width=0.45\textwidth, trim={0cm 2.2cm 0cm 2cm}, clip]{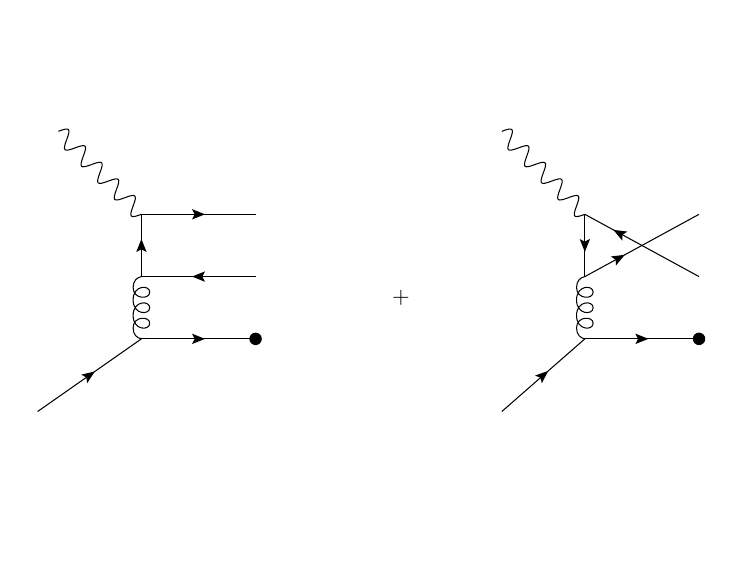}
\caption{RR diagrams: C type as defined in the text.}
\label{fig:nnlorrC}
\end{figure}
%
\begin{figure}[!ht]
\includegraphics[width=0.45\textwidth, trim={0cm 2.2cm 0cm 1.6cm}, clip]{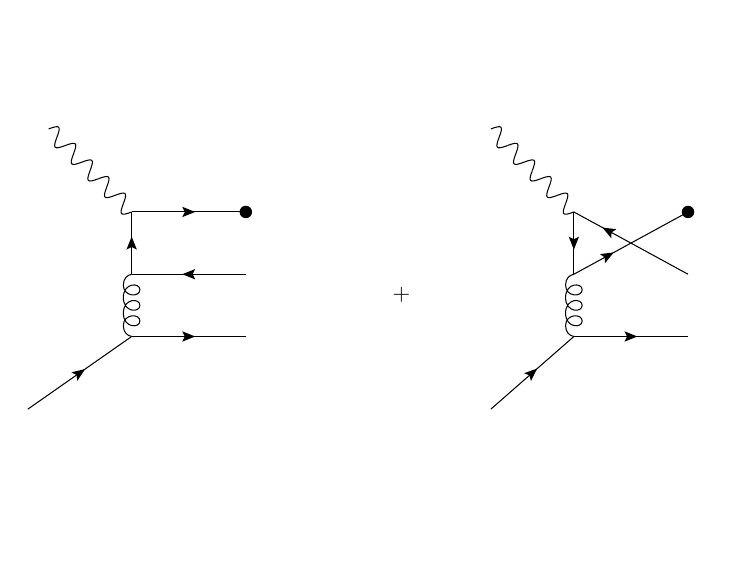}
\caption{RR diagrams: D type as defined in the text.}
\label{fig:nnlorrD}
\end{figure}

At NNLO, the VV contributions are given by the two-loop virtual amplitudes, Fig.~\ref{fig:nnlovv}, and the square of one-loop virtual correction to the LO sub-process, cf. upper left diagram in Fig.~\ref{fig:nlo}.
The RV contributions arise from the interference of one loop-one real emission amplitudes, Fig.~\ref{fig:nnlorv} with one real emission amplitudes Fig.~\ref{fig:nlo}.

The two-parton real emissions (RR) have three distinct sub-processes, namely the ones with a pair of gluons in the final state, cf.\ Fig.~\ref{fig:q2qgg}, 
those with one gluon along with a quark anti-quark pair Fig.~\ref{fig:g2gqQ}, and the ones that have a final state quark anti-quark pair from gluon splitting.
The latter type contains
four different sub-processes denoted by A, B, C and D, shown in Figs.~\ref{fig:nnlorrA} -- \ref{fig:nnlorrD}.

Having obtained the amplitudes and the squared matrix elements in $d$ dimensions,
we extract the SFs
$F_i$, $g_1$ using appropriate projectors and perform
loop as well as phase space integration as described in the next section.

\subsection{Master Integrals}
The $n+1$-body phase space for scattering processes with momenta $p_a+q\rightarrow \sum_{i=1}^{n}k_i+p_b$ and a constraint on the momentum $p_b$ of the fragmenting parton is given by
\begin{align}
\int_{z'} \text{[dPS]}_{n+1}
&=\prod_{i=1}^n
\left(\int
\frac{d^d k_i}{(2 \pi)^{d-1}} \delta(k_i^2)
\right)
\int  d^{d}p_b  (2 \pi) \delta(p_b^2) \nonumber\\
         & \times \delta^{d}(p_b+K-p_a-q) \delta\left(z'- \frac{p_{a}.p_{b}}{p_{a}.q}\right)\, ,
\end{align}
where $K=\sum_i^n k_i$ and $q = k_l-k_l'$.  
Note that we have introduced an additional delta function to define
scaling variable $z'=z/z_1$ corresponding to the fragmenting parton.
In center-of-mass frame of $p_a$ and $q$, the two-body phase space takes the simple form:
\begin{align}
\int_{z'} \text{[dPS]}_2&= \frac{2\pi}{(4\pi)^{\frac{d}{2}}\Gamma_{\frac{d}{2} -1}}\bigg(\frac{Q^2(1-x')}{x'}\bigg)^{\frac{d}{2}-2} \int_{0}^{1}d w \nonumber\\
         & ~~~~~~~\times(w(1-w))^{\frac{d}{2} -2} \delta(w- z')\, ,
\end{align}
where the integral $\int_{z'}$ implies the constraint on $z'$ through $\delta(z'-p_a\cdot p_b/p_a\cdot q)$, which has been omitted for brevity, here $x'=x/x_1$. 
The Gamma function $\Gamma(n)$ is presented as $\Gamma_n$.
Similarly, the three-body phase space for the scattering $p_a+q\rightarrow k_1 + k_2 +p_b$ in the center-of-mass frame of the momenta $k_1$ and $k_2$ is found to be 
(see \cite{Matsuura:1988sm,Zijlstra:1992qd,Rijken:1996ns,Ravindran:2003um}) 
\begin{align}
 \int_{z'} &\text{[dPS]}_3 = \frac{1}{(4\pi)^{d}\Gamma_{d-3}}\bigg(\frac{Q^2(1-x')}{x'}\bigg)^{d-3} \int_{0}^{\pi} d\theta (\sin\theta)^{d-3}\nonumber\\
&\times \int_{0}^{\pi} d\phi(\sin\phi)^{d-4} \int_{0}^{1} dw \int_{0}^{1}dz ~w^{\frac{d-4}{2}}(1-w)^{d-3} \nonumber\\
&\times\big(z( 1 - z)\big)^{\frac{d-4}{2}}\delta(w-z')\, .
\end{align}

The computation of the phase space integrals with the constraint $z' = p_a.p_b/{p_a.q}$ is technically challenging compared to fully inclusive DIS. 
The significant number of parton level sub-processes for the RV and RR cases leave us with large number of phase space integrals, most of them related to each other, though.  
In order to find a set of basis integrals, called master integrals (MIs), we use integration by parts identities (IBP) \cite{Chetyrkin:1981qh,Laporta:2000dsw}.  
To obtain IBP identities for phase space integrals, we use the method of reverse unitarity \cite{Anastasiou:2003gr,Anastasiou:2012kq},  
convert all of them into loop integrals with the help of the identity,
\begin{eqnarray}
(2 \pi i) \delta(p^2) = \frac{1}{p^2+i \epsilon} - \frac{1}{p^2-i \epsilon}\, ,
\end{eqnarray}
and then follow the standard approach that one uses for loop integrals to generate IBP identities.
With the help of these identities, we can express all
the phase space integrals in terms of a few phase space MIs.
We group the integrals into families and generate IBP identities using the \textit{Mathematica} package \texttt{LiteRed} \cite{Lee:2012cn,Lee:2013mka}.

In the case of RR at NNLO level, we have 8 independent propagators which give 13 families. Each family is denoted by $\mathds{A}_i, i=1,\cdots,13$ and an ordered set,
\begin{widetext}
%
\begin{center}
\begin{tabular}{cccccccccccccccccc}
$\mathds{A}_1$: & $\{D^\mathds{A}_1,D^\mathds{A}_5,D^\mathds{A}_6 \}$ & \ & 
$\mathds{A}_2$: &  $\{D^\mathds{A}_1,D^\mathds{A}_5,D^\mathds{A}_7 \}$ & \ &
$\mathds{A}_3$:& $\{D^\mathds{A}_1,D^\mathds{A}_5,D^\mathds{A}_8 \}$ & \  & 
$\mathds{A}_4$: & $\{D^\mathds{A}_1,D^\mathds{A}_6,D^\mathds{A}_8 \}$ & \ &
$\mathds{A}_5$:& $\{D^\mathds{A}_1,D^\mathds{A}_2,D^\mathds{A}_4\}$
\\[0.5ex]
$\mathds{A}_6$: & $\{D^\mathds{A}_1,D^\mathds{A}_2,D^\mathds{A}_8\}$ & \ &
$\mathds{A}_7$: & $\{D^\mathds{A}_1,D^\mathds{A}_4,D^\mathds{A}_5 \}$ & \ & 
$\mathds{A}_{8}$: &$\{D^\mathds{A}_1,D^\mathds{A}_4,D^\mathds{A}_7 \}$  & \ &
$\mathds{A}_{9}$: & $\{D^\mathds{A}_1,D^\mathds{A}_4,D^\mathds{A}_8 \}$ & \ & 
$\mathds{A}_{10}$:& $\{D^\mathds{A}_2,D^\mathds{A}_4,D^\mathds{A}_5\}$
\\
$\mathds{A}_{11}$:& $\{D^\mathds{A}_2,D^\mathds{A}_5,D^\mathds{A}_7 \}$ & \  &
$\mathds{A}_{12}$:& $\{D^\mathds{A}_2,D^\mathds{A}_5,D^\mathds{A}_8\}$ & \ &
$\mathds{A}_{13}$:& $\{D^\mathds{A}_2,D^\mathds{A}_5,D^\mathds{A}_6 \}$
\end{tabular}
\end{center}
where the propagators ($D^\mathds{A}_i$) are defined below.
\begin{center}
\begin{tabular}{ccccccccccc}
$D^\mathds{A}_1$ :& \hspace{-0.7cm} $(k_1-p_a)^2$ &\hspace{0.5cm} &
$D^\mathds{A}_2$ :& \hspace{-0.7cm}$(k_1-q)^2$ & \hspace{0.5cm} &
$D^\mathds{A}_3$ :& $(k_2-p_a)^2$ & \hspace{0.5cm} &
$D^\mathds{A}_4$ :& \hspace{-0.9cm}$(k_2-q)^2$\\
$D^\mathds{A}_5$ :& $(k_1-p_a-q)^2$ & \hspace{0.5cm} &
$D^\mathds{A}_6$ :& $(k_2-p_a-q)^2$ & \hspace{0.5cm} &
$D^\mathds{A}_7$ :& $(k_1+k_2)^2$ & \hspace{0.5cm} &
$D^\mathds{A}_8$ :& $(k_1+k_2-p_a)^2$
\, .
\end{tabular}
\end{center}
\end{widetext}
An integral $J\left(\mathds{A}_n,n_1,n_2,n_3\right)$ in the family $\mathds{A}_n$ is given by
\begin{align}
J\left(\mathds{A}_n,n_1,n_2,n_3\right) = \int_{z'} \text{[dPS]}_3\frac{1}{ 
\left(D^\mathds{A}_{a_1}\right)^{n_1} \left(D^\mathds{A}_{a_2}\right)^{n_2} \left(D^\mathds{A}_{a_3}\right)^{n_3}}
\, ,
\end{align}
where the 
propagators $D^\mathds{A}_{a_1}, D^\mathds{A}_{a_2}, D^\mathds{A}_{a_3}$ are the elements
from the ordered set of family $\mathds{A}_n$. 
The solution of the IBP identities leads to the following 20 MIs,
\begin{align}
 &J\left(\mathds{A}_1;0,0,0\right),   ~J\left(\mathds{A}_1;0,0,1\right),       
 ~J\left(\mathds{A}_1;1,1,1\right),\nonumber  \\
 &J\left(\mathds{A}_2;1,1,1\right),   ~J\left(\mathds{A}_3;0,0,1\right),    ~J\left(\mathds{A}_3;1,1,1\right), \nonumber \\
 &J\left(\mathds{A}_4;1,1,1\right),   ~J\left(\mathds{A}_5;0,0,1\right),    ~J\left(\mathds{A}_5;0,0,2\right),\nonumber \\
 &J\left(\mathds{A}_5;0,1,1\right),   ~J\left(\mathds{A}_5;1,1,1\right),    ~J\left(\mathds{A}_6;0,1,1\right), \nonumber \\
 &J\left(\mathds{A}_6;1,1,1\right),    ~J\left(\mathds{A}_7;0,1,1\right),
 ~J\left(\mathds{A}_7;1,1,1\right),\nonumber\\
 &J\left(\mathds{A}_{8};1,1,1\right),      ~J\left(\mathds{A}_{10};1,1,1\right),  J\left(\mathds{A}_{11};1,1,1\right),\nonumber\\
 &J\left(\mathds{A}_{12};1,1,1\right),     J\left(\mathds{A}_{13};1,1,1\right) 
 \, .
 \label{tab:RR}
\end{align}

The VV part up to the NNLO level requires the two-loop contributions to the Born sub-process ($p_a + q \rightarrow p_b$).  All integrals are known up to desired accuracy in $a_s$ and $\varepsilon$, see \cite{Matsuura:1988sm,Matsuura:1987wt,Gehrmann:2005pd} and details of their computation are listed here for completeness. 
At two loops, one encounters two families of integrals
\begin{center}
\begin{tabular}{cc}
$\mathds{B}_1$:    $\{D^\mathds{B}_1, D^\mathds{B}_2, D^\mathds{B}_3, D^\mathds{B}_4, D^\mathds{B}_6, D^\mathds{B}_7, D^\mathds{B}_8 \}$\, , \\
$\mathds{B}_2$:    $\{D^\mathds{B}_1, D^\mathds{B}_2, D^\mathds{B}_3, D^\mathds{B}_5, D^\mathds{B}_6, D^\mathds{B}_7, D^\mathds{B}_9 \}$\, ,
\end{tabular}
\end{center}
where the propagators $D^\mathds{B}_i$ are defined as,
\begin{flushleft}
\begin{tabular}{lllllll}
$D^\mathds{B}_1$ : & $(l_1)^2$ & \hspace{3mm} &
$D^\mathds{B}_2$ : & $(l_2)^2$ &  & \\[0.5ex]
$D^\mathds{B}_3$ : & $(l_1-l_2)^2$ & & 
$D^\mathds{B}_4$ : & $(l_1+p_a)^2$ & & \\[0.5ex]
$D^\mathds{B}_5$ : & $(l_1-p_b)^2$ & &
$D^\mathds{B}_6$ : & $(l_1+p_a-p_b)^2$ & & \\[0.5ex]
$D^\mathds{B}_7$ : &$(l_2+p_a)^2$ & &
$D^\mathds{B}_8$ : & $(l_2+p_a-p_b)^2$& & \\[0.5ex]
$D^\mathds{B}_9$ : &$(l_1-l_2-p_b)^2$ & . & 
& & &
\end{tabular}
\end{flushleft}
Here, $D^\mathds{B}_7$ and $D^\mathds{B}_5$ serve as auxiliary propagators in the families $\mathds{B}_1$ and $\mathds{B}_2$, respectively.
The two-loop integrals  $J\left(\mathds{B}_n,n_1,\cdots,n_7\right)$ for the family $\mathds{B}_n$ are given by
\begin{align}
J\left(\mathds{B}_n,n_1,\cdots,n_7\right) &= \int \frac{d^d l_1 }{(2\pi)^d} \frac{d^d l_2}{(2\pi)^d} 
\frac{1}{\left(D^\mathds{B}_{a_1}\right)^{n_1}  \cdots \left(D^\mathds{B}_{a_7}\right)^{n_7}}
\, ,
\end{align}
where again the
propagators are to be taken from the ordered set of family $\mathds{B}_n$. 
An IBP reduction leads to the following 4 MIs,
\begin{center}
\begin{tabular}{lll}
  $J\left(\mathds{B}_1;0,1,1,0,1,0,0\right)$ & \hspace{0.7cm}  &  
  $J\left(\mathds{B}_1;1,1,0,0,1,0,1\right)$ \\
  $J\left(\mathds{B}_1;0,1,1,1,0,0,1\right)$ & \hspace{0.7cm}  &  $J\left(\mathds{B}_2;1,1,1,0,1,1,1\right)$\, .
\end{tabular}
\end{center}

The RV case consists of 3 families. They are given by
\begin{center}
\begin{tabular}{ll}
$\mathds{C}_1$: &$\{D^\mathds{C}_1,D^\mathds{C}_2,D^\mathds{C}_3,D^\mathds{C}_6 \} $ \\
$\mathds{C}_2$: &$\{D^\mathds{C}_1,D^\mathds{C}_2,D^\mathds{C}_3,D^\mathds{C}_5 \} $ \\
$\mathds{C}_3$: &$\{D^\mathds{C}_1,D^\mathds{C}_2,D^\mathds{C}_4,D^\mathds{C}_5 \}$\, ,
\end{tabular}
\end{center}
where the propagators $D^\mathds{C}_i$ are defined as,
\begin{center}
\begin{tabular}{lllllll}
$D^\mathds{C}_1$ :& $(l_1)^2$ & \hspace{0.5cm} &
$D^\mathds{C}_2$ :& $(l_1-p_a)^2$ & \hspace{0.5cm} & \\
$D^\mathds{C}_3$ :& $(l_1+q)^2$ & \hspace{0.5cm} &
$D^\mathds{C}_4$ :& $(l_1-p_a-q+k_1)^2$ & \hspace{0.5cm} &\\
$D^\mathds{C}_5$ :& $(l_1-p_a+k_1)^2$  & \hspace{0.5cm} &
$D^\mathds{C}_6$ :& $(l_1+q-k_1)^2$ & . &
\end{tabular}
\end{center}
We define an integral in the family $\mathds{C}_n$ as
\begin{align}
{\lefteqn{
J\left(\mathds{C}_n,n_1,\cdots,n_4\right) = }}
\nonumber \\ & 
\int \frac{d^d l_1}{(2\pi)^d} \int_{z'} \text{[dPS]}_2\frac{1}{
\left(D^\mathds{C}_{a_1}\right)^{n_1} \cdots \left(D^\mathds{C}_{a_4}\right)^{n_4}}
\, ,
\end{align}
for propagators $D^\mathds{C}_{a_1},\cdots, D^\mathds{C}_{a_4}$ taken from the ordered set of family $\mathds{C}_n$.
After IBP reduction using \texttt{LiteRed}, we obtain 7  MIs:
\begin{center}
\begin{tabular}{cccccccc}
   $J(\mathds{C}_1;0,1,1,0)$ &  &  & $J(\mathds{C}_1;1,0,0,1)$ &  &  & $J(\mathds{C}_1;1,0,1,0)$ \\
   $J(\mathds{C}_1;1,1,1,1)$ &  &  & $J(\mathds{C}_2;1,0,0,1)$ &  &  & $J(\mathds{C}_2;1,1,1,1)$ \\
   $J(\mathds{C}_3;1,1,1,1)$ &  &  &  &  &  &
\end{tabular}
\end{center}
The task to evaluate the MIs for the VV, RV and RR type  diagrams in dimensional regularization to the desired accuracy in $\varepsilon$ will be discussed next.

\subsection{Results of MIs}
The two-loop MIs for the VV case to the required accuracy in $\varepsilon$ can be found in~\cite{Matsuura:1988sm,Matsuura:1987wt,Gehrmann:2005pd}.
The MIs for the RV case have been listed in~\cite{Matsuura:1988sm}.  
We also note that these phase-space integrals have been recently discussed in a publication \cite{Bonino:2024adk}, 
which expands on the previous findings presented in \cite{Gehrmann:2022cih}.
Additionally, in \cite{Ahmed:2024owh}, we also have presented these integrals
along with the complete computational details.

Unlike the pure virtual case (VV), the RV integrals require special care when using them in the physical regions spanned by the scaling variables $x'$ and $z'$. 
We list these integrals below for arbitrary $x'$ and $z'$, and, subsequently, we express them in a form that can be used in different physical regions with the help of analytic continuation and analyticity properties of the integrals.
\begin{align}
J\left(\mathds{C}_1;0,1,1,0\right) &= -
 i\frac{2}{\varepsilon}\frac{C_{\varepsilon}}{(1+\varepsilon)}(-s)^{\varepsilon/2} \, ,\\
J(\mathds{C}_1;1,0,0,1) &= -i\frac{2}{\varepsilon}\frac{C_{\varepsilon}}{(1+\varepsilon)}(-u)^{\varepsilon/2}\, ,\\
J\left(\mathds{C}_1;1,0,1,0\right) &= -i\frac{2}{\varepsilon}\frac{C_{\varepsilon}}{(1+\varepsilon)}(Q^2)^{\varepsilon/2}\, ,\\
J\left(\mathds{C}_2;1,0,0,1\right) &= -i\frac{2}{\varepsilon}\frac{C_{\varepsilon}}{(1+\varepsilon)}(-t)^{\varepsilon/2}\, ,\\
J\left(\mathds{C}_1;1,1,1,1\right) &= -i\frac{8C_\varepsilon}{\varepsilon^2}\biggl\{\frac{(Q^2)^{\varepsilon/2}}{su} F_{\varepsilon}\bigg(\frac{Q^2t}{su}\bigg)\nonumber\\
&  \hspace{-1.3cm}
-\frac{(-u)^{\varepsilon/2}}{su}F_{\varepsilon}\bigg(\frac{-t}{s}\biggr) -\frac{(-s)^{\varepsilon/2}}{su} F_{\varepsilon}\bigg(\frac{-t}{u}\bigg)\biggr\}
\label{eq:JC1}
\, ,\\
J\left(\mathds{C}_2;1,1,1,1\right) &= -i\frac{8C_\varepsilon}{\varepsilon^2}\biggl\{\frac{(Q^2)^{\varepsilon/2}}{st} F_{\varepsilon}\bigg(\frac{Q^2u}{st}\bigg)\nonumber\\
& \hspace{-1.3cm}
-\frac{(-t)^{\varepsilon/2}}{st}F_{\varepsilon}\bigg(\frac{-u}{s}\biggr)-\frac{(-s)^{\varepsilon/2}}{st} F_{\varepsilon}\bigg(\frac{-u}{t}\bigg)\biggr\}
\label{eq:JC2}
\, ,\\
 J\left(\mathds{C}_3;1,1,1,1\right) &= -i\frac{8C_\varepsilon}{\varepsilon^2}\biggl\{\frac{(Q^2)^{\varepsilon/2}}{tu} F_{\varepsilon}\bigg(\frac{Q^2s}{tu}\bigg)\nonumber\\
& \hspace{-1.7cm}
-\frac{(-u)^{\varepsilon/2}}{tu}F_{\varepsilon}\bigg(\frac{-s}{t}\biggr)-\frac{(-t)^{\varepsilon/2}}{tu} F_{\varepsilon}\bigg(\frac{-s}{u}\bigg)\biggr\}
\label{eq:JC3}
\, ,
\end{align}
where,
\begin{align}
&{F}_{\varepsilon}(\xi) = {}_2F_1\left(1,\tfrac{\varepsilon}{2};1+\tfrac{\varepsilon}{2};\xi\right) 
= 1-\sum_{n=1}^{\infty} \left(-\tfrac{\varepsilon}2\right)^n  \text{Li}_n(\xi)\, ,
\nonumber\\
&{C_{\varepsilon} =\frac{1}{16\pi^2}\left(\frac{\Gamma{(1-\frac{\varepsilon}{2})}\Gamma^2{(1+\frac{\varepsilon}{2})}}{ (4\pi)^{\frac{\varepsilon}{2}}\Gamma{(1+\varepsilon)}}\right)} \,,
\nonumber\\
&s = \frac{Q^2(1-x')}{x'},t = \frac{-Q^2(1-z')}{x'} , u = \frac{-Q^2 z'}{x'}
\,.
\end{align}
In the above formulas we imply that $$(-s)^{\varepsilon/2}=(-s-i0)^{\varepsilon/2}=e^{-i\pi\varepsilon/2}s^{\varepsilon/2}.$$

The expressions \eqref{eq:JC1}--\eqref{eq:JC3} contain hypergeometric functions with arguments which can be larger than one when $x'\geqslant \mathrm{min}(z',1-z')$.
Therefore, in order to extend the applicability of these formulas to the whole physical region, one should perform an analytical continuation.
Fortunately, it is easy to rewrite eqs.~\eqref{eq:JC1}--\eqref{eq:JC3} in the form which is explicitly analytic in the whole physical region. For this purpose, we represent $\text{F}_{\varepsilon}(\xi)$ for positive $\xi$ as,
\begin{multline}\label{eq:tildeF}
	\text{F}_{\varepsilon}(\xi)=\frac{\varepsilon ^2(1-\xi)}{4\xi ^{1+\frac{\varepsilon }{2}} }  \widetilde{F}_{\varepsilon}(1-\xi^{-1})
	\\-\frac{\varepsilon}{2\xi^{\varepsilon /2}}    \left[\psi \left({\varepsilon }/{2}\right)+\ln \left(\xi^{-1}-1\right)+\gamma_{E} \right]\,,
\end{multline}
where $\widetilde{F}_{\varepsilon}(z)={}_3F_2\left(1,1,1+\varepsilon/2;2,2;z\right)$, $\psi(x)=\Gamma'(x)/\Gamma(x)$, and $\gamma_{E}=-\psi(1)=0.577\ldots$ is the Euler constant.

Note that the only function in the right-hand side of eq.~\eqref{eq:tildeF} which has branching point on the interval $(0,+\infty)$ at $\xi=1$ is $\ln \left(\xi^{-1}-1\right)$. 
Then it is easy to check that the branching logarithms cancel in eqs.~\eqref{eq:JC1}--\eqref{eq:JC3}. 
We have
\begin{widetext}
\begin{align}
J\left(\mathds{C}_1;1,1,1,1\right)
&=
-\frac{2i C_{\varepsilon }  }{(Q^2)^{2-\varepsilon/2}}\left(\frac{(1-{z'}){x'} }{(1-{x'}) {z'}}\right)^{1-\frac{\varepsilon }{2}}
\bigg\{\frac{ {x'}-{z'}}{(1-{z'})^2}\left[
\widetilde{F}_{\varepsilon }\left(\frac{{x'}-{z'}}{ (1-{z'}){x'}}\right)
-{x'}\widetilde{F}_{\varepsilon }\left(\frac{{x'}-{z'}}{1-{z'}}
\right)\right]
\nonumber\\
& \hspace{5cm}
+\frac{4e^{-\frac{i \pi  \varepsilon }{2} } {x'} (1-{z'})^{\varepsilon /2-1} }{\varepsilon ^2{z'}^{\varepsilon /2}} F_{\varepsilon }\left(\frac{{z'}-1}{{z'}}\right)
-\frac{2{x'} \ln ({x'})}{\varepsilon  (1-{z'})}\bigg\}
\, ,
\\
J\left(\mathds{C}_2;1,1,1,1\right)
&= J\left(\mathds{C}_1;1,1,1,1\right)\vert_{z'\to1-z'}
\, ,
\\
J\left(\mathds{C}_3;1,1,1,1\right)
&=
-\frac{2 i C_{\varepsilon }  }{(Q^2)^{2-\varepsilon/2}}\left(\frac{\left(1-x'\right) x'}{\left(1-z'\right) z'}\right)^{1-\frac{\varepsilon }{2}}
\bigg\{
\frac{x' \left(1-z'-x'\right)}{\left(1-x'\right)^2} \, \widetilde{F}_\varepsilon\left(\frac{1-z'-x'}{1-x'}\right)
+\frac{x' \left(z'-x'\right)}{\left(1-x'\right)^2} \, \widetilde{F}_\varepsilon\left(\frac{z'-x'}{1-x'}\right)
\nonumber\\
& \hspace{-2cm}
+\frac{\left(z'-x'\right) \left(1-z'-x'\right)}{\left(1-x'\right)^2} \, \widetilde{F}_\varepsilon\left(\frac{\left(z'-x'\right) \left(1-z'-x'\right)}{\left(x'-1\right) x'}\right)
+\frac{2  x' }{\varepsilon \left(1-x'\right)}
\left[\psi\left({\varepsilon }/{2}\right)+\gamma_E -\ln \left({x'}^{-1}-1\right)\right]
\bigg\}
\, .
\end{align}
These new expressions are explicitly analytic in the whole physical region.
It is well known, that the 
hypergeometric function $\widetilde{F}_{\varepsilon}(\xi)$ can be expanded in a power series in $\varepsilon$ in terms of 
generalized~\cite{Goncharov:1998kja} or harmonic~\cite{Remiddi:1999ew} polylogarithms using,
\begin{equation}
	{}_3F_2\left(1,1,1+\varepsilon/2;2,2;z\right)=
	-\frac1{z}\sum_{n=0}^\infty\left(-\varepsilon/2\right)^n G(\underbrace{0,1,\ldots, 1}_{n+2}|z)=
	\frac1{z}\sum_{n=0}^\infty\left(\varepsilon/2\right)^n H(2,\underbrace{1,\ldots, 1}_{n}|z)
\, .
\end{equation}
\end{widetext}

Our next task is to perform the mass factorization
for the partonic cross sections $d(\Delta) \hat{\sigma}_{i,ab}$ in eq.~(\ref{eq:massfact}) to obtain finite CFs. 
This removes collinear singularities arising from radiations off the incoming parton as well as off the fragmenting parton. 
The AP kernels $(\Delta) \Gamma_{c\leftarrow a}$ and
$\tilde \Gamma_{b\leftarrow d}$ are pure counter-terms, containing only poles in $\varepsilon$ which multiply '+'-distributions ${\cal D}_j(w)=(\ln^j(1-w)/(1-w))_+$, delta functions $\delta(1-w)$,
and regular terms in $w$, where $w=x',z'$ (see, e.g.~\cite{Goyal:2023zdi}).
In order to cancel the collinear singularities in $d (\Delta )\hat{\sigma}_{1,ab}$ against those from AP kernels, 
we need to extract from the former ones the  poles in $\varepsilon$ in terms of the same `+'-distributions and regular functions.

The IR divergences in the partonic sub-cross section show up as $(1-x')^{-1}$ and/or $(1-z')^{-1}$. 
These terms originate either from MIs or their coefficients at the level of squared matrix elements and diverge in the respective threshold regions $x'\rightarrow 1$ and/or $z' \rightarrow 1$. 
In $d=4+\varepsilon$ dimensions,
they are regulated by $(1-x')^{a \varepsilon}$ and $(1-z')^{b\varepsilon}$ respectively which originate from phase space and loop integrals.

At NNLO, we encounter IR singularities in sub-processes for the RR case through terms of the form
\begin{equation}
\frac{(1-x')^{  \varepsilon} (1-z')^{  \varepsilon}}
{(1-x')(1-z')(z'-x')((1+x')^2-4 x' z')} f_1(x',z',\varepsilon)
\, .
\end{equation}
After partial fractioning, it is easy to see that these terms contain only simple poles at the thresholds i.e.\ $x' \rightarrow 1$ and $z' \rightarrow 1$ when $\varepsilon \rightarrow 0$.
Similarly, in integrals of RV type we find IR singular terms of the form
\begin{align}
\frac{(1-x')^{a \varepsilon}   (1-z')^{b \varepsilon}|z'-x'|^{c \varepsilon}|1-z'-x'|^{d \varepsilon} }{(1-x') (1-z')(z'-x')(1-z'-x')} f_2(x',z',\varepsilon)
\, .\nonumber
\end{align}
Here, $a,b$ can take the values $1/2$ or $1$ and $c,d$ can be $0$ or $-1/2$, while 
$f_i,i=1,2$ are regular in all their arguments.

After partial fractioning, we end up with the following combinations of denominators $\frac{1}{(1-x')(1-z')}$, $\frac{1}{(1-x')(z'-x')}$, $\frac{1}{(1-x')((1+x')^2-4x'z')}$, $\frac{1}{(1-x')}$, $\frac{1}{(1-z')}$ etc.

The extraction of the pole structure proceeds sequentially, first separating the singularity at $x'\rightarrow 1$ and then $z'\rightarrow 1$, or vice versa. 
For the RV case, the following form of IR singular terms is suitable for mass factorization.
\begin{widetext}
\begin{align}
\label{eq:etract-IRpoles}
\frac{(1-x')^{a\varepsilon}(1-z')^{b\varepsilon}|z'-x'|^{c\varepsilon}}{(1-x')(1-z')} f(x',z')&=  \frac{(1-x')^{a\varepsilon}_E}{1-x'}\frac{(1-z')^{b\varepsilon}_E}{1-z'}\bigg(|z'-x'|^{c\varepsilon}_E ~ f(x',z') -(1-x')^{c\varepsilon}_E f(x',1) \nonumber \\
& -(1-z')^{c\varepsilon}_E \big(f(1,z')-f(1,1) \big)\bigg)  +\frac{(1-x')^{(a+c)\varepsilon}_E}{1-x'}\bigg[\frac{(1-z')^{b\varepsilon}}{1-z'}\bigg]_P f(x',1) \nonumber \\
& + \frac{(1-x')^{a\varepsilon}_E}{1-x'}\bigg[\frac{(1-z')^{(b+c)\varepsilon}}{1-z'}\bigg]_P f(1,1)  +  \bigg[\frac{(1-x')^{a\varepsilon}}{1-x'}\bigg]_P\bigg[\frac{(1-z')^{(b+c)\varepsilon}}{1-z'}\bigg]_P f(1,1)  \nonumber\\
&+ \bigg[\frac{(1-x')^{a\varepsilon}}{1-x'}\bigg]_P\frac{(1-z')^{(b+c)\varepsilon}_E}{1-z'}\big(f(1,z')-f(1,1)\big) \nonumber \\
\nonumber \\
\frac{(1-x')^{a\varepsilon}(1-z')^{b\varepsilon}|z'-x'|^{c\varepsilon}}{(1-x')(z'-x')} f(x',z') &=
\frac{(1-x')^{a\varepsilon}_E}{1-x'}\frac{(1-z')^{b\varepsilon}_E}{z'-x'}~\big|z'-x'\big|^{c\varepsilon}_E ~ f(x',z') \nonumber\\
&+\frac{(1-x')^{a\varepsilon}_E}{1-x'}\frac{(1-z')^{(b+c)\varepsilon}_E}{1-z'}\big(f(1,z')-f(1,1)\big) \nonumber\\
&-\bigg[\frac{(1-x')^{a\varepsilon}}{1-x'}\bigg]_P\frac{(1-z')^{(b+c)\varepsilon}_E}{1-z'}\big(f(1,z')-f(1,1)\big) \nonumber \\
& +\frac{(1-x')^{a\varepsilon}_E}{1-x'}\bigg[\frac{(1-z')^{(b+c)\varepsilon}}{1-z'}\bigg]_P f(1,1) -\bigg[\frac{(1-x')^{a\varepsilon}}{1-x'}\bigg]_P\bigg[\frac{(1-z')^{(b+c)\varepsilon}}{1-z'}\bigg]_P f(1,1)
\end{align}
\end{widetext}
In the RR case, the numerator $|z'-x'|^{c \varepsilon}$ is absent. 
Hence, we use the same identities after setting $c=0$ in eq.~\eqref{eq:etract-IRpoles}.
The subscripts $P$ and $E$ in eq.~\eqref{eq:etract-IRpoles} imply the shorthands as follows
\begin{align}
(\kappa)^{\eta\varepsilon}_E &= \sum_{i=0}^{\infty}\frac{(\eta\varepsilon)^i}{i!}\ln^{i}(\kappa) \, ,\\
\bigg[\frac{(1-\kappa)^{\eta\varepsilon}}{1-\kappa}\bigg]_P &= \frac{1}{\eta\varepsilon}\delta(1-\kappa) + \bigg[\frac{(1-\kappa)^{\eta\varepsilon}_E}{1-\kappa}\bigg]_{+} \nonumber\\
&= \frac{1}{\eta\varepsilon}\delta(1-\kappa) + \sum_{i=0}^{\infty}\frac{(\eta\varepsilon)^{i}}{i!}
\bigg[\frac{\ln^{i}(1-\kappa)}{1-\kappa}\bigg]_{+}
\, ,
\end{align}
where the expansions should be properly truncated, since the IR singularities show up explicitly as poles in $\varepsilon$.

\section{Mass Factorization}
\label{sec:massfact}
At this stage all UV and IR divergences in the perturbative expansion of the partonic cross sections have been extracted as poles in $\varepsilon$ and we use the short-hand $a_s = {\alpha_s}/(4\pi)$.
The UV divergences present in the loop corrections (VV and RV-contributions) are removed by renormalization of the strong coupling,
\begin{equation*}
\hat{a}_s S_{\varepsilon} \bigg(\frac{1}{\mu^2}\bigg)^{\varepsilon/2} = a_s{(\mu_R^2)} \bigg(\frac{1}{\mu_R^2}\bigg)^{\varepsilon/2} Z_a(\mu_R^2)
\, ,
\end{equation*}
where the renormalization constant $Z_a$ to order $a_s$ is given by
\begin{equation}
 Z_a(\mu_R^2)= 1+ a_s(\mu_R^2) \bigg(\frac{2\beta_0}{\varepsilon} \bigg) 
 + O(a_s^2)\, ,
\end{equation}
with the one-loop coefficient of the QCD beta-function 
$\beta_{0} = \Big(\frac{11}{3} C_A - \frac{2}{3} n_f\Big)$.

The RV and RR sub-processes involving emissions of on-shell partons contain soft and collinear divergences, with the soft ones canceling in the sum of all real emission and virtual contributions. 
The collinear divergences due to emissions from initial(final)-state partons are removed by space(time)-like AP-kernels. 
We apply eq.~(\ref{eq:massfact}) to extract the collinear finite CFs $(\Delta){\cal C}_i$ order by order in perturbation theory from the partonic sub-process cross sections computed in the previous section. 
The AP kernels can be expanded in powers of $a_s$ as,
\begin{align}
\label{eq:APkernels}
\BbbGamma_{c\la d}(\xi,\mu_F^2,\varepsilon)&=\delta_{cd}\delta(1-\xi)+   \sum_{l=1}^{\infty}a_s^l(\mu_F^2)~\BbbGamma^{(l)}_{c\la d}(\xi,\varepsilon)
\, ,
\end{align}
where $\BbbGamma$ denotes the set $\BbbGamma \in \{ \Gamma, \Delta \Gamma, \tilde\Gamma\}$, subject to a  renormalization group evolution equation,
\begin{align}
\mu_F^2 \frac{d }{d\mu_F^2}\BbbGamma_{c\la d} = \frac{1}{2}
\mathds{P}_{c e}(a_s(\mu_F^2))\otimes \BbbGamma_{e\la d}
\, ,\nonumber
\end{align}
where we abbreviate the splitting functions collectively by the set $\mathds{P} \in \{ P, \Delta P, \tilde P \}$.
Their perturbative expansion reads $\mathds{P}_{ce} =  \sum_{l=1}^{\infty} a_s^{l}(\mu_F^2) \mathds{P}_{ce}^{(l-1)}$
and they are all available in the literature to the required order and beyond, cf.\ e.g.\
refs.~\cite{Moch:2004pa,Vogt:2004mw} for the space-like unpolarized splitting functions $P$, 
refs.~\cite{Mertig:1995ny,Vogelsang:1995vh,Vogelsang:1996im,Moch:2014sna,Blumlein:2021enk,Blumlein:2021ryt,Blumlein:2022gpp} for the space-like polarized ones $\Delta P$, 
and refs.~\cite{Almasy:2011eq,Chen:2020uvt} for the time-like unpolarized ones $\tilde P$.

The AP kernels $\BbbGamma^{(l)}_{c\la d}$ at NLO ($l=1$ in eq.~(\ref{eq:APkernels})) are given by,
\begin{center}
\begin{tabular}{lll}
  $\BbbGamma_{q\la q}^{(1)}$ = $\frac{1}{\varepsilon} {\mathds{P}}^{(0)}_{qq},$   &  \hspace{0.6cm} & $\BbbGamma_{g\la g}^{(1)} = \frac{1}{\varepsilon} {\mathds{P}}^{(0)}_{gg},$\\
  & & \\
$\BbbGamma_{\overline{q}\la g}^{(1)} = \BbbGamma_{q\la g}^{(1)} = \frac{1}{\varepsilon} {\mathds{P}}^{(0)}_{qg}$, &  \hspace{0.6cm} &
$\BbbGamma_{q\la \overline{q}}^{(1)} = 0, $\\
& & \\
$\BbbGamma_{g\la \overline{q}}^{(1)} =\BbbGamma_{g\la q}^{(1)} = \frac{1}{\varepsilon} {\mathds{P}}^{(0)}_{gq},$ & \hspace{0.6cm}  &
\end{tabular}
\end{center}
and at NNLO ($l=2$ in eq.~(\ref{eq:APkernels})),
\begin{center}
  \begin{tabular}{lllll}
  $\BbbGamma_{q\la q}^{(2)} = \BbbGamma_{q\la q}^{(2),\text{S}} + \BbbGamma_{q\la q}^{(2),\text{NS}},$     & \hspace{0.3cm} & $\BbbGamma_{ \overline{q}\la q }^{(2)} =\BbbGamma_{q\la \overline{q} }^{(2)},$ \\
  & &\\
 $\BbbGamma_{q\la \overline{q}}^{(2)} = \BbbGamma_{q\la \overline{q}}^{(2),\text{S}} + \BbbGamma_{q\la \overline{q}}^{(2),\text{NS}},$ & \hspace{0.3cm} & $\BbbGamma_{q\la \overline{q}}^{(2),\text{S}} =\BbbGamma_{q\la q}^{(2),\text{S}},$  \\
   & &\\
 $\BbbGamma_{\overline{q}'\la q}^{(2)} = \BbbGamma_{q'\la q}^{(2)} = \BbbGamma_{q\la q}^{(2),\text{S}},$&  &  \text{ \{for $q'\neq q$\} }
  \end{tabular}
\end{center}
and are related to the splitting functions as,
\begin{widetext}
\begin{align*}
\BbbGamma_{q\la \overline{q}}^{(2),\text{NS}} &= \frac{1}{2\varepsilon} {\mathds{P}}^{(1),-}_{qq},
\hspace{1.6cm}
\BbbGamma_{q\la q}^{(2),\text{S}} = \frac{1}{2\varepsilon}{\mathds{P}}_{qq}^{(1),\text{S}} +\frac{1}{2\varepsilon^2}\biggl( {\mathds{P}}_{qg}^{(0)}\otimes {\mathds{P}}_{gq}^{(0)}  \biggr), \nonumber\\
\BbbGamma_{q\la q}^{(2),\text{NS}} &= \frac{1}{2\varepsilon}{\mathds{P}}_{qq}^{(1),\text{NS}} +\frac{1}{2\varepsilon^2}\bigg( 2\beta_{0}{\mathds{P}}_{qq}^{(0)} +{\mathds{P}}_{qq}^{(0)}\otimes {\mathds{P}}_{qq}^{(0)}   \bigg) ,\nonumber\\
\BbbGamma_{g\la g}^{(2)} &=
\frac{1}{2\varepsilon}{\mathds{P}}^{(1)}_{gg} +
\frac{1}{2\varepsilon^2}\biggl(2\beta_{0} {\mathds{P}}_{gg}^{(0)} + 2 n_f\Big({\mathds{P}}_{gq}^{(0)}\otimes {\mathds{P}}_{qg}^{(0)}\Big)+{\mathds{P}}_{gg}^{(0)}\otimes {\mathds{P}}_{gg}^{(0)} \biggr), \nonumber\\
\BbbGamma_{\overline{q}\la g}^{(2)} = \BbbGamma_{q\la g}^{(2)} &= \frac{1}{2\varepsilon}{\mathds{P}}^{(1)}_{qg} + \frac{1}{2\varepsilon^2}\biggl(2\beta_{0} {\mathds{P}}_{qg}^{(0)} +{\mathds{P}}_{qg }^{(0)}\otimes {\mathds{P}}_{gg}^{(0)}+{\mathds{P}}_{qq}^{(0)}\otimes {\mathds{P}}_{ qg}^{(0)} \biggr), \nonumber\\
\BbbGamma_{g\la \overline{q}}^{(2)} = \BbbGamma_{g\la q}^{(2)} &= \frac{1}{2\varepsilon}{\mathds{P}}^{(1)}_{gq} +\frac{1}{2\varepsilon^2}\biggl(2\beta_{0} {\mathds{P}}_{gq}^{(0)} +{\mathds{P}}_{gg}^{(0)}\otimes {\mathds{P}}_{gq}^{(0)}
+{\mathds{P}}_{gq }^{(0)}\otimes {\mathds{P}}_{qq}^{(0)} \biggr) \, . \nonumber
\end{align*}
\end{widetext}
The required expressions for the splitting functions ${\mathds{P}}^{(l)}_{ab}$ up to two loops are collected in Appendixes~\ref{Appendix:A}, \ref{Appendix:B} and \ref{Appendix:C}. 
The perturbative expansion of the parton level cross sections $(\Delta)\hat{\mathcal{C}}_{i,ab}$ (the short notation for $x^{1-i}d (\Delta) \hat \sigma_{i,ab}/dx dy dz$, $i=1,2$), i.e.\ the left-hand side of eq.~(\ref{eq:massfact}) reads, 
\begin{align}
(\Delta)\hat{\mathcal{C}}_{i,ab}&=(\Delta)\hat{\mathcal{C}}_{i,ab}^{(0)}+  a_s(\mu_R^2)  \left(\frac{Q^2}{\mu_R^2}\right)^{\frac{\varepsilon}{2}}(\Delta)\hat{\mathcal{C}}_{i,ab}^{(1)}\nonumber\\
&+ a_s^2(\mu_R^2) \left(\frac{2\beta_0}{\varepsilon}\right)  \left(\frac{Q^2}{\mu_R^2}\right)^{\frac{\varepsilon}{2}} (\Delta)\hat{\mathcal{C}}_{i,ab}^{(1)}\nonumber\\
&+ a^2_s(\mu_R^2) \left(\frac{Q^2}{\mu_R^2}\right)^{\varepsilon} (\Delta)\hat{\mathcal{C}}_{i,ab}^{(2)}+O(a_s^3(\mu_F^2))\, .
\end{align}
Upon substitution of the AP kernels $\BbbGamma^{(l)}_{c\la d}$ in eq.~(\ref{eq:massfact}) we can extract order by order the finite CFs $(\Delta) {\cal C}_i$ entering eq.~(\ref{eq:StrucCoeff}),
\begin{align}
(\Delta){\mathcal{C}}_{i,ab}&=(\Delta){\mathcal{C}}_{i,ab}^{(0)}+a_s(\mu_F^2)~(\Delta){\mathcal{C}}_{i,ab}^{(1)}\nonumber\\
&+  a_s^2(\mu_F^2)~(\Delta){\mathcal{C}}_{i,ab}^{(2)} + O(a_s^3(\mu_F^2))
\, .
\end{align}
For equal scales, setting $\mu_R=\mu_F$, the CFs $(\Delta){\mathcal{C}}_{i,ab}^{(l)}$ for $l=0,1,2$ are found to be the following:
\begin{widetext}
\begin{align} \label{eq:CoeffNiLO}
&\hspace{-1.2cm}\text{at LO ($l = 0$):}\nonumber\\ \nonumber\\
&\hspace{-1.2cm}\text{ $q(\overline{q})\rightarrow q(\overline{q})$}\nonumber\\
& (\Delta){\mathcal{C}}_{i,qq}^{(0)}(x',z')=(\Delta)\hat{\mathcal{C}}_{i,qq}^{(0)}(x',z') =\delta(1-x')~\delta(1-z'),\nonumber\\  \nonumber\\
&\hspace{-1.2cm} \text{at NLO ($l =1$):}\nonumber\\ \nonumber\\
&\hspace{-1.2cm}\text{ $q (\overline{q}) \rightarrow q(\overline{q})$ }\nonumber\\
& (\Delta){\mathcal{C}}_{i,qq}^{(1)}(x',z') = \left(\frac{Q^2}{\mu_F^2}\right)^{\frac{\epsilon}{2}} (\Delta)\hat{\mathcal{C}}_{i,qq}^{(1)}(x',z') -\delta(1-x')~\tilde{\Gamma}_{q\la q}^{(1)}(z')- (\Delta){\Gamma}_{q \la q}^{(1)}(x')~\delta(1-z'),\nonumber\\
&\hspace{-1.2cm}\text{ $q(\overline{q})\rightarrow g$}\nonumber\\
&(\Delta){\mathcal{C}}_{i,qg}^{(1)}(x',z') = \left(\frac{Q^2}{\mu_F^2}\right)^{\frac{\epsilon}{2}}
(\Delta)\hat{\mathcal{C}}_{i,qg}^{(1)}(x',z')
-\delta(1-x')~\tilde{\Gamma}_{g\la q}^{(1)}(z'),\nonumber\\
&\hspace{-1.2cm}\text{ $g \rightarrow q(\overline{q})$}\nonumber\\
&(\Delta){\mathcal{C}}_{i,gq}^{(1)}(x',z')
=\left(\frac{Q^2}{\mu_F^2}\right)^{\frac{\epsilon}{2}}
(\Delta)\hat{\mathcal{C}}_{i,gq}^{(1)}(x',z')- (\Delta){\Gamma}_{q\la g}^{(1)}(x')~\delta(1-z'),\nonumber\\  \nonumber\\
&\hspace{-1.2cm}\text{at NNLO ($l=2$):} \nonumber\\ \nonumber\\
&\hspace{-1.2cm}\text{ $q(\overline{q})\rightarrow q(\overline{q}) $}\nonumber\\
&(\Delta){\mathcal{C}}^{(2)}_{i,qq}(x',z')	=\left(\frac{Q^2}{\mu_F^2}\right)^{\varepsilon}(\Delta)\hat{\mathcal{C}}_{i,qq}^{(2)}(x',z')+  \frac{2\beta_0}{\varepsilon} \left(\frac{Q^2}{\mu_F^2}\right)^{\frac{\varepsilon}{2}}(\Delta)\hat{\mathcal{C}}_{i,qq}^{(1)}(x',z')  -(\Delta){\mathcal{C}}^{(1)}_{i,qq}(x',z')\tilde \otimes\tilde{\Gamma}^{(1)}_{q\la q}(z') \nonumber\\
&\hspace{2.3cm} - (\Delta){\mathcal{C}}^{(1)}_{i,qg}(x',z')\tilde\otimes\tilde{\Gamma}^{(1)}_{q\la g}(z')-\delta(1-x')~\tilde{\Gamma}^{(2)}_{q\la q}(z') -(\Delta)\Gamma^{(2)}_{q\la q}(x')~\delta(1-z')\nonumber\\
&\hspace{2.3cm} -(\Delta)\Gamma^{(1)}_{q\la q}(x')~\tilde{\Gamma}^{(1)}_{q\la q}(z')-(\Delta)\Gamma^{(1)}_{q\la q}(x')\otimes(\Delta){\mathcal{C}}^{(1)}_{i,qq}(x',z')  \nonumber\\
&\hspace{2.3cm}- (\Delta)\Gamma^{(1)}_{g\la q}(x')\otimes(\Delta){\mathcal{C}}^{(1)}_{i,gq}(x',z'),\nonumber\\
&\hspace{-1.2cm}\text{ $q(\overline{q})\rightarrow g$}\nonumber\\
&(\Delta){\mathcal{C}}^{(2)}_{i,qg}(x',z') =\left(\frac{Q^2}{\mu_F^2}\right)^{\varepsilon}(\Delta)\hat{\mathcal{C}}_{i,qg}^{(2)}(x',z')+                 \frac{2\beta_0}{\varepsilon}\left(\frac{Q^2}{\mu_F^2}\right)^{\frac{\varepsilon}{2}}(\Delta)\hat{\mathcal{C}}_{i,qg}^{(1)} (x',z')-(\Delta){\mathcal{C}}^{(1)}_{i,qq}(x',z')\tilde\otimes\tilde{\Gamma}^{(1)}_{g\la q}(z') \nonumber\\
&\hspace{2.3cm} -(\Delta){\mathcal{C}}^{(1)}_{i,qg}(x',z')\tilde\otimes\tilde{\Gamma}^{(1)}_{g\la g}(z')  -\delta(1-x')~\tilde{\Gamma}^{(2)}_{g\la q}(z')  -(\Delta)\Gamma^{(1)}_{q\la q}(x')\otimes(\Delta){\mathcal{C}}^{(1)}_{i,qg}(x',z')\nonumber\\
&\hspace{2.3cm} -(\Delta)\Gamma^{(1)}_{q\la q}(x')~\tilde{\Gamma}^{(1)}_{g\la q}(z'),
\nonumber\\
&\hspace{-1.2cm}\text{ $q(\overline{q})\rightarrow \overline{q}(q)$}\nonumber\\
&(\Delta){\mathcal{C}}^{(2)}_{i,q\overline{q}}(x',z') =\left(\frac{Q^2}{\mu_F^2}\right)^{\varepsilon}(\Delta)\hat{\mathcal{C}}_{i,q\overline{q}}^{(2)}(x',z') -(\Delta){\mathcal{C}}^{(1)}_{i,qg}(x',z' )\tilde\otimes\tilde{\Gamma}^{(1)}_{\overline{q}\la g}(z') - \delta(1-x')~\tilde{\Gamma}^{(2)}_{\overline{q}\la q}(z')  \nonumber\\
&\hspace{2.3cm} -(\Delta)\Gamma^{(1)}_{g\la q}(x')\otimes(\Delta){\mathcal{C}}^{(1)}_{i,g\overline{q}}(x',z')-(\Delta)\Gamma^{(2)}_{\overline{q}\la q}(x')~\delta(1-z'),
\nonumber\\
&\hspace{-1.2cm}\text{ $q(\overline{q}) \rightarrow q'(\overline{q}'),$ \quad for \{$q\neq q'$\}}\nonumber\\
&(\Delta){\mathcal{C}}^{(2)}_{i,qq'}(x',z')=\left(\frac{Q^2}{\mu_F^2}\right)^{\varepsilon}
(\Delta)\hat{\mathcal{C}}_{i,qq'}^{(2)}(x',z')
-(\Delta){\mathcal{C}}^{(1)}_{i,qg}(x',z')\tilde\otimes\tilde{\Gamma}^{(1)}_{q\la g}(z') - \delta(1-x')~\tilde{\Gamma}^{(2)}_{q'\la q}(z') \nonumber\\
&\hspace{2.3cm} -(\Delta)\Gamma^{(1)}_{g\la q}(x')\otimes(\Delta){\mathcal{C}}^{(1)}_{i,gq}(x',z') -(\Delta)\Gamma^{(2)}_{q'\la q}(x')~\delta(1-z'),\nonumber\\
&\hspace{-1.2cm}\text{ $q(\overline{q}) \rightarrow \overline{q}'(q'),$ \quad for \{$q\neq q'$\}   }  \nonumber\\
&(\Delta){\mathcal{C}}^{(2)}_{i,q\overline{q}'}(x',z')=\left(\frac{Q^2}{\mu_F^2}\right)^{\varepsilon}
(\Delta)\hat{\mathcal{C}}_{i,q\overline{q}'}^{(2)}(x',z')
-(\Delta){\mathcal{C}}^{(1)}_{i,qg}(x',z')\tilde\otimes\tilde{\Gamma}^{(1)}_{\overline{q}\la g}(z') - \delta(1-x')~\tilde{\Gamma}^{(2)}_{\overline{q}'\la q}(z') \nonumber\\
&\hspace{2.3cm} -(\Delta)\Gamma^{(1)}_{g\la q}(x')\otimes(\Delta){\mathcal{C}}^{(1)}_{i,g\overline{q}}(x',z') -(\Delta)\Gamma^{(2)}_{\overline{q}'\la q}(x')~\delta(1-z'),\nonumber\\
&\hspace{-1.2cm}\text{ $g\rightarrow q(\overline{q})$} \nonumber\\
&(\Delta){\mathcal{C}}^{(2)}_{i,gq}(x',z')   =\left(\frac{Q^2}{\mu_F^2}\right)^{\varepsilon} (\Delta)\hat{\mathcal{C}}_{i,gq}^{(2)}(x',z')+  \frac{2\beta_0}{\varepsilon} \left(\frac{Q^2}{\mu_F^2}\right)^{\frac{\varepsilon}{2}}(\Delta)\hat{\mathcal{C}}_{i,gq}^{(1)}(x',z') -(\Delta){\mathcal{C}}^{(1)}_{i,gq}(x',z')\tilde\otimes\tilde{\Gamma}^{(1)}_{q\la q}(z')  \nonumber\\
&\hspace{2.3cm}- (\Delta)\Gamma^{(1)}_{q\la g}(x')~\tilde{\Gamma}^{(1)}_{q\la q}(z') -(\Delta)\Gamma^{(1)}_{q\la g}(x')\otimes(\Delta){\mathcal{C}}^{(1)}_{i,qq}(x',z')\nonumber\\
& \hspace{2.3cm}-    (\Delta)\Gamma^{(1)}_{g\la g}(x')\otimes(\Delta){\mathcal{C}}^{(1)}_{i,gq}(x',z') -(\Delta)\Gamma^{(2)}_{q\la g}(x')~\delta(1-z'),\nonumber\\
&\hspace{-1.2cm}\text{ $g \rightarrow g$}\nonumber\\
&(\Delta){\mathcal{C}}^{(2)}_{i,gg}(x',z')=\left(\frac{Q^2}{\mu_F^2}\right)^{\varepsilon}
(\Delta)\hat{\mathcal{C}}_{i,gg}^{(2)}(x',z')
-(\Delta){\mathcal{C}}^{(1)}_{i,gq}(x',z')\tilde\otimes\tilde{\Gamma}^{(1)}_{g\la q}(z') -(\Delta){\mathcal{C}}^{(1)}_{i,g\overline{q}}(x',z')\tilde\otimes\tilde{\Gamma}^{(1)}_{g\la \overline{q}}(z')\nonumber\\
& \hspace{2.3cm} -(\Delta)\Gamma^{(1)}_{q\la g}(x')~\tilde{\Gamma}^{(1)}_{g\la q}(z') - (\Delta)\Gamma^{(1)}_{\overline{q}\la g}(x')~\tilde{\Gamma}^{(1)}_{g\la \overline{q}}(z')  \nonumber \\
&\hspace{2.3cm}-(\Delta)\Gamma^{(1)}_{q\la g}(x')\otimes(\Delta){\mathcal{C}}^{(1)}_{i,qg}(x',z') - (\Delta)\Gamma^{(1)}_{\overline{q}\la g}(x')\otimes(\Delta){\mathcal{C}}^{(1)}_{i,\overline{q}g}(x',z').
\end{align}

\end{widetext}
Mass factorization for the polarized SF $g_1$ needs additional care due to the prescription for $\gamma_5$ in Larin's scheme, which requires the use of the spin-dependent AP kernels computed in same scheme.
Given that the hadronic cross sections (here, $g_1$) are scheme independent, an additional finite scheme transformation is in order to convert the AP kernels and the finite CF from Larin's to the conventional $\overline{ \rm{MS}}$ scheme.  
With PDFs $\Delta f_{a}^{L}$ and CF $\Delta {\cal C}_{1,ab}^{L}(\mu_F^2)$ defined in Larin's scheme (denoted by the superscript $L$), the mass factorization for $g_1$ reads
\begin{align}
\label{eq:g1L}
g_1 = \sum_{a,b} \Delta f_{a}^{L}(\mu_F^2)\otimes \Delta {\cal C}_{1,ab}^{L}(\mu_F^2) \tilde \otimes D_b(\mu_F^2)
\, .
\end{align}
Here the FFs $D_b$ and the time-like AP kernels ($\tilde \Gamma_{c\leftarrow d}$) for handling the final state collinear singularities are already taken in the $\overline{\rm{MS}}$ scheme, since the final state spins are summed.
With a scheme transformation~\cite{Moch:2014sna} through the finite renormalization constants $Z_{ab}$, 
known in the literature~\cite{Matiounine:1998re,Ravindran:2003gi,Moch:2014sna}, for PDFs
\begin{eqnarray}
 \Delta f_{a}(\mu_F^2)= Z_{ca}(\mu_F^2) \otimes \Delta f_{c}^{L}(\mu_F^2)
\end{eqnarray}
and for CF,
\begin{equation}\label{eq:larinG}
 \Delta  {\cal C}_{1,ab}(\mu_F^2)= \big(Z^{-1}(\mu_F^2)\big)_{ad}\otimes \Delta {\cal C}_{1,db}^{L} (\mu_F^2)\,.
\end{equation}
$g_1$ in eq.~(\ref{eq:g1L}) can be expressed in terms of $\overline{\rm{MS}}$ quantities:
\begin{equation}
 g_1 = \sum_{a,b} \Delta f_{a} (\mu_F^2)\otimes  \Delta {\cal C}_{1,ab} (\mu_F^2) \tilde \otimes D_b(\mu_F^2)
\, .
\end{equation}
For extraction of the CF $\Delta {\cal C}^L_{1,ab}$ in Eq.~(\ref{eq:g1L})  requires the polarized AP kernels in Larin's scheme, 
which can derived from the polarized space-like AP kernels ($\Delta \Gamma_{c\leftarrow d}$) in the $\overline{\rm{MS}}$ scheme through the relation $Z_{ab}$ by $\Delta \Gamma^L = Z^{-1} \otimes \Delta \Gamma$. 
This leads to the scheme transformation for the splitting functions
\begin{align}
\Delta P^{L} &= Z^{-1} \otimes \Delta P \otimes Z + 2
\beta(a_s)\ Z\otimes \frac{dZ^{-1}}{d a_s} \, ,
\label{1a}
\end{align}
where $\beta(a_s) =-\sum_{l=0}^{\infty} a_s^{l+2}(\mu_F^2) \beta_{l}$. 

The symmetric matrix valued $Z$-factor, e.g. given in \cite{Matiounine:1998re,Ravindran:2003gi,Moch:2014sna}, reads
\begin{align}
Z_{ab} = 1 + \sum_{l=1}^{\infty} a_s^{l} z_{ab}^{(l)}\,,
\end{align}
and with $z_{qg}^{(l)} =z_{gq}^{(l)} =z_{gg}^{(l)} =0$, we have
$Z_{qg} =1,Z_{gq} =1,Z_{gg} =1$. 
The use of
\begin{align}
Z_{q_{i}q_{j}} &= 1 + a_s z^{(1)}_{q_{i}q_{j}}+a_s^2 z^{(2)}_{q_{i}q_{j}} +{\mathcal{O}}(a_s^3)\ ,\nonumber\\
Z_{q_{i}\overline{q}_{j}} &= 1 + a_s z^{(1)}_{q_{i}\overline{q}_{j}}+a_s^2 z^{(2)}_{q_{i}\overline{q}_{j}} +{\mathcal{O}}(a_s^3)\ ,\nonumber\\
\Delta P^{L}_{cd} &=  a_s \Delta P^{L,(0)}_{cd}+a_s^{2} \Delta P^{L,(1)}_{cd} +{\mathcal{O}}(a_s^3)
\end{align}
and eq.~(\ref{1a}) provides the spin dependent space-like splitting functions in Larin's scheme:
\begin{align}
  \Delta P^{L,(0)}_{cd} &= \Delta P^{(0)}_{cd} \, ,\nonumber\\
 \Delta P^{L,(1)}_{qq} &= \Delta P^{(1)}_{qq} + 2 \beta_0
z^{(1)}_{qq}  \, ,\nonumber\\
\Delta P^{L,(1)}_{qg} &=\Delta P^{(1)}_{qg}  - \Delta P^{
(0)}_{qg} \otimes z^{(1)}_{qq}  \, ,\nonumber\\
\Delta P^{L,(1)}_{gq} &= \Delta P^{(1)}_{gq}   + \Delta P^{
(0)}_{gq} \otimes z^{(1)}_{qq}  \, ,\nonumber\\
\Delta P^{L,(1)}_{gg} &= \Delta P^{(1)}_{gg}\, .
\end{align}
Likewise, cf. eq.~(\ref{eq:larinG}), the CF in the $\overline{\rm{MS}}$ scheme read,
\begin{align}
\label{eq:CFs-1}
\Delta {\cal{C}}^{(0)}_{1,qq} &=\Delta {\cal{C}}^{L,(0)}_{1,qq} = \delta{(1-x')}\delta{(1-z')} \, ,\nonumber\\
\Delta{\cal{C}}^{(1)}_{1,qq} &=\Delta {\cal{C}}^{L,(1)}_{1,qq} + (-z^{(1)}_{qq})\otimes \Delta {\cal{C}}^{L,(0)}_{1,qq} \, ,\nonumber\\
\Delta{\cal{C}}^{(1)}_{1,qg} &= \Delta{\cal{C}}^{L,(1)}_{1,qg} \, ,\nonumber\\
\Delta{\cal{C}}^{(1)}_{1,gq} &= \Delta{\cal{C}}^{L,(1)}_{1,gq} \, ,\nonumber\\
\Delta{\cal{C}}^{(2)}_{1,qq} &= \Delta{\cal{C}}^{L,(2)}_{1,qq} + (-z^{(1)}_{qq})\otimes \Delta{\cal{C}}^{L,(1)}_{1,qq}  \nonumber\\&+ \Big( z^{(1)}_{qq} \otimes z^{(1)}_{qq}  -  z^{(2)}_{qq}  \Big) \otimes \Delta{\cal{C}}^{L,(0)}_{1,qq} \, ,\nonumber\\
\Delta{\cal{C}}^{(2)}_{1,q\overline{q}}  &=  \Delta{\cal{C}}^{L,(2)}_{1,q\overline{q}} + \Big(-  z^{(2)}_{q\overline{q}}  \Big) \otimes \Delta{\cal{C}}^{L,(0)}_{1,\overline{q}\overline{q}} \, ,\nonumber\\
 \Delta{\cal{C}}^{(2)}_{1,qq'}  &=  \Delta{\cal{C}}^{L,(2)}_{1,qq'} + \Big(-  z^{(2)}_{qq'} \Big) \otimes \Delta{\cal{C}}^{L,(0)}_{1,q'q'} \, ,\nonumber\\
\Delta{\cal{C}}^{(2)}_{1,qg} &= \Delta{\cal{C}}^{L,(2)}_{1,qg} + (-z^{(1)}_{qq})\otimes \Delta{\cal{C}}^{L,(1)}_{1,qg} \, ,\nonumber\\
\Delta{\cal{C}}^{(2)}_{1,gq} &= \Delta{\cal{C}}^{L,(2)}_{1,gq}\, ,
\hspace{0.3cm} \Delta{\cal{C}}^{(2)}_{1,gg} = \Delta{\cal{C}}^{L,(2)}_{1,gg}
\, .
\end{align}
Expressions for  $z_{ab}^{(l)}$  are provided in Appendix-(\ref{Appendix:D}), 
and the  CF in the $\overline{\rm{MS}}$ scheme in an ancillary file.  
Note, that the flavor-nonsinglet CF of polarized SIDIS agree with those of the SF $F_3$, cf.~\cite{Goyal:2024xxx}. 
The latter requires an additional renormalization of the axial-vector current and a kinematics independent finite renormalization transformation from the Larin to the $\overline{\rm{MS}}$ scheme~\cite{Larin:1993tq,Ahmed:2015qpa}.
We find full agreement for the respective CFs, which provides a strong independent check on our computation.
For the subsequent discussion of the CFs with initial/final quarks of same or different flavor, it is useful to factor the dependence on the quark electric charges $e_q$ as follows:
\begin{align}
\label{eq:CFs-2}
(\Delta)\mathcal{C}^{(0)}_{i,qq} &= (\Delta)\mathcal{C}^{(0)}_{i,\overline{q}\overline{q}} =  e_q^2 (\Delta)\text{C}_{i,qq}^{(0)}\, ,
\nonumber\\
(\Delta)\mathcal{C}^{(1)}_{i,qq} &= (\Delta)\mathcal{C}^{(1)}_{i,\overline{q}\overline{q}} =  e_q^2 (\Delta)\text{C}_{i,qq}^{(1)}\, ,
\nonumber\\
(\Delta)\mathcal{C}^{(1)}_{i,qg} &= (\Delta)\mathcal{C}^{(1)}_{i,\overline{q}g} = e_q^2
(\Delta)\text{C}_{i,qg}^{(1)} \, ,\nonumber\\
(\Delta)\mathcal{C}^{(1)}_{i,gq} &= (\Delta)\mathcal{C}^{(1)}_{i,g\overline{q}} = e_q^2
(\Delta)\text{C}_{i,gq}^{(1)} \, ,\nonumber\\
(\Delta)\mathcal{C}^{(2)}_{i,qq} &= (\Delta)\mathcal{C}^{(2)}_{i,\overline{q}\overline{q}} =  e_q^2 (\Delta)\text{C}_{i,qq}^{(2),\text{NS}}
\nonumber\\
&+ \left(\sum_{q_k} e_{q_k}^2\right) (\Delta)\text{C}_{i,qq}^{(2),\text{PS}} \, ,\nonumber\\
(\Delta)\mathcal{C}^{(2)}_{i,qg} &= (\Delta)\mathcal{C}^{(2)}_{i,\overline{q}g} = e_q^2
(\Delta)\text{C}_{i,qg}^{(2)} \, ,\nonumber\\
(\Delta)\mathcal{C}^{(2)}_{i,gq} &= (\Delta)\mathcal{C}^{(2)}_{i,g\overline{q}} = e_q^2
(\Delta)\text{C}_{i,gq}^{(2)} \, ,\nonumber\\
(\Delta)\mathcal{C}^{(2)}_{i,q\overline{q}} &= (\Delta)\mathcal{C}^{(2)}_{i,\overline{q}q} = e_q^2
(\Delta)\text{C}_{i,q\overline{q}}^{(2)} \, ,\nonumber\\
(\Delta)\mathcal{C}^{(2)}_{i,gg} &= \left(\sum_{q_k}e_{q_k}^2\right)
(\Delta)\text{C}_{i,gg}^{(2)} \, ,\nonumber\\
(\Delta)\mathcal{C}^{(2)}_{i,qq'} &=(\Delta)\mathcal{C}^{(2)}_{i,\overline{q}\overline{q}'} = e_q^2(\Delta) \text{C}^{(2),[1]}_{i,qq'}+e_q'^2(\Delta) \text{C}^{(2),[2]}_{i,qq'}
\nonumber \\
&+e_q e_q'(\Delta)\text{C}^{(2),[3]}_{i,qq'}\, ,\nonumber\\
(\Delta)\mathcal{C}^{(2)}_{i,q\overline{q}'}&=(\Delta)\mathcal{C}^{(2)}_{i,\overline{q}q'} = e_q^2 (\Delta)\text{C}^{(2),[1]}_{i,qq'}+e_q'^2 (\Delta)\text{C}^{(2),[2]}_{i,qq'} \nonumber \\
&- e_q e_q'(\Delta)\text{C}^{(2),[3]}_{i,qq'} \, .
\end{align}
where, $e_{q}'$ is the electric charge of quark ($q'$) of different flavor from quark $q$ and summation is over the number of active flavor light quarks.
\begin{table}[t!]
\vspace*{5mm}
    \centering
\renewcommand{\arraystretch}{2.2}
    \begin{tabular}{|c|c|}
        \hline
        CFs~ & ~ Diagrams \\
        \hline
        $(\Delta){\text{C}}_{i,qq}^{(2),\text{NS}}$ & 
        VV [Figs.~\ref{fig:lo},\ref{fig:nlo},\ref{fig:nnlovv}], 
        RV [Figs.~\ref{fig:nlo},\ref{fig:nnlorv}], 
        RR [Figs.~\ref{fig:q2qgg}],\\
       & A$^2$ [Fig.~\ref{fig:nnlorrA}], B$^2$ [Fig.~\ref{fig:nnlorrB}], D$^2$ [Fig.~\ref{fig:nnlorrD}],\\
       & AB [Figs.~\ref{fig:nnlorrA}  \ref{fig:nnlorrB}], AC [Figs.~\ref{fig:nnlorrA},\ref{fig:nnlorrC}], AD [Figs.~\ref{fig:nnlorrA},\ref{fig:nnlorrD}], \\
    & BC [Figs.~\ref{fig:nnlorrB},\ref{fig:nnlorrC}], BD [Figs.~\ref{fig:nnlorrB},\ref{fig:nnlorrD}], CD [Figs.~\ref{fig:nnlorrC},\ref{fig:nnlorrD}]   \\
         \hline
        $(\Delta){\text{C}}_{i,qq}^{(2),\text{PS}}$ & C$^2$ [Fig.~\ref{fig:nnlorrC}]  \\
        \hline
        $(\Delta){\text{C}}_{i,q\bar{q}}^{(2)}$ &  B$^2$ [Fig.~\ref{fig:nnlorrB}], D$^2$ [Fig.~\ref{fig:nnlorrD}], BD [Figs.~\ref{fig:nnlorrB},\ref{fig:nnlorrD}] \\
                \hline
        $(\Delta)\text{C}^{(2),[1]}_{i,qq'}$ &  B$^2$ [Fig.~\ref{fig:nnlorrB}] \\
                \hline
        $(\Delta)\text{C}^{(2),[2]}_{i,qq'}$ &  D$^2$ [Fig.~\ref{fig:nnlorrD}] \\
                        \hline
        $(\Delta)\text{C}^{(2),[3]}_{i,qq'}$ &  BD [Figs.~\ref{fig:nnlorrB},\ref{fig:nnlorrD}] \\
                 \hline
        $(\Delta){\text{C}}_{i,qg}^{(2)}$    &  RV($q \gamma^{*} \rightarrow g$) [Figs.~\ref{fig:nlo},\ref{fig:nnlorv}], RR [Fig.~\ref{fig:q2qgg}]\\
                 \hline
        $(\Delta){\text{C}}_{i,gq}^{(2)}$    &  RV($g \gamma^{*}\rightarrow q$) [Figs.~\ref{fig:nlo},\ref{fig:nnlorv}], RR [Fig.~\ref{fig:g2gqQ}]\\
                 \hline
        $(\Delta){\text{C}}_{i,gg}^{(2)}$    &  RR($g \gamma^{*}\rightarrow g$) [Fig.~\ref{fig:g2gqQ}]\\
    \hline
    \end{tabular}
    \caption{Feynman diagrams contributing to the CFs of individual partonic channels.}
    \label{tab:cftable}
\end{table}

In Table~\ref{tab:cftable} we have listed the individual Feynman diagrams contributing to the CFs for the different channels at NNLO level. 
The non-singlet CFs, i.e.\ $(\Delta){\text{C}}_{i,qq}^{(2),\text{NS}}$ receive contributions from VV diagrams in Figs.~\ref{fig:lo}, \ref{fig:nlo} and \ref{fig:nnlovv}, from RV diagrams in Figs.~\ref{fig:nlo} and \ref{fig:nnlorv}, and from RR real emission diagrams. 
The latter consist of diagrams with gluon emissions (Fig:~\ref{fig:q2qgg}) as well as the squared diagrams 
A$^2$, B$^2$, D$^2$ and their interference among themselves along with their interference with C type diagrams,
cf.\ Figs.~\ref{fig:nnlorrA}, \ref{fig:nnlorrB}, \ref{fig:nnlorrC} and \ref{fig:nnlorrD}.
The pure-singlet CFs, i.e.\ $(\Delta)\text{C}_{i,qq}^{(2),\text{PS}}$, are obtained from 
the RR diagrams C$^2$ in Figs.~\ref{fig:nnlorrC}.
CFs in partonic channels with other flavor combinations for initial and fragmenting quarks/anti-quarks arise from RR diagrams.
$(\Delta)\text{C}^{(2)}_{i,q\overline{q}}$ originates from diagrams of type B and D in Figs.~\ref{fig:nnlorrB} and  \ref{fig:nnlorrD}, while $(\Delta)\mathcal{C}^{(2)}_{i,qq'}$ where $q\neq q'$ is sub-divided into three parts 
according to the interaction of the virtual photon with different quark flavors:
$(\Delta)\text{C}^{(2),[1]}_{i,qq'}$ with B$^2$ in Fig.~\ref{fig:nnlorrB},
$(\Delta)\text{C}^{(2),[2]}_{i,qq'}$ with D$^2$ in Fig.~\ref{fig:nnlorrD},
$(\Delta)\text{C}^{(2),[3]}_{i,qq'}$ with interference between diagrams B and D.

Finally, for the assembly of all partonic CFs from eqs.~(\ref{eq:CFs-1}) and (\ref{eq:CFs-2}) to the SIDIS SFs, cf.\ eq.~(\ref{eq:StrucCoeff}), 
it is useful to introduce functions $(\Delta)H_{ab}$ for the combination of PDFs and FFs in as follows:
\begin{align}
(\Delta)H_{qq} &= (\Delta) f_q(x) D_q(z) + (\Delta) f_{\bar{q}} (x)D_{\bar{q}}(z)\, , \nonumber\\
(\Delta) H_{q\bar{q}} &= (\Delta) f_q (x) D_{\bar{q}} (z)+ (\Delta) f_{\bar{q}}(x) D_q(z)\, , \nonumber\\
(\Delta)H_{qg} &= (\Delta) f_q (x) D_g (z)+(\Delta) f_{\bar{q}}(x) D_g (z)\, , \nonumber\\
(\Delta)H_{gq} &= (\Delta) f_g (x) D_q (z) +  (\Delta) f_g D_{\bar{q}}(z)\, ,\nonumber\\
(\Delta) H_{gg} &= (\Delta) f_g (x)D_g(z)\, , \nonumber\\
(\Delta) H^{\pm}_{qq'} &= (\Delta) f_q (x)D_{q'}(z) \pm (\Delta) f_q (x)D_{\bar{q}'}(z) \nonumber\\
&\pm  (\Delta) f_{\bar{q}} (x)D_{q'}(z)+  (\Delta) f_{\bar{q}} (x) D_{\bar{q}'}(z)\ .
\end{align}
With the perturbative expansion of the SFs in eq. (\ref{eq:StrucCoeff}) as 
$(g_1)F_i = {\displaystyle \sum_{l=0}} a_s^l ~ (g_1^{(l)})F_i^{(l)}$  
we obtain then to NNLO,
\begin{widetext}
\begin{align}
(g_1^{(0)})F_i^{(0)}& =\sum_{q} e_{q}^{2} ~(\Delta)H_{qq}\ ,\\
(g_1^{(1)})F_i^{(1)} &= \sum_{q} e_{q}^{2}~\bigg(  (\Delta)H_{qq}\hat \otimes (\Delta){\text{C}}_{i,qq}^{(1)} + (\Delta)H_{qg} \hat \otimes (\Delta){\text{C}}_{i,qg}^{(1)} + (\Delta)H_{gq} \hat \otimes (\Delta){\text{C}}_{i,gq}^{(1)} \bigg)
\ ,\label{eq:nlo-g1} \\
(g_1^{(2)})F_i^{(2)}
&= \sum_{q} e_{q}^{2}\bigg(  (\Delta)H_{qq}\hat \otimes (\Delta){\text{C}}_{i,qq}^{(2),\text{NS}} + (\Delta)H_{q\bar{q}}\hat \otimes  (\Delta){\text{C}}_{i,q\bar{q}}^{(2)} + (\Delta)H_{qg} \hat \otimes (\Delta){\text{C}}_{i,qg}^{(2)} + (\Delta)H_{gq} \hat \otimes (\Delta){\text{C}}_{i,gq}^{(2)} \bigg) \nonumber\\
&+ \Big(\sum_{q_k} e_{q_k}^{2}\Big)\bigg( (\Delta)H_{qq}\hat \otimes (\Delta){\text{C}}_{i,qq}^{(2),\text{PS}}  + (\Delta)H_{gg} \hat \otimes (\Delta){\text{C}}_{i,gg}^{(2)} \bigg) \nonumber\\
&+ \sum_{q} \sum_{q'\neq q}\bigg( e_{q}^{2}~(\Delta)H^{+}_{qq'} \hat \otimes  (\Delta){\text{C}}_{i,qq'}^{(2),[1]} + e_{q'}^{2}~(\Delta)H^{+}_{qq'}\hat \otimes  (\Delta){\text{C}}_{i,qq'}^{(2),[2]} + e_{q}e_{q'} (\Delta)H^{-}_{qq'} \hat \otimes (\Delta){\text{C}}_{i,qq'}^{(2),[3]}\bigg)
\, .\label{eq:nnlo-g1}
\end{align}
\end{widetext}
where $\hat\otimes$ denotes the convolution of $(\Delta){\text{C}}^{(l)}_{i,ab}$ with $(\Delta)H_{ab}$ in both variables $x$ and $z$.

The CFs for the unpolarized SFs $F_{1,2}$ presented here extend our previous work~\cite{Goyal:2023zdi}, which had focused on the non-singlet contributions only. 
Our results for the CFs have been subject to the following checks.
In the threshold expansion around $x', z' \rightarrow 1$ our results for $(\Delta)\text{C}_{i,ab}^{(l)}$, $i=1,2$ are in complete agreement with predictions for all 
`+'-distributions in $w=x', z'$, ${\cal D}_j(w)=(\ln^j(1-w)/(1-w))_+$ and terms to proportional delta functions $\delta(1-w)$ 
in the unpolarized SFs $F_{1,2}$ in refs.~\cite{Abele:2021nyo,Goyal:2024xxx} (see also \cite{Ravindran:2006bu,Ahmed:2014uya}).
This also holds for the CF  $\Delta \text{C}_{ab}(x',z')$ of the polarized SF $g_{1}$, since the soft and virtual terms are same as for unpolarized SIDIS.
Also, the complete results have also been checked by the independent computation of refs.~\cite{Bonino:2024qbh,Bonino:2024wgg}, finding also perfect agreement for the full NNLO CFs.

\section{Phenomenology}
\label{sec:pheno}
In this section, we demonstrate the phenomenological impact of the new higher-order QCD corrections to SIDIS for the EIC at the center-of-mass energy of $\sqrt{s}=140$ GeV.  
Beyond LO in QCD, the cross section receives contributions from various partonic channels, i.e.\ all individual sub-processes discussed in Sec.~\ref{sec:partonlevel}.

\onecolumngrid

\vspace{0.2cm}

\begin{figure*}[h]
\includegraphics[width=0.47\textwidth]{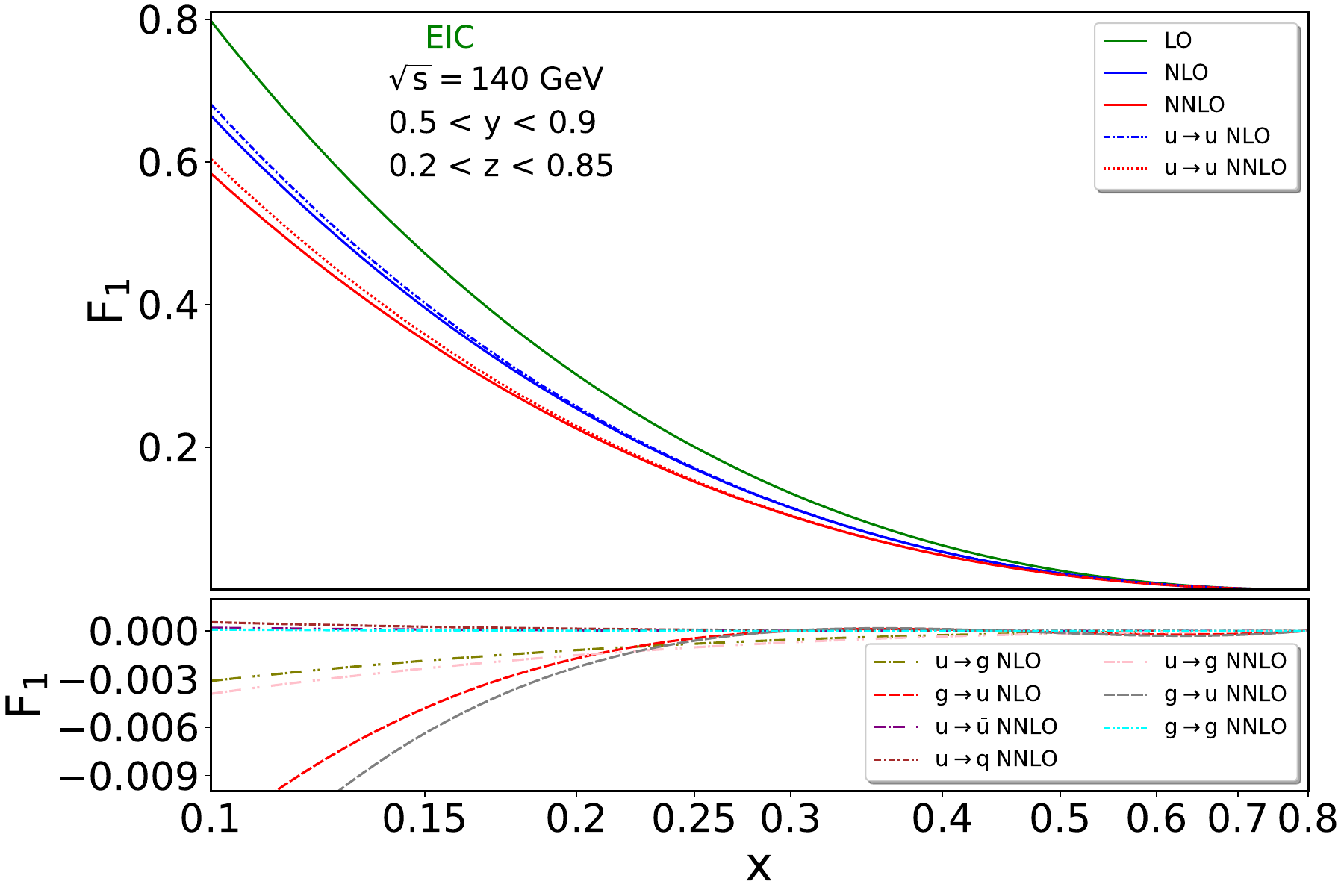}
\includegraphics[width=0.47\textwidth]{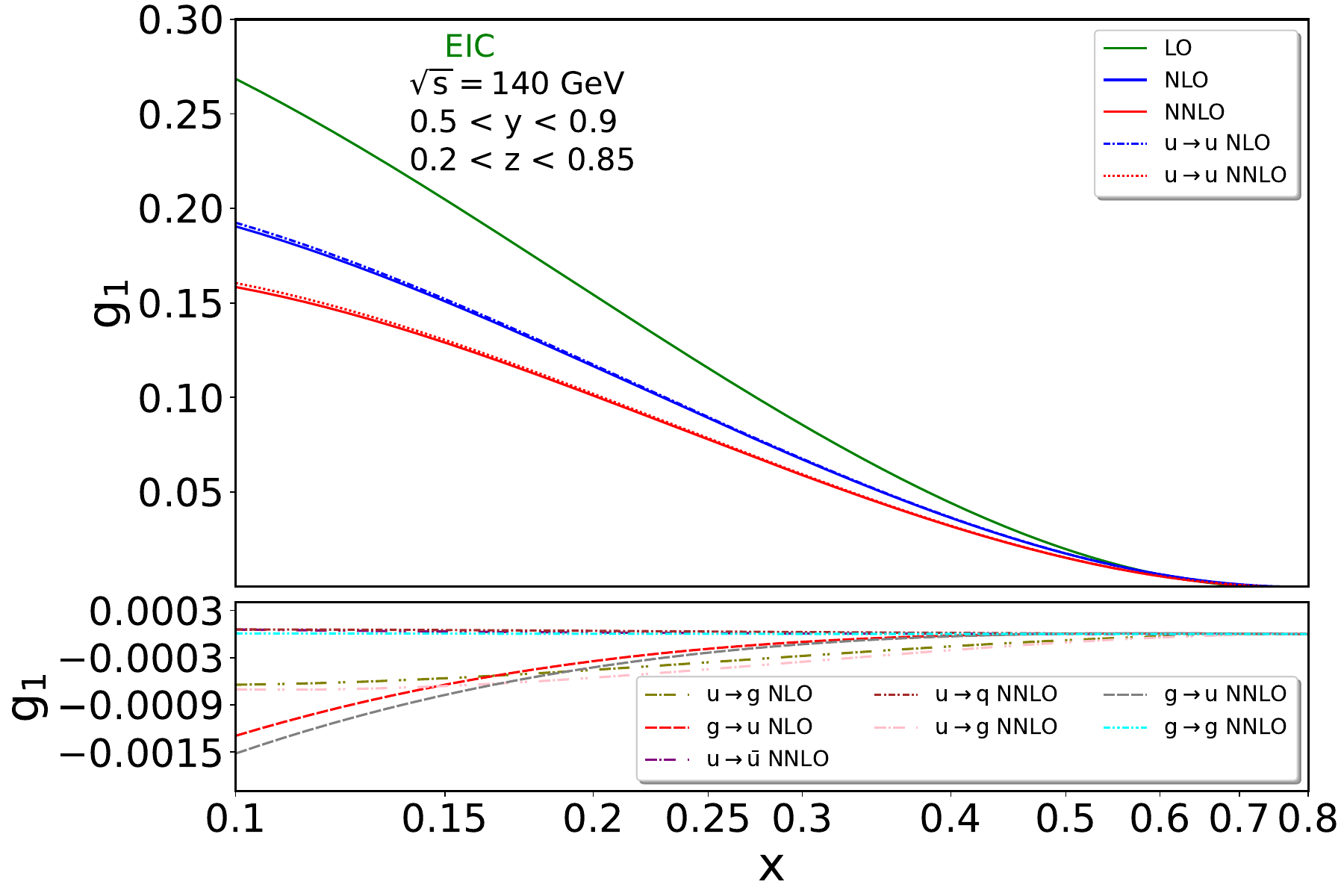}
\caption{Contributions from all partonic channels to the SFs $\text{F}_1$ (left panel) and $\text{g}_1$ (right panel) as a function of $x$ for the EIC at $\sqrt{s}=140$ GeV.}
\label{fig:1FGx}
\end{figure*}

\twocolumngrid

The hadron level SFs $F_1$ and $g_1$ are computed according to eq.~(\ref{eq:StrucCoeff}) as a convolution of the CFs with the corresponding PDFs and FFs for the specific parton-level sub-process.
The SIDIS SFs are functions of the three scaling variables $x, y$ and $z$.
The momentum squared transferred, $Q^2$, depends on the center-of-mass energy 
of the scattering ($s$) through $Q^2 = x y s$.
In the following, we plot the SFs $F_1$ and $g_1$ as a function of one of these scaling variables, either by fixing the other two variables or by integrating the SFs in some kinematic range of them.

\begin{figure*}[h]
\vspace*{2mm}
\includegraphics[width=0.495\textwidth]{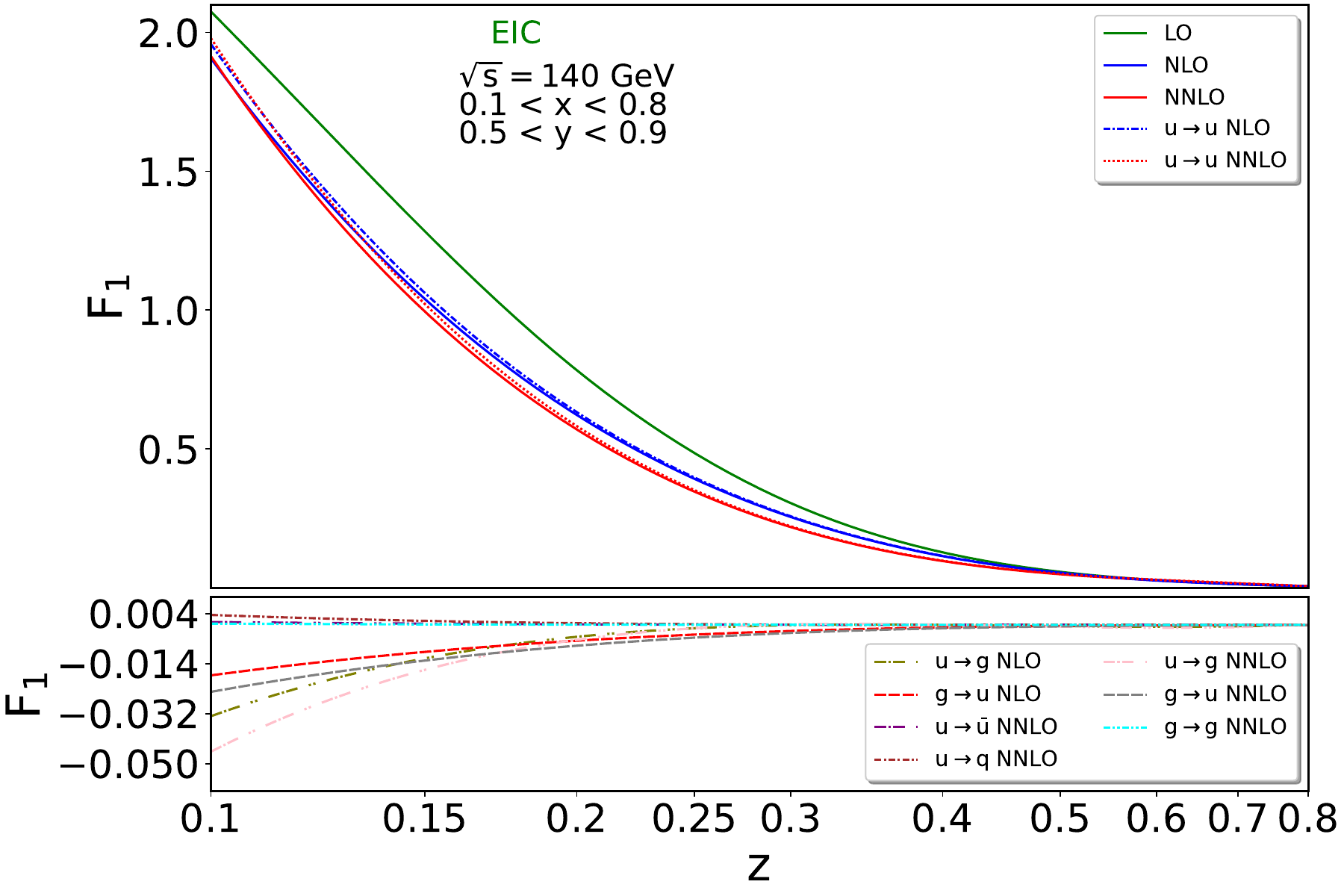}
\includegraphics[width=0.495\textwidth]{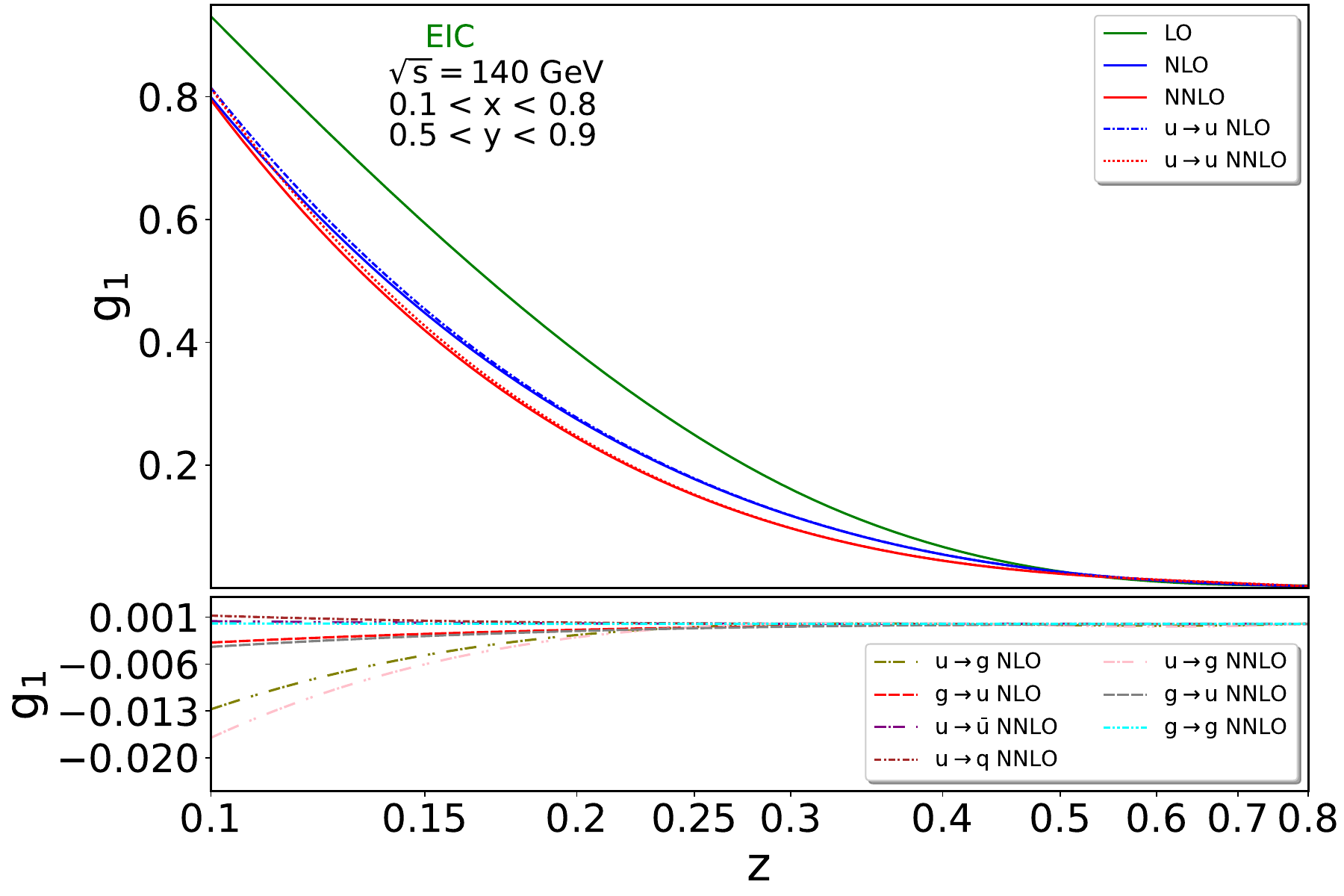}
\caption{
Same as Fig.~\ref{fig:1FGx}, now as a function of $z$. 
}
\label{fig:1FGz} 
\end{figure*}

Starting with the relative contributions of the various partonic channels to SFs at the EIC with $\sqrt{s}= 140$ GeV, 
we present in the left (right) panel of the Fig.~\ref{fig:1FGx} results for $F_1(g_1)$ at successive perturbative orders as a function of $x$, after integrating over $z$ in the range between $0.2$ and $0.85$ and $y$ in the range $0.5$ to $0.9$.
Similarly, in Fig.~\ref{fig:1FGz}, the left (right) panel contains contributions from various sub-processes to $F_1(g_1)$ through NNLO, as a function of $z$, after integration of $x$ between $0.1$ and $0.8$ and of $y$ between $0.5$ and $0.9$.
We observe that the contributions from the CF $(\Delta)C_{1,qq}^{(l)}$, $l=1,2$ are dominating, i.e.\ they are much larger than those from the CFs of the other partonic sub-processes. 
We have set both the renormalization and the factorization scale equal to the central scale $\mu_F=\mu_R=Q$.  
For $F_1$, we have used the \texttt{NNPDF31} PDF sets~\cite{NNPDF:2017mvq} at LO, NLO and NNLO, respectively.
In contrast, the predictions for $g_1$ have been obtained using 
the \texttt{BDSSV24NLO} PDF set at LO and NLO, and the \texttt{BDSSV24NNLO} PDF set~\cite{Borsa:2024mss} at the NNLO level. 
In both cases, for $F_1$ and $g_1$, the \texttt{NNFF10PIp} FF sets~\cite{Bertone:2017tyb} have been utilized at the respective orders (LO, NLO, NNLO).

\onecolumngrid

\begin{figure*}[h]
\includegraphics[width=0.495\textwidth]{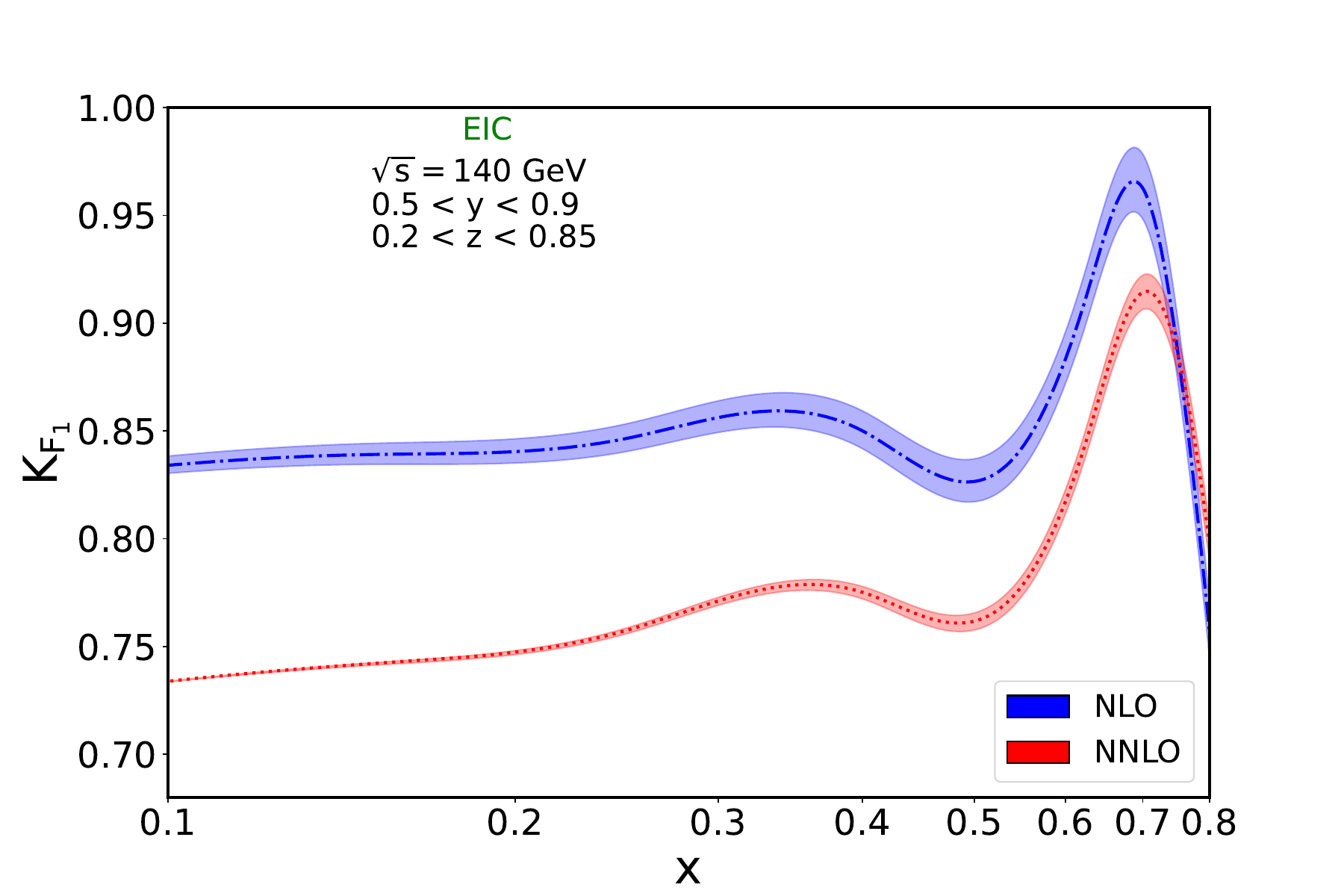}
\includegraphics[width=0.495\textwidth]{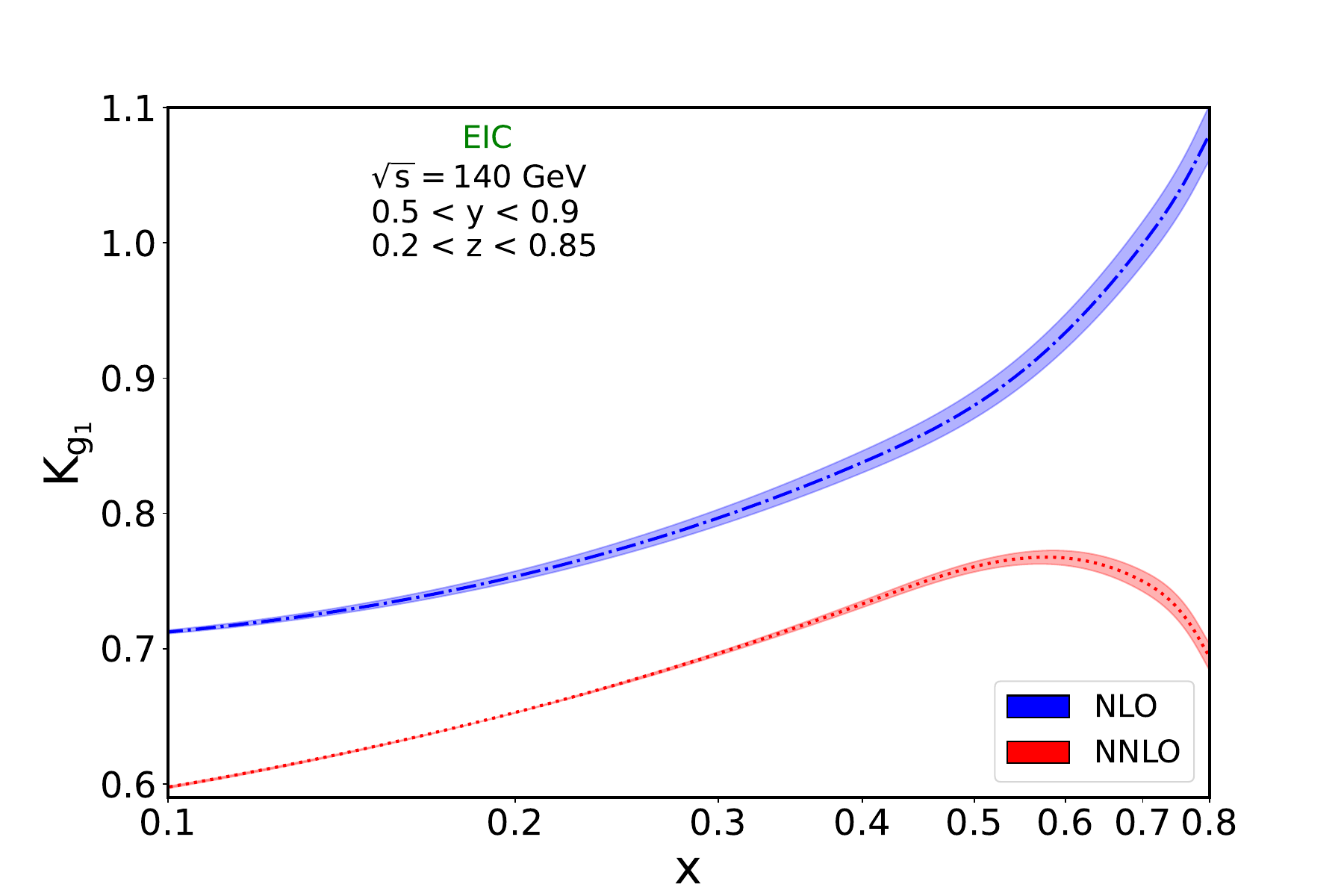}
\caption{NLO and NNLO $K$-factors for $F_1$ and $g_1$ as a function of $x$ for the EIC at $\sqrt{s}=140$ GeV.}
\label{fig:2FGx}
\end{figure*}

\begin{figure*} 
\includegraphics[scale=0.27]{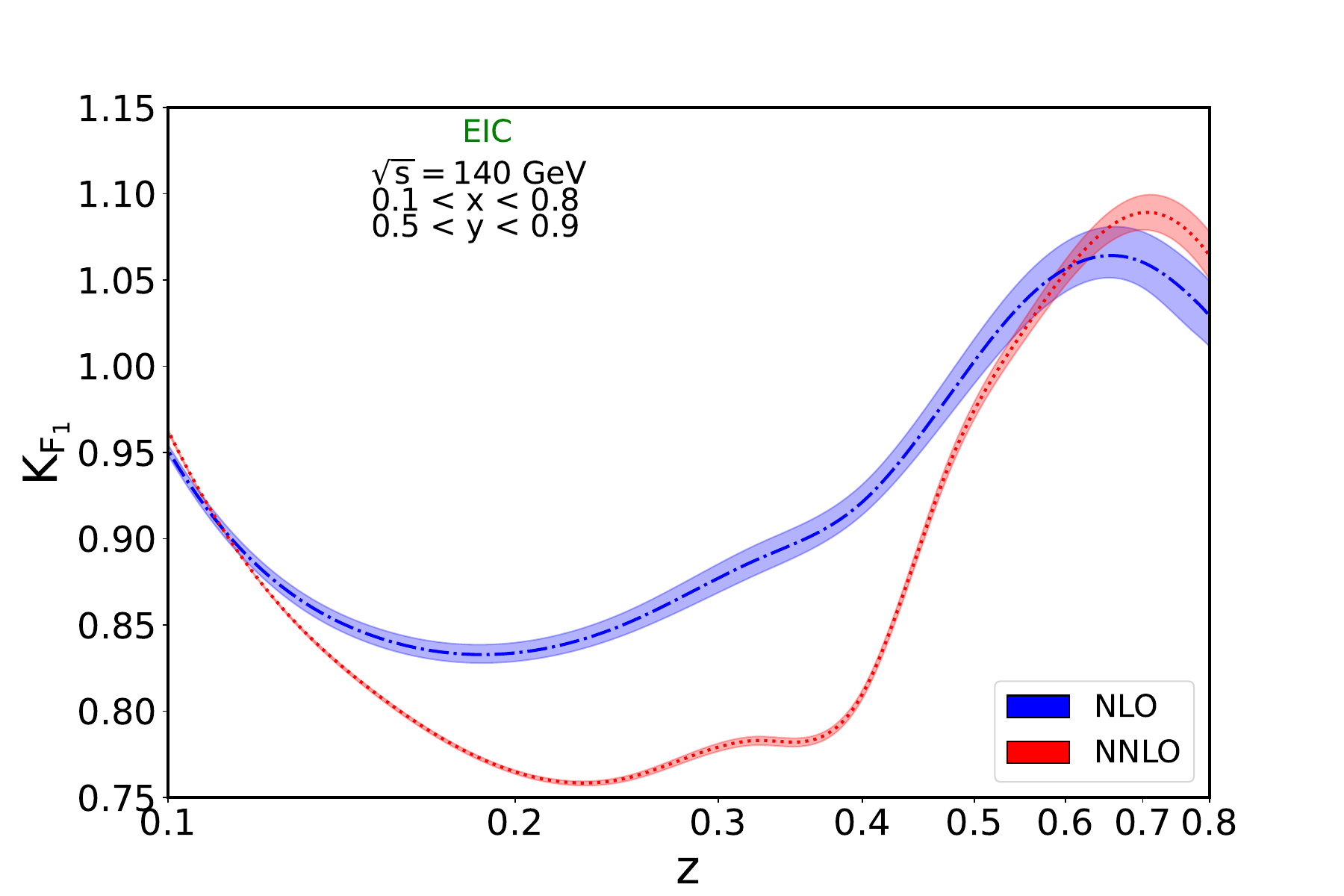}
\includegraphics[scale=0.27]{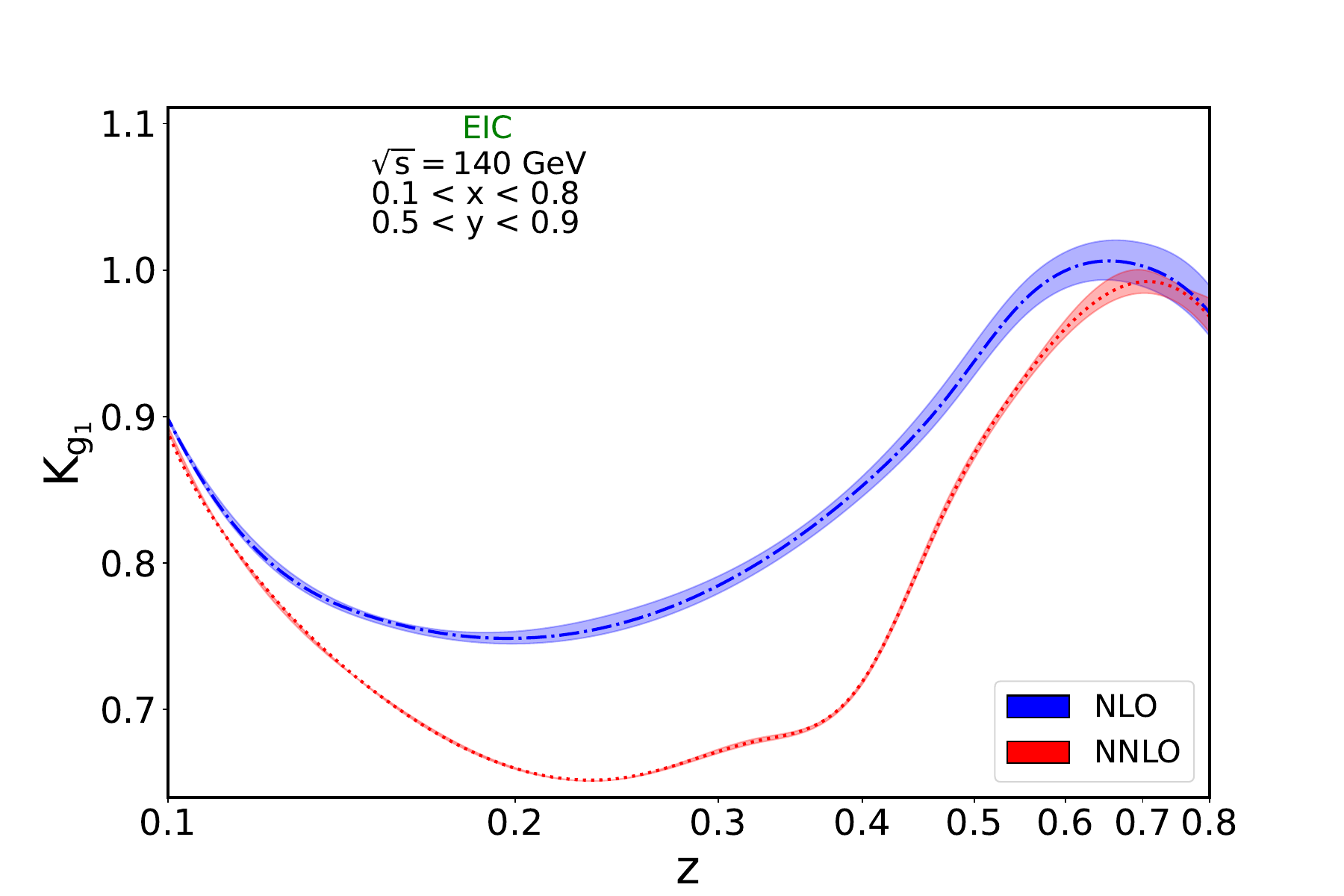}
\caption{Same as Fig.~\ref{fig:2FGx}, now as a function of $z$.
}
\label{fig:2FGz}
\end{figure*}

\twocolumngrid

From Figs.~\ref{fig:1FGx} and \ref{fig:1FGz}, we find that the total perturbative corrections at order $\alpha_s$ as well as $\alpha_s^2$ are negative over a wide range of $x$ and $z$. 
As a result, the NLO contribution is smaller than that of the LO, and the NNLO prediction is less than that of the NLO in the ranges of $x$ (see Fig.~\ref{fig:1FGx}) and $z$ (see Fig.~\ref{fig:1FGz}). 
For larger values of $x$ and $z$ however, the NLO contribution is larger than the LO one ($x$ is above $0.7$).  
Similarly, both the NLO and NNLO corrections become larger than LO beyond $z=0.7$. 
The exact values of $x$ and $z$ at which this transition occurs depends on the choice of PDFs and of FFs as well as the choices of the integration regions for $x$, $y$, and $z$.

In order to illustrate these findings, it is useful to quantify the impact of the NLO and NNLO contributions with respect to LO through the $K$-factor, defined as 
\begin{eqnarray}
\label{eq:Kfactor}
   K_{(g_1)F_1}^{\rm (N)NLO} = \frac{(g_1)F_1^{\rm (N)NLO}}{(g_1)F_1^{\rm LO}} \, ,
\end{eqnarray}
cf.~eq.~(\ref{eq:StrucCoeff}).
These $K$-factors for $F_1$ ($g_1$) are plotted as a function of $x$ on the left (right) hand side of Fig.~\ref{fig:2FGx},  after integrating $y$ and $z$ in the ranges $0.5<y<0.9$ and $0.2<z<0.85$. 
Similarly, Fig.~\ref{fig:2FGz} shows the $K$-factors as a function of $z$, after integration over $0.5<y<0.9$ and $0.1<x<0.8$. 
We have used the same sets of PDFs and FFs as in Figs.~\ref{fig:1FGx} and \ref{fig:1FGz}. 
As mentioned above, in some regions of $x$ and $z$, the QCD corrections are positive. 
The $K$-factors of $F_1$, both at NLO and NNLO, become larger than unity beyond $z=0.6$ and start decreasing for smaller values of $z$.  
Also, the NLO $K$-factor for $g_1$ is larger than unity when $x$ is above $0.7$.
The particular shapes of the $K$-factors in certain regions of $x$ or $z$ are due to choices of PDFs and of FFs.

The bands for $K$-factors correspond to variation of the renormalization scale $\mu_R^2$ in the interval  $[Q^2_{avg}/2,2Q^2_{avg}]$ while keeping the factorization scale $\mu_F= Q_{avg}$, where $Q^2_{avg} = x y_{avg} s$ for Fig. 13 and $Q^2_{avg} = x_{avg}  y_{avg}  s$ for Fig. 14. 
We find that theory uncertainties from scale variation drop when we go from NLO to NNLO accuracy, even though the sensitivity to scales is already
small at both these orders. 
Numerical values for the $K$-factors in eq.~(\ref{eq:Kfactor}) for $F_1$ and $g_1$ at NLO and NNLO are presented in Tabs.~\ref{tab:FG_x} and \ref{tab:FG_z} 
with respect to $x$ and $z$, respectively.
Theory uncertainties (in percent) of SFs due to the variation of the renormalization scale $\mu_R$ around the central 
value $\mu_R=Q_{avg}$ have been given as well.
Note that the $K$-factor for $F_1$ is less than unity for both, the NLO and NNLO predictions in a wider range of $x$ and $z$. 
For $F_1$ at NLO, it ranges from 0.84 at $x=0.15$ to 0.76 at $x=0.8$. 
Similarly at NNLO, the corresponding $K$-factors are 0.741 and 0.79. 
However, the corresponding $K$-factor for $g_1$ takes values 0.73 and 1.1 at NLO and 0.62 and 0.69 at NNLO.

\onecolumngrid

\begin{table*}[ht]
\vspace*{2mm}
\begin{tabular}{|p{2cm}|p{3cm}|p{3cm}||p{3cm}|p{3cm}|}
\hline
\centering
 $K$-Factor & \multicolumn{2}{c}{$F_1$} & \multicolumn{2}{|c|}{$g_1$}\\
\cline{2-3} \cline{4-5}
\centering
  $x$ &  NLO & NNLO & NLO & NNLO\\
\hline
 $0.15$  & $0.83879^{+0.61\%}_{-0.54\%}$ & $0.7414^{+0.0471\%}_{-0.1035\%}$
       & $0.7329^{+0.3657\%}_{-0.3241\%}$ &  $0.6245^{+0.0189\%}_{-0.068\%}$  \\
 \hline
 $0.25$   & $0.8469^{+0.808\%}_{-0.72\%}$ & $0.7579^{+0.1534\%}_{-0.2097\%}$
        & $0.7735^{+0.6539\%}_{-0.5826\%}$ & $0.6769^{+0.05465\%}_{-0.1123\%}$\\
  \hline
 $0.4$   & $0.84849^{+1.092\%}_{-0.9775\%}$ & $0.7746^{+0.3502\%}_{-0.3994\%}$
       &  $0.8367^{+1.1176\%}_{-0.9055\%}$&  $0.7372^{+0.2956\%}_{-0.3485\%}$ \\
  \hline
 $0.6$   & $0.8854^{+1.4469\%}_{-1.2998\%}$ & $0.8187^{+0.6832\%}_{-0.70886\%}$
       & $0.93456^{+1.446\%}_{-1.2996\%}$ & $0.7646^{+0.7157\%}_{-0.740\%}$\\
  \hline
 $0.8$   & $0.7565^{+1.833\%}_{-1.651\%}$ & $0.7948^{+1.074\%}_{-1.0644\%}$
    &$1.0779^{+1.9330\%}_{-1.74115\%}$ & $0.6938^{+1.4313\%}_{-1.38546\%}$ \\
  \hline\hline
\end{tabular}
\caption{$K$-factors of eq.~(\ref{eq:Kfactor}) as function of $x$ (see text for integration ranges of $y$, $z$).}
\label{tab:FG_x}
\end{table*}
%
%
\begin{table*}[h]
\begin{tabular}{|p{2cm}|p{3cm}|p{3cm}||p{3cm}|p{3cm}|}
\hline
\centering
 $K$-Factor & \multicolumn{2}{c}{$F_1$} & \multicolumn{2}{|c|}{$g_1$}\\
\cline{2-3} \cline{4-5}
\centering
 $z$  &  NLO & NNLO & NLO & NNLO\\
\hline
 $0.15$  & $0.84544^{+0.6176\%}_{-0.5535\%}$ &  $0.8137^{+0.0925\%}_{-0.144\%}$
       & $0.7634^{+0.4531\%}_{-0.406\%}$ & $0.71467^{+0.0092\%}_{-0.0592\%}$   \\
 \hline
  $0.25$  & $0.8497^{+0.8086\%}_{-0.7246\%}$ &  $0.7576^{+0.156\%}_{-0.2111\%}$
       & $0.761^{+0.6438\%}_{-0.5769\%}$ & $0.6513^{+0.0439\%}_{-0.101\%}$   \\
 \hline
  $0.4$  & $0.922^{+1.1041\%}_{-0.9895\%}$ & $0.8062^{+0.2819\%}_{-0.3349\%}$
       & $0.8492^{+0.95411\%}_{-0.855\%}$ & $0.7164^{+0.1596\%}_{-0.2189\%}$   \\
 \hline
  $0.6$  & $1.0558^{+1.4558\%}_{-1.305\%}$ & $1.0584^{+0.6802\%}_{-0.707\%}$
       & $0.9982^{+1.335\%}_{-1.197\%}$ & $0.9637^{+0.537\%}_{-0.577\%} $  \\
 \hline
  $0.8$  & $1.028^{+1.952\%}_{-1.749\%}$ & $1.0667^{+1.3069\%}_{-1.274\%}$
       & $0.97^{+1.8417\%}_{-1.65\%}$ & $0.97^{+1.33\%}_{-1.1208\%}$   \\
  \hline\hline
\end{tabular}
\caption{Same as Tab.~\ref{tab:FG_x} as function of $z$.
}
\label{tab:FG_z}
\end{table*}

\twocolumngrid

\begin{figure*}[h]
\includegraphics[width=0.97\textwidth]{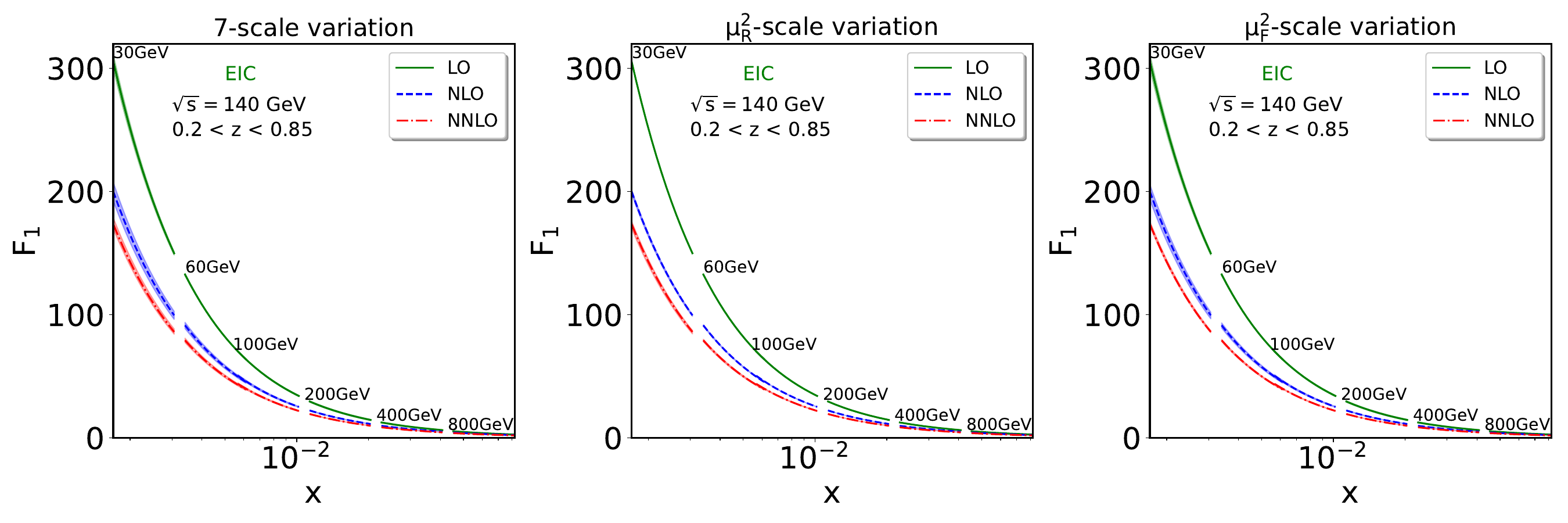}
\caption{Scale variation of $F_1$ as a function $x$ for six different values of $Q^2$.  
}
\label{fig:4Fx}
\end{figure*}

\begin{figure*}[h]
\includegraphics[width=0.97\textwidth]{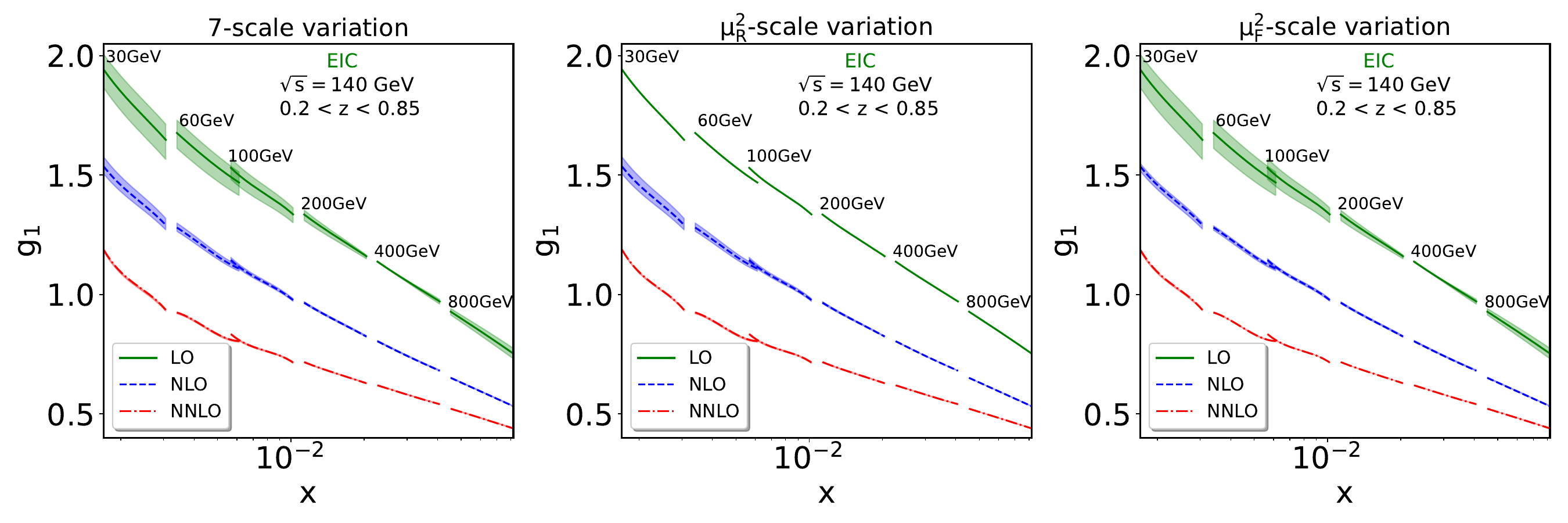}
\caption{Same as Fig.~\ref{fig:4Fx} for $g_1$.
}
\label{fig:4Gx}
\end{figure*}

In order to understand the dependence on the choice of scales for various values of $Q^2$, or equivalently values of $y = Q^2/(x s)$, 
we present in Figs.~\ref{fig:4Fx} and \ref{fig:4Gx} the variation of the renormalization and factorization scales $\mu_{R}$ and $\mu_{F}$ for $F_1$ and $g_1$ as functions of $x$. 
We choose six different values of $Q^2$, namely $Q^2=\{30, 60, 100, 200, 400, 800\}$~GeV.  
The allowed range of $x$ for a given $Q^2$ is obtained by demanding $y$ to be in the range between 0.5 and 0.9.  
On the left hand side of Figs.~\ref{fig:4Fx} and \ref{fig:4Gx}, scales are varied independently by a factor two, i.e.\, $\mu_{R}^2$, $\mu_{F}^2$ $\in [Q^2/2, 2 Q^2]$ with the constraint $1/2 \leq \mu_R^2/\mu_F^2 \leq 2$ in the so-called seven-point scale variation. 
Plots in the center columns of Figs.~\ref{fig:4Fx} and \ref{fig:4Gx} display the impact of the renormalization scale variation alone in the range $\mu_{R}^2 \in [Q^2/2, 2 Q^2$], keeping $\mu_F=Q$. 
Finally, on the right hand side, the factorization scale $\mu_F$ is varied in the range $\mu_F^2 \in [Q^2/2, 2 Q^2$] with fixed $\mu_R =Q$. 

The large uncertainty bands of the LO predictions, in particular for the $g_1$ SF, are due to the significant factorization scale dependence of the PDFs and FFs.
The plots in Figs.~\ref{fig:4Fx} and \ref{fig:4Gx} clearly demonstrate how higher order contributions decrease the $\mu_F$ dependence.  
The $\mu_R$ dependence can be assessed in a meaningful way starting from NLO, cf.\ plots on the left and in the center, and a comparison of uncertainty bands for the NLO and NNLO predictions show a significant reduction of the $\mu_R$ dependence in the latter case. 
Together with the observed apparent convergence of perturbative series for SIDIS, the reduction of the scale sensitivity give support to the robustness of our theoretical framework.

\onecolumngrid


\twocolumngrid

\onecolumngrid

\begin{figure*}[!h] 
\includegraphics[width=0.87\textwidth]{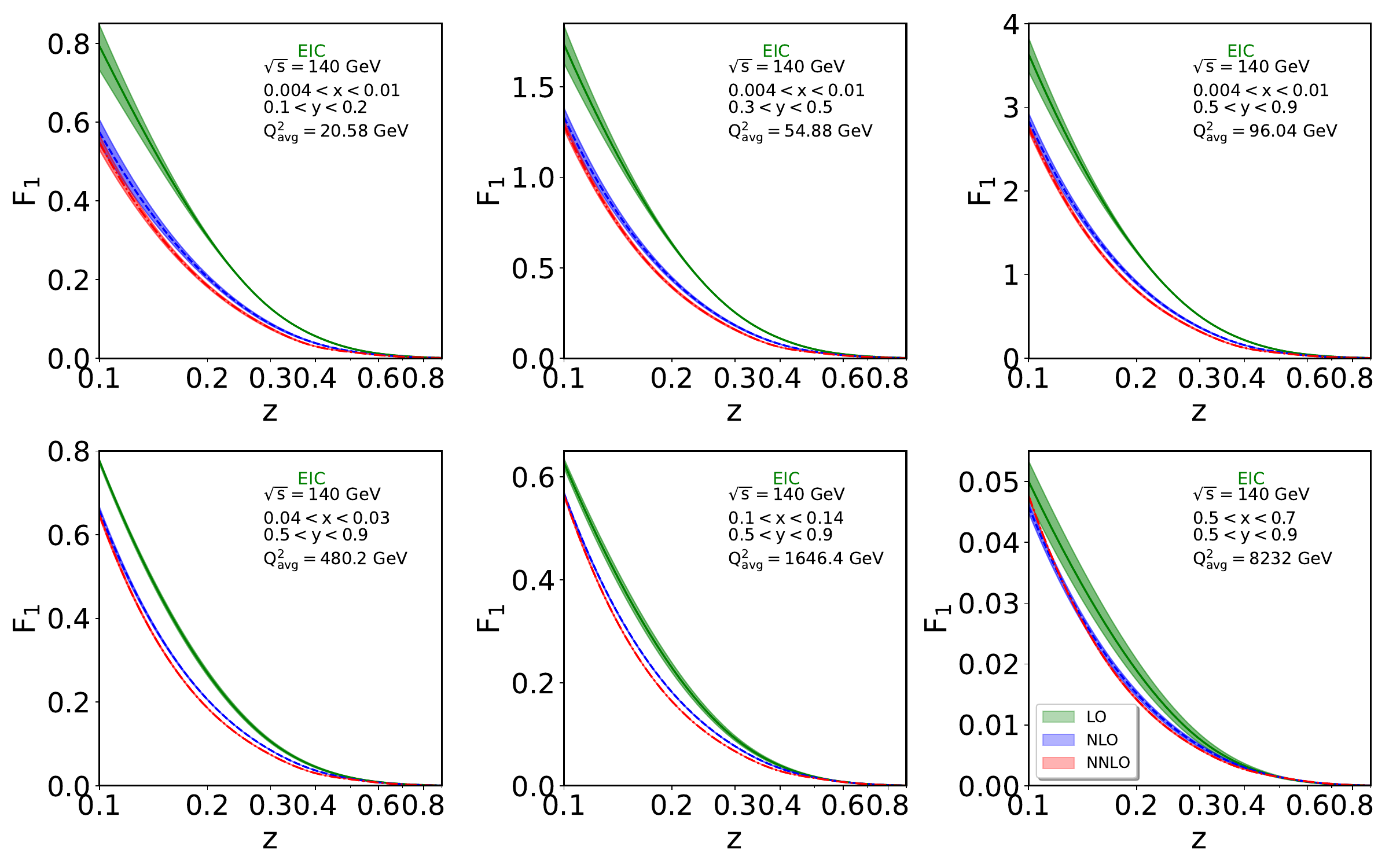}
\caption{Scale variation of $F_1$ as a function of $z$ for six different values of $Q_{avg}^2$.
}
\label{fig:4Fz}
\end{figure*}
\begin{figure*}[!h]
\includegraphics[width=0.87\textwidth]{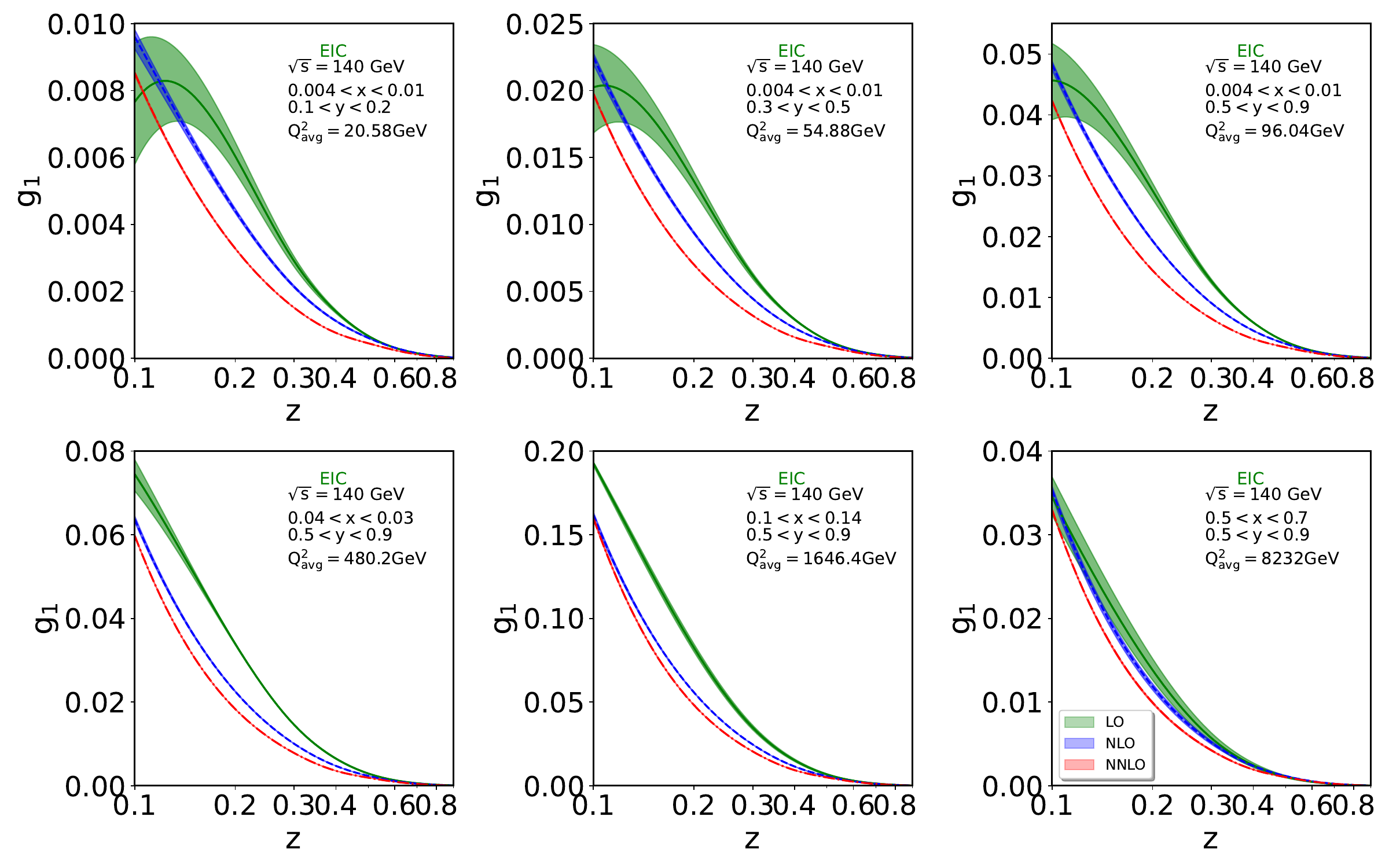}
\caption{Same as Fig.~\ref{fig:4Fz} for $g_1$.
}
\label{fig:4Gz}
\end{figure*}
 
\twocolumngrid

In Figs.~\ref{fig:4Fz} and \ref{fig:4Gz}, we show the SFs $F_1$ and $g_1$ as a function of $z$ at different values of $Q_{avg}$, after integration over $x$ and $y$ in different ranges, such that each range gives a fixed $Q_{avg}$, as indicated in each plot. The bands in the plots denote the seven-point scale variation around $Q^2_{avg}$ and, again, the sensitivity to the renormalization and factorization scales decreases significantly as we include the NLO and NNLO QCD corrections.

\begin{figure}[ht]
\includegraphics[width=0.478\textwidth]{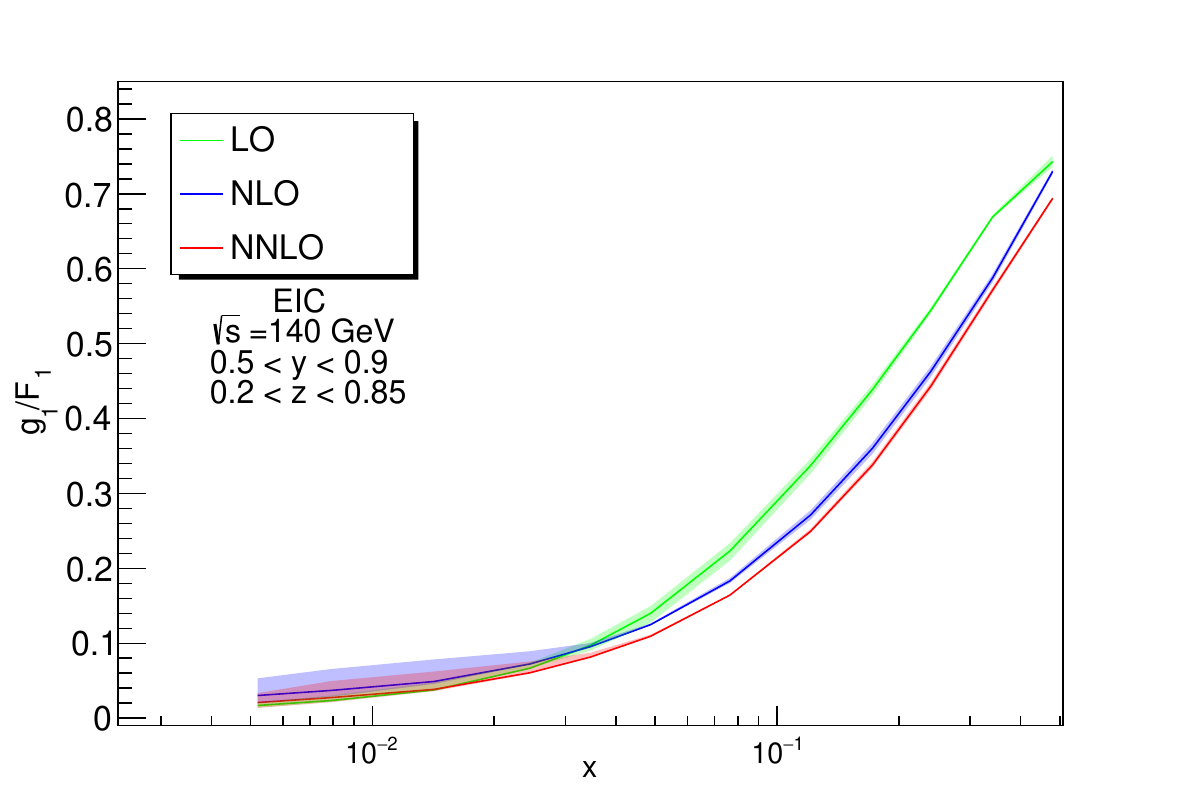}
\caption{Asymmetry $g_{1}/F_{1}$ as a function of $x$ for the EIC at $\sqrt{s}=140$ GeV.}
\label{fig:3x}
\end{figure}

The COMPASS experiment \cite{COMPASS:2010hwr} at CERN has measured the SIDIS asymmetry $g_1/F_1$ as a function of $x$.  
The comparison of our predictions for this asymmetry with the data from COMPASS has already been reported in a previous publication~\cite{Goyal:2024tmo}. 
Here, in Fig.~\ref{fig:3x}, we provide the NNLO predictions as a function $x$ and $Q^2$ for the EIC at $\sqrt{s} = 140$ GeV. 
We have integrated over $z$ in the range $0.2 < z < 0.85$, 
used the unpolarized \texttt{NNPDF31} PDFs for $F_1$, 
the \texttt{NNFF10} FFs for $\pi^{+}$ production at their respective orders 
and the polarized \texttt{BDSSV24NLO} PDFs at LO, NLO and 
\texttt{BDSSV24NNLO} PDFs at NNLO for $g_1$. 
The bands correspond to seven-point scale variation of $\mu_F$ and $\mu_R$ around the nominal scale $\mu_R=\mu_F=Q$. 
The value of the asymmetry decreases with increasing perturbative order and,
as expected, the scale uncertainty also improves.

\begin{figure*}[ht!]
\includegraphics[width=0.496\textwidth]{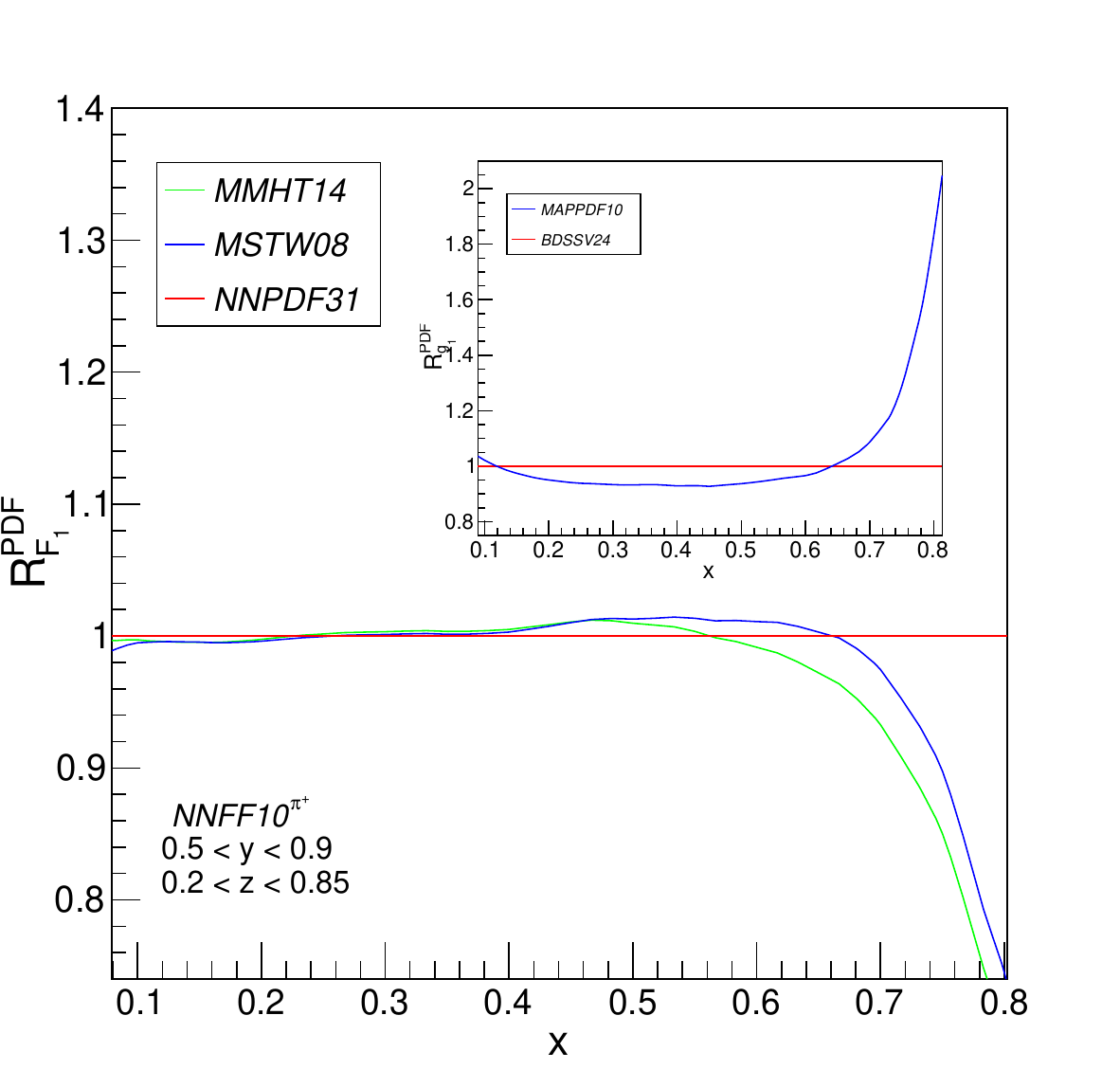}
\includegraphics[width=0.496\textwidth]{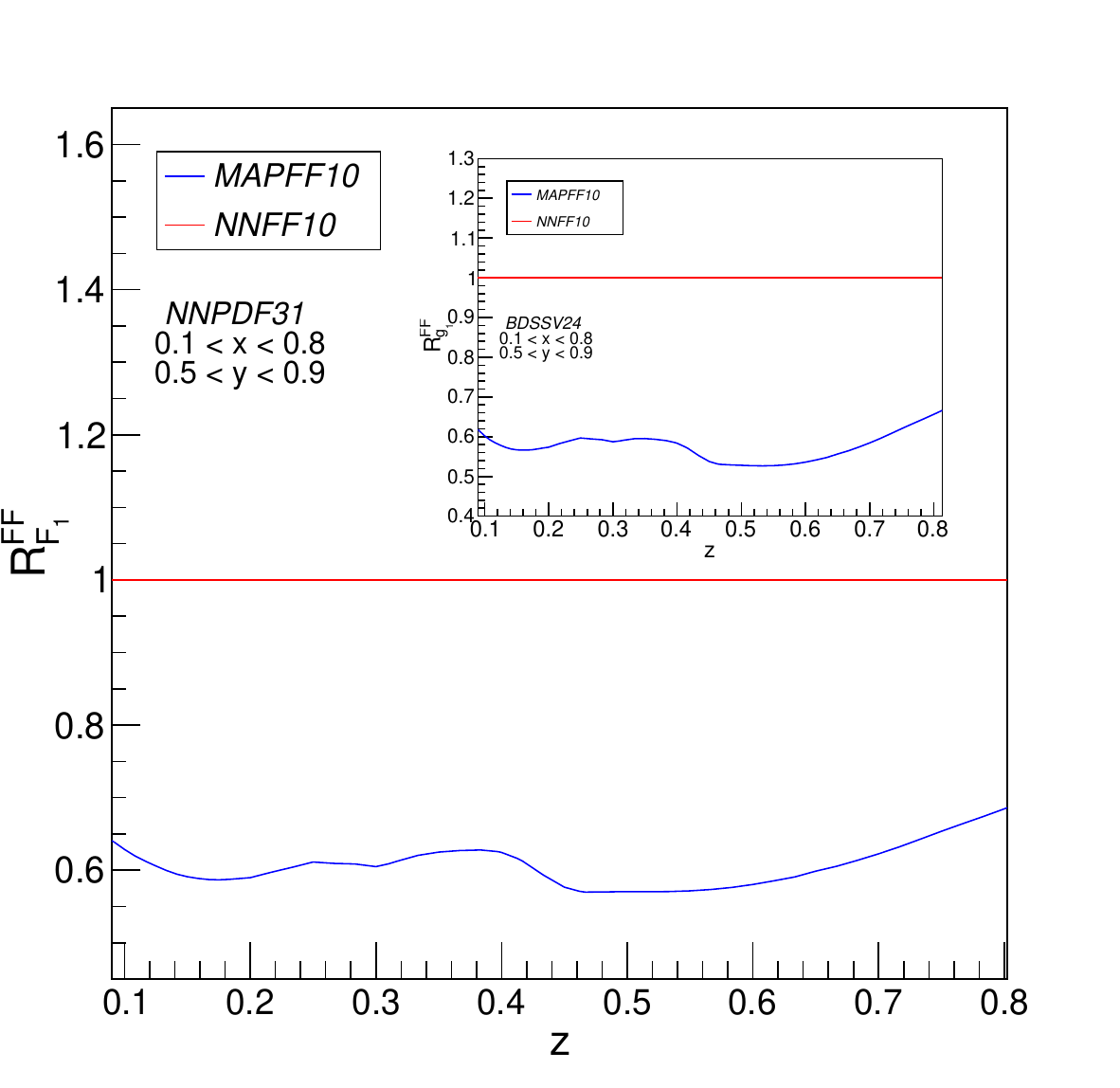}
\caption{NNLO QCD predictions for SIDIS SFs $F_1$ (and $g_1$ in inlay) for ${\pi^{+}}$-production.
Left panel: Different choices of PDF sets in ratio to predictions with \texttt{NNPDF31} and with FFs \texttt{NNFF10} fixed.
Right panel: Same for different choices of FF sets in ratio to predictions with \texttt{NNFF10} with \texttt{NNPDF31} PDFs fixed.
}
\label{fig:5xz}
\end{figure*}

Lastly, in order to understand the impact of various PDF and FF sets on the predictions for $F_1$ and $g_1$ at NNLO, 
we have plotted two ratios in Fig.~\ref{fig:5xz}, namely 
\begin{equation}
R^{\rm PDF}_{(g_1)F_1} = \frac{(g_1)F_1^{\rm PDF}}{(g_1)F_1^{\rm NNPDF}} \, ,
\end{equation}
for PDF variation as a function $x$ and
\begin{equation}
R^{\rm FF}_{g_1(F_1)} = \frac{(g_1)F_1^{\rm FF}}{(g_1)F_1^{\rm NNFF}}   ,
\end{equation}

for FF variation as a function of $z$. 
The PDF variation is illustrated in the left panel of Fig.~\ref{fig:5xz}, where we have integrated over the ranges $0.2 < z < 0.85$ and $0.5 < y < 0.9$. 
Similarly on the right hand side of Fig.~\ref{fig:5xz}, for the FF variation, we have integrated over $0.1 < x < 0.8$ and $0.5 < y < 0.9$. 
For both the ratios, we have set  $\mu_F^2=\mu_R^2=Q_{avg}^2$ where, $Q^2_{avg} = x y_{avg} s$ for left panel and $Q^2_{avg} = x_{avg}  y_{avg}  s$ for the right panel. 
The SFs $F_1$ and $g_1$ do not differ too much over a wide range of $x$ for the different choices of PDF sets, except for large $x$. 

\onecolumngrid

\twocolumngrid


There the predictions for $F_1$ from the unpolarized PDF sets of \texttt{MMHT}~\cite{Harland-Lang:2014zoa} and \texttt{MSTW}~\cite{Martin:2009iq} deviate from the \texttt{NNPDF} ones significantly, as well as the predictions for $g_1$ comparing the polarized PDFs of 
\texttt{MAPPDF}~\cite{Bertone:2024taw} and \texttt{BDSSV24}.
For the FFs on the other hand, there is a sizable difference between predictions prepared with 
the paramtrizations of \texttt{MAPFF10}~\cite{Abdul_Khalek_2021} and \texttt{NNFF10}. 
This illustrates the importance of future EIC measurements of SIDIS SFs and the great potential to constrain FFs in the relevant kinematic range.


 

\section{Discussion and Conclusion}
\label{sec:conclusion}

The SIDIS scattering of a lepton off a proton (or lighter ion), which produces another identified hadron in the final state, is a fundamental reaction in QCD. 
It holds great promise for the determination of PDFs and FFs, which are accessible through standard QCD factorization. Given the accuracy of currently available SIDIS data, and also in light of the EIC, which opens the possibility for SIDIS with polarized beams, we have undertaken the effort to push the precision of QCD theory predictions to the next quantum loop beyond the state of the art, i.e., to NNLO. 
In the current article, we have provided extensive documentation of the computation of these NNLO QCD corrections for both polarized and unpolarized SIDIS, of which only first results have been reported previously~\cite{Goyal:2023zdi,Goyal:2024tmo}, and we have systematically included contributions from all the partonic channels to CFs in the $\overline {\rm{MS}}$ scheme.

A detailed study on the impact of these corrections to hadronic SFs demonstrates that the NNLO corrections are not only significant but also relevant in reducing the inherent theoretical uncertainty through the dependence on the $\mu_R$ and $\mu_F$ scales.
At NNLO accuracy, both the apparent convergence and the scale stability of the perturbation expansion are greatly improved.
The results can now be used to improve the extraction of PDFs and FFs, facilitating a consistent fit of these functions using NNLO QCD theory predictions for the CFs together with the NNLO QCD evolution equations, the latter having been known for a long time.
Beyond these immediate applications, our research opens new avenues for further studies, particularly regarding the resummation of threshold corrections at large $x$ and $z$. However, the available analytic results also provide an opportunity to gain new structural insights into SFs at small $x$ and $z$. We leave these topics for future studies.

The analytical results for the unpolarized CFs, $\mathcal{C}_{i,ab}$ and polarized CF, $\Delta\mathcal{C}_{1,ab}$ in $\overline{\text{MS}}$ scheme, up to NNLO are provided as ancillary files in \texttt{Mathematica} format with this article submission. They are also available from the authors upon request.

\section*{Acknowledgments}
We thank authors of \cite{Bonino:2025bqa,Bonino:2024wgg} for correspondence regarding the finite renormalization constant $Z_{ab}$.
The phenomenological results presented in the original paper are unaffected, as numerical impact of the modifications is not numerically significant.

This work has been supported through a joint Indo-German research grant by
the Department of Science and Technology (DST/INT/DFG/P-03/2021/dtd.12.11.21). 
S.M. acknowledges the ERC Advanced Grant 101095857 {\it Conformal-EIC}.
N.R. is partially supported by the SERB-SRG under Grant No. SRG/2023/000591.
The work of R.L. was supported via RSF grant No. 20-12-00205.
In addition we would also like to thank the computer administrative unit of IMSc for their help and support.

\begin{widetext}
\appendix
\section{Space-like Unpolarized Splitting Function}\label{Appendix:A}
\noindent
In this appendix, we present the space-like unpolarized splitting functions at the 
leading and next-to leading orders. The leading order functions are
\begin{align}
     P_{qq}^{(0)}(x) &= C_F \Bigg\{
          6~\delta\big(1-x\big) 
       - 4 +  \frac{8}{1-x} - 4 x\Bigg\}\,.\\
     P_{qg}^{(0)}(x) &=  T_F \Bigg\{
        4 - 8 x + 8 x^2\Bigg\}\,.\\
     P_{gq}^{(0)}(x) &= C_F \Bigg\{
       - 8 + \frac{8}{x} + 4 x\Bigg\}\,.\\
     P_{gg}^{(0)}(x) &= C_A \Bigg\{
          \frac{22}{3} \delta\big(1-x\big)  
       - 16 +  \frac{8}{x} +  \frac{8}{1-x} + 8 x - 8 x^2\Bigg\}
  - n_fT_F \Bigg\{
            \frac{8}{3}\delta\big(1-x\big) \Bigg\}\,. 
 \end{align}
The next-to-leading order functions are, 
\begin{align}
P_{qq}^{(1),\text{NS}}(x) &= C_Fn_fT_F \Bigg\{
        \frac{16}{3}\bigg( 1 - \frac{2}{1-x} +  x \bigg) \ln(x)  
       -   \frac{4}{3}\Big(   1+ 8 \zeta_2 \Big)\delta\big(1-x\big)  
       - \frac{16}{9} - \frac{160}{9} \frac{1}{1-x} + \frac{176}{9} x\Bigg\}
\nonumber\\& 
 +
   C_AC_F \Bigg\{
          \bigg( \frac{17}{3} - 24 \zeta_3 + \frac{88}{3} \zeta_2 \bigg) \delta\big(1-x\big) 
       - 4\bigg(   1 -  \frac{2}{1-x} +  x \bigg) \ln^2(x)        
       -  \frac{4}{3}\bigg( 5  -  \frac{22}{1-x} + 5 x \bigg) \ln(x)  
\nonumber\\&       
       + \frac{212}{9}  - \frac{748}{9} x + \bigg(\frac{536}{9} - 16 \zeta_2\bigg) \frac{1}{1-x} + 8 \zeta_2 \Big(1+x\Big)\Bigg\}
 +
   C_F^2 \Bigg\{ 
       16\bigg( 1 -  \frac{2}{1-x} + x \bigg) \ln(x) \ln(1-x)   
\nonumber\\&        
       -  8\bigg(   \frac{3}{1-x} + 2 x \bigg) \ln(x)  
       -   4\Big(   1 +  x \Big) \ln^2(x)
       + 3\Big( 1 + 16 \zeta_3 - 8 \zeta_2 \Big) \delta\big(1-x\big)  
       - 40\Big(1-x \Big)\Bigg\}  \, .
\\        
 P_{qq}^{(1),\text{S}}(x) &=  C_FT_F \Bigg\{
        8\bigg( 1 + 5 x + \frac{8}{3} x^2 \bigg) \ln(x)  
       -  8\Big(  1 +  x \Big) \ln^2(x) 
       - 16 + \frac{160}{9} \frac{1}{x} + 48 x - \frac{448}{9} x^2\Bigg\} \,. 
\\
P_{qq}^{(1),-}(x) &=\bigg(C_F^2-\frac{1}{2} C_AC_F \bigg) \Bigg\{ \bigg(   \frac{1+x^2}{1+x} \bigg)\bigg( 8 \ln^2(x)   
       - 32\ln(x) \ln(1+x)      
       -  32 \text{Li}_{2}(-x) -  16 \zeta_2  \bigg)
       +16\Big(1+x\Big)  \ln(x)
\nonumber\\&       
        + 32 \Big(1-x\Big)  \Bigg\}\, .
\\
P_{qg}^{(1)}(x)  &= C_A T_F \Bigg\{
         8\bigg( 1 + 8 x + \frac{44}{3} x^2 \bigg) \ln(x)
       -  32\Big(    x - x^2 \Big)\ln(1-x) 
       - 8 \Big(   1 - 2 x + 2 x^2 \Big)\ln^2(1-x)    
\nonumber\\&       
       -  16\Big(   1 + 2 x + 2 x^2 \Big)\bigg( \text{Li}_{2}(-x) +\ln(x) \ln(1+x)  \bigg)
       - 8\Big(   1 + 2 x \Big) \ln^2(x)             
       - 16 + \frac{160}{9} \frac{1}{x}       
       + 200 x - \frac{1744}{9} x^2 
\nonumber\\&         
       - 32 \zeta_2 x\Bigg\} 
 +
   C_F T_F \Bigg\{
         8\Big( 1 - 2 x + 2 x^2 \Big) \ln^2(1-x)  
       + 4\Big( 1 - 2 x + 4 x^2 \Big) \ln^2(x)  
       +32\Big( x - x^2 \Big) \ln(1-x)   
\nonumber\\&        
       - 16\Big(   1 - 2 x + 2 x^2 \Big) \ln(x) \ln(1-x)    
       + 4\Big( 3 - 4 x + 8 x^2 \Big) \ln(x) 
       + 56 - 116 x + 80 x^2      
       - 16\Big(1 - 2 x + 2 x^2\Big)\zeta_2\Bigg\} \, .
\\
P_{gq}^{(1)}(x)  &= C_F n_f T_F \Bigg\{
           \frac{32}{3} \bigg(2 -  \frac{2}{x} - x \bigg)\ln(1-x)
       + \frac{320}{9} - \frac{320}{9} \frac{1}{x} - \frac{256}{9} x\Bigg\}
+
   C_AC_F \Bigg\{
         8\Big( 2 +  x \Big) \ln^2(x)        
\nonumber\\&       
-  \frac{8}{3} \bigg(   22 -  \frac{22}{x} - 17 x \bigg) \ln(1-x)
       - 8 \bigg(   2 - \frac{2}{x} -  x \bigg) \ln^2(1-x)          
       + 16\bigg( 2 -  \frac{2}{x} -  x \bigg) \ln(x) \ln(1-x)  
\nonumber\\&       
       - \bigg(   96 + 40 x + \frac{64}{3} x^2 \bigg) \ln(x)
       +  16\bigg( 2 + \frac{2}{x} +  x \bigg) \bigg(\text{Li}_{2}(-x)+\ln(x) \ln(1+x)  \bigg)   
       + \frac{152}{9} + \frac{8}{x} + 32 \zeta_2
\nonumber\\&          
       + \frac{296}{9} x + \frac{352}{9} x^2 \Bigg\}
 +
   C_F^2 \Bigg\{
        8\bigg( 6 - \frac{6}{x} - 5 x \bigg) \ln(1-x)  
       + 8\bigg( 2 -  \frac{2}{x} -  x \bigg) \ln^2(1-x)         
       + 4\Big( 4 + 7 x \Big) \ln(x)  
\nonumber\\&       
       -  4\Big(   2 - x \Big) \ln^2(x) 
       - 20 - 28 x\Bigg\}    \, .
\\
P_{gg}^{(1)}(x)  &=    C_Fn_fT_F \Bigg\{
       -   16\Big(   3 + 5 x \Big)\ln(x) 
       - 16\Big(   1 +  x \Big) \ln^2(x)  
       -  \Big(   8 \Big)\delta\big(1-x\big)  
       - 128 + \frac{32}{3} \frac{1}{x} + 64 x + \frac{160}{3} x^2\Bigg\}
 \nonumber\\&
 +
   C_An_fT_F \Bigg\{
       - \frac{32}{3}\Big( 1   +  x \Big) \ln(x)  
       - \bigg(   \frac{32}{3} \bigg) \delta\big(1-x\big)  
       + \frac{464}{9} - \frac{368}{9} \frac{1}{x} - \frac{160}{9} \frac{1}{1-x} - \frac{304}{9} x + \frac{368}{9} x^2\Bigg\}
\nonumber\\& 
 +
   C_A^2 \Bigg\{
          32\bigg( 2 +  \frac{1}{x} -  \frac{1}{1+x} + x + x^2 \bigg)\bigg(\text{Li}_{2}(-x) + \ln(x) \ln(1+x) \bigg)
       - \frac{8}{3}\bigg(  25  - 11 x + 44 x^2 \bigg) \ln(x)            
\nonumber\\&
       + 32\bigg( 2 -  \frac{1}{x} - \frac{1}{1-x} -  x + x^2 \bigg) \ln(x) \ln(1-x)  
       +8\bigg(  \frac{1}{1-x} +  \frac{1}{1+x} + 4 x - 2 x^2 \bigg) \ln^2(x)   
\nonumber\\&         
       +  8\bigg( \frac{8}{3} + 3 \zeta_3 \bigg) \delta\big(1-x\big) 
      + \bigg(\frac{536}{9} - 16 \zeta_2\bigg)
         \frac{1}{1-x}    - \frac{100}{9}   
        - \frac{436}{9} x - \frac{16}{1+x} \zeta_2 
         + 64 \zeta_2 + 32 \zeta_2 x^2\Bigg\}\, .
 \end{align}
\section{Space-like Polarized Splitting Function}
\label{Appendix:B}
\noindent
Here we present the space-like polarized splitting functions at leading and next-to leading orders.
The leading order splitting function are,
\begin{align}
\Delta P_{qq}^{(0)}(x) &= C_F \Bigg\{
         6~\delta\big(1-x\big)  + \frac{8}{1-x} - 4\Big( 1+x \Big)\Bigg\}\, .\\
\Delta P_{qg}^{(0)}(x) &=  T_F \bigg\{
       - 4 + 8 x\bigg\}\, .\\
\Delta P_{gq}^{(0)}(x) &= C_F \bigg\{
         8 - 4 x\bigg\}\, .\\
\Delta P_{gg}^{(0)}(x) &=  C_A \Bigg\{
         \frac{22}{3} \delta\big(1-x\big)
       + 8 +  \frac{8}{1-x} - 16 x\Bigg\}
-       
 n_fT_F \Bigg\{
         \frac{8}{3} \delta\big(1-x\big)  \Bigg\}\, .
\end{align}
The next-to leading order functions are,
\begin{align}
\Delta P_{qq}^{(1),\text{NS}}(x) &= C_F n_f T_F \Bigg\{
        \frac{16}{3}\bigg( 1 -  \frac{2}{1-x} +  x \bigg)\ln(x)   
       -    \frac{4}{3}\Big(  1 + 8\zeta_2 \Big)\delta\big(1-x\big) 
       - \frac{16}{9} - \frac{160}{9} \frac{1}{1-x} + \frac{176}{9} x\Bigg\}
\nonumber\\& 
+
   C_AC_F \Bigg\{
          \bigg( \frac{17}{3} - 24 \zeta_3 + \frac{88}{3} \zeta_2 \bigg)\delta\big(1-x\big)     
       -  \frac{4}{3}\bigg( 5 -  \frac{22}{1-x} + 5 x \bigg)\ln(x)  
       -  4 \bigg(   1 -  \frac{2}{1-x} +  x \bigg) \ln^2(x)
\nonumber\\&       
       + \frac{212}{9}  - \frac{748}{9} x + \bigg(\frac{536}{9} - 16 \zeta_2 \bigg)\frac{1}{1-x} + 8 \zeta_2 \Big(1+x\Big)\Bigg\}
 +
   C_F^2 \Bigg\{
        16 \bigg( 1 -  \frac{2}{1-x} +  x \bigg)\ln(x) \ln(1-x) 
 \nonumber\\&       
       - 8\bigg( \frac{3}{1-x} + 2 x \bigg)\ln(x)  
       - 4 \Big( 1 +  x \Big)\ln^2(x)       
       + 3 \Big( 1 + 16 \zeta_3 - 8 \zeta_2 \Big)\delta\big(1-x\big)  
       - 40 \Big(1-x\Big)\Bigg\}\, .   \\
\Delta P_{qq}^{(1),\text{S}}(x) &= C_F T_F \Bigg\{
       -  8\Big(   1 - 3 x \Big)\ln(x)  
       - 8\Big(   1 +  x \Big)\ln^2(x)   
       + 8\Big( 1-  x\Big)\Bigg\}\, . \\
\Delta P_{qq}^{(1),-}(x) &= \bigg(C_F^2-\frac{1}{2}C_AC_F \bigg)\Bigg\{ 
         \bigg( \frac{1+x^2}{1+x} \bigg)\bigg(32 \ln(x) \ln(1+x)  
       - 8\ln^2(x) 
       +   32 \text{Li}_{2}(-x)  +  16 \zeta_2 \bigg)
    -16\Big( 1 +  x \Big)\ln(x) 
\nonumber\\&    
    - 32 \Big(1- x \Big) \Bigg\}\, .
\\
\Delta P_{qg}^{(1)}(x) &= C_AT_F \bigg\{
         8\Big( 1 - 2 x \Big) \ln^2(1-x) 
       - 32\Big(  1  - x \Big)\ln(1-x) 
       + 8\Big( 1 + 8 x \Big)\ln(x) 
       - 8\Big(  1 + 2 x \Big)\ln^2(x)  
\nonumber\\&       
       - 16\Big( 1 + 2 x \Big) \bigg(\text{Li}_{2}(-x)+\ln(x) \ln(1+x)\bigg)
       + 96 - 88 x - 16 \zeta_2\bigg\}
\nonumber\\& 
 +
   C_FT_F \bigg\{
           32\Big( 1 -  x \Big)\ln(1-x)      
       -  8\Big(   1 - 2 x \Big)\ln^2(1-x) 
       -  \Big(   36 \Big)\ln(x) 
       +  16\Big( 1 - 2 x \Big)\ln(x) \ln(1-x)
\nonumber\\&       
       -  4\Big(   1 - 2 x \Big)\ln^2(x)  
       - 88 + 108 x + 16 \zeta_2\Big(1-2 x\Big)\bigg\}\, .
\\
\Delta P_{gq}^{(1)}(x) &=    C_AC_F \Bigg\{
          \frac{8}{3} \Big( 10 + x \Big)\ln(1-x)  
       +  8 \Big( 2 -  x \Big)\ln^2(1-x) 
       +   8\Big( 4 - 13 x \Big) \ln(x)
       -  16\Big(   2 -  x \Big)\ln(x) \ln(1-x)
\nonumber\\&        
       +   8\Big( 2 +  x \Big) \ln^2(x) 
       +  16\Big( 2 +  x \Big)\bigg(\text{Li}_{2}(-x) +\ln(x) \ln(1+x)\bigg)
       + \frac{328}{9} + \frac{280}{9} x + 16 \zeta_2 x\Bigg\}
\nonumber\\& 
-
   C_F^2 \Bigg\{
         8\Big(   2 +  x \Big)\ln(1-x)  
       +  8\Big(   2 -  x \Big)\ln^2(1-x)  
       +  4\Big(   4 -  x \Big)\ln(x)  
       -  4\Big( 2 -  x \Big) \ln^2(x) 
       + 68 - 32 x\Bigg\}
\nonumber\\&       
    - C_Fn_fT_F \Bigg\{
           \frac{32}{3}\Big(  2 -  x \Big)\ln(1-x)
       + \frac{128}{9} + \frac{32}{9} x\Bigg\} \, .  \\
\Delta P_{gg}^{(1)}(x) &=C_F n_f T_F \Bigg\{
       -   16\Big(  5 -  x \Big)\ln(x) 
       - 16\Big(   1 +  x \Big)\ln^2(x)   
       -  \Big(  8 \Big)\delta\big(1-x\big)  
       - 80\Big(1-x\Big)\Bigg\}
 \nonumber\\&
 +   C_A n_f T_F \Bigg\{
       -  \frac{32}{3}\bigg(  1  + x \bigg)\ln(x)  
       -  \bigg(   \frac{32}{3} \bigg) \delta\big(1-x\big) 
       - \frac{448}{9} - \frac{160}{9} \frac{1}{1-x} + \frac{608}{9} x\Bigg\}
\nonumber\\&
+
   C_A^2 \Bigg\{
         \bigg( \frac{64}{3} + 24 \zeta_3 \bigg)\delta\big(1-x\big)     
       +  \bigg( \frac{232}{3} - \frac{536}{3} x \bigg)\ln(x)  
       -   32\bigg(   1 + \frac{1}{1-x} - 2 x \bigg)\ln(x) \ln(1-x) 
\nonumber\\&       
       +  32\bigg( 1 +  \frac{1}{1+x} + 2 x \bigg)\bigg( \text{Li}_{2}(-x) +\ln(x) \ln(1+x) \bigg)   
       +  8\bigg( 4 +  \frac{1}{1-x} -  \frac{1}{1+x} \bigg) \ln^2(x) + 
         \frac{16}{1+x}\zeta_2
\nonumber\\&       
       - \frac{148}{9}  - \frac{388}{9} x + \bigg(\frac{536}{9} - 16 \zeta_2\bigg) \frac{1}{1-x}  + 64 \zeta_2 x\Bigg\}  \, .
\end{align}

\section{Time-like Unpolarized Splitting Function}
\label{Appendix:C}
\noindent
In this appendix, we present the time-like unpolarized splitting functions at leading and next-to-leading orders. The leading order splitting functions are,
\begin{align}
\tilde{P}_{qq}^{(0)}(z) &=  C_F \Bigg\{
           6~\delta\big(1-z\big) 
       - 4 +  \frac{8}{1-z} - 4 z\Bigg\}\, .\\
\tilde{P}_{qg}^{(0)}(z) &=T_F \Bigg\{
        4 - 8 z + 8 z^2\Bigg\}\, .\\
\tilde{P}_{gq}^{(0)}(z) &=  C_F \Bigg\{
       - 8 +  \frac{8}{z} + 4 z\Bigg\}\, .\\
\tilde{P}_{gg}^{(0)}(z) &=   C_A \Bigg\{
        \frac{22}{3} \delta\big(1-z\big)  
       - 16 +  \frac{8}{z} +  \frac{8}{1-z} + 8 z - 8 z^2\Bigg\}
- n_fT_F \Bigg\{
         \frac{8}{3}  \delta\big(1-z\big)   \Bigg\}\, .
\end{align}
The next-to leading order functions are,
\begin{align}
\tilde{P}_{qq}^{(1),\text{NS}}(z) &=C_Fn_fT_F \Bigg\{
       -  \frac{16}{3} \bigg(\frac{1+z^2}{1-z} \bigg)\ln(z)   
       -  \frac{4}{3} \bigg(  1 + 8 \zeta_2 \bigg)\delta\big(1-z\big)  
       - \frac{16}{9} - \frac{160}{9} \frac{1}{1-z} + \frac{176}{9} z\Bigg\}
\nonumber\\& 
 +
   C_AC_F \Bigg\{
       \bigg( \frac{44}{3} \bigg(\frac{1+z^2}{1-z}\bigg) + 8\Big(1 + z\Big) \bigg) \ln(z)   
       + 4\bigg(   \frac{1+z^2}{1-z} \bigg) \ln^2(z)  
       +  \bigg( \frac{17}{3} - 24 \zeta_3 + \frac{88}{3} \zeta_2 \bigg) \delta\big(1-z\big) 
\nonumber\\&       
       + \frac{212}{9}  - \frac{748}{9} z + \bigg(\frac{536}{9}  - 16 \zeta_2\bigg) \frac{1}{1-z} + 8 \zeta_2 \Big(1+z\Big)\Bigg\}
 +
   C_F^2 \Bigg\{
        \bigg( 12 \bigg(\frac{1+z^2}{1-z}\bigg)  - 28 - 12 z  \bigg) \ln(z)
\nonumber\\&       
       +  16\bigg(  \frac{1+z^2}{1-z} \bigg) \ln(z) \ln(1-z) 
       - \bigg( 16 \bigg(\frac{1+z^2}{1-z}\bigg) - 4\Big(1+ z\Big)  \bigg) \ln^2(z)  
       +  3\Big( 1 + 16\zeta_3 - 8 \zeta_2 \Big) \delta\big(1-z\big) 
\nonumber\\&       
       - 40 \Big(1 - z\Big)\Bigg\}\, .
\\
\tilde{P}_{qq}^{(1),\text{S}}(z) &= C_FT_F \Bigg\{
           8\Big( 1 +  z \Big) \ln^2(z)
         -\Big(   40 + 72 z + \frac{64}{3} z^2 \Big)\ln(z)    
       - 64 - \frac{160}{9} \frac{1}{z} + 32 z + \frac{448}{9} z^2\Bigg\}\,.
\\
\tilde{P}_{qq}^{(1),-}(z) &=  \bigg(C_F^2-\frac{1}{2}C_AC_F\bigg) \Bigg\{\bigg(  \frac{1+z^2}{1+z} \bigg)
        \bigg(8 \ln^2(z)
        - 32\text{Li}_{2}(-z) 
        - 16 \zeta_2 
        - 32 \ln(z) \ln(1+z)\bigg)
        + 16\Big(1+z\Big) \ln(z)
\nonumber\\&        
       + 32 \Big(1-z\Big) \Bigg\}\, .
\\
\tilde{P}_{gq}^{(1)}(z) &= C_AC_F \Bigg\{
         16\bigg( 2 +  \frac{2}{z} +  z \bigg) \bigg(\text{Li}_{2}(-z) +\ln(z) \ln(1+z) \bigg)         
       + 64 \bigg( 2 -  \frac{2}{z} - z \bigg) \text{Li}_{2}(1-z) 
       - \Big(   16 z \Big) \ln(1-z)  
\nonumber\\&
       + 8 \bigg( 2 - \frac{2}{z} -  z \bigg)\ln^2(1-z)  
       + 8\bigg( 8 -  \frac{6}{z} + 9 z + \frac{8}{3} z^2 \bigg) \ln(z)  
       + 16 \bigg( 2 -  \frac{2}{z} -  z \bigg) \ln(z) \ln(1-z)  
\nonumber\\&              
       - 8\bigg(   2 + \frac{4}{z} + 3 z \bigg) \ln^2(z)  
       + 40 + \frac{136}{9} \frac{1}{z} - 8 z - \frac{352}{9} z^2 
       +\bigg( \frac{128}{z} - 96  +
         64 z\bigg)\zeta_2 \Bigg\}
\nonumber\\&        
 +
   C_F^2 \Bigg\{
          4\Big( 2 -  z \Big) \ln^2(z)   
       - 64\bigg(   2 -  \frac{2}{z} - z \bigg)\text{Li}_{2}(1-z)   
       -  8\bigg(   2 -  \frac{2}{z} -  z \bigg) \ln^2(1-z) 
       - 4\Big(   16 -  z \Big) \ln(z)
\nonumber\\&              
       + \Big( 16 z \Big) \ln(1-z)  
       -  32\bigg(  2 - \frac{2}{z} - z \bigg)\ln(z) \ln(1-z)  
       - 4 + 36 z - 64\bigg(  \frac{2}{z} - 2  + z\bigg) \zeta_2 \Bigg\} \,.   
\\
\tilde{P}_{gg}^{(1)}(z) &= C_Fn_fT_F \Bigg\{
         \bigg( 80 + \frac{128}{3} \frac{1}{z} + 112 z + \frac{128}{3} z^2 \bigg) \ln(z) 
       +  16\Big( 1 +  z \Big) \ln^2(z) 
       -  \Big(   8 \Big) \delta\big(1-z\big) 
       - 32   
       + \frac{736}{9} \frac{1}{z} + 96 z 
\nonumber\\&            
       - \frac{1312}{9} z^2\Bigg\}
 +
   C_An_f T_F \Bigg\{
        \frac{32}{3}\bigg( 3 -  \frac{2}{z} -  \frac{2}{1-z} - 3 z + 2 z^2 \bigg) \ln(z)  
       -  \bigg(   \frac{32}{3} \bigg) \delta\big(1-z\big)    
       + \frac{464}{9} - \frac{368}{9} \frac{1}{z}
\nonumber\\&           
       - \frac{160}{9} \frac{1}{1-z} - \frac{304}{9} z + \frac{368}{9} z^2\Bigg\}
 +
   C_A^2 \Bigg\{
        32\bigg( 2 +  \frac{1}{z} -  \frac{1}{1+z} +  z +  z^2 \bigg) \bigg(\text{Li}_{2}(-z)  + \ln(z) \ln(1+z)\bigg)
\nonumber\\&        
       -\bigg(   88 + \frac{176}{3} \frac{1}{z} - \frac{176}{3} \frac{1}{1-z} + 8 z + \frac{176}{3}z^2 \bigg) \ln(z)           
       - 32\bigg(   2 - \frac{1}{z} -  \frac{1}{1-z} -  z +z^2 \bigg) \ln(z) \ln(1-z)  
\nonumber\\&       
       - \bigg(    \frac{32}{z} + \frac{24}{1-z} -  \frac{8}{1+z} + 64 z - 16 z^2 \bigg) \ln^2(z)   
       + \bigg( \frac{64}{3} + 24 \zeta_3 \bigg) \delta\big(1-z\big)  
       + \bigg(\frac{536}{9}  - 16 \zeta_2 \bigg)\frac{1}{1-z}       
\nonumber\\&              
        - \frac{100}{9}  - \frac{436}{9} z 
        - \frac{16}{1+z}\zeta_2
        + 64 \zeta_2 + 32 \zeta_2 z^2\Bigg\}\, .
\end{align}

\section{Z-factor} 
\label{Appendix:D}
\noindent
Here, we present the finite renormalization constant $Z_{ab}$ which enters into the transformation of the splitting and coefficient functions from the Larin to $\overline{\text{MS}}$ scheme,
\begin{align}
Z_{ab}(x) &= \delta(1-x) + \sum_{l=1}^{\infty}a_s^{l} z^{(l)}_{ab}(x)\,,
\end{align}
with
\begin{align}
z_{q_{i}q_{j}}^{(l)} &= z_{\overline{q}_{i}\overline{q}_{j}}^{(l)} = \delta_{ij}z_{qq}^{(l),\rm V} + z_{qq}^{(l),\rm S}\, ,\nonumber\\
z_{q_{i}\overline{q}_{j}}^{(l)} &= z_{\overline{q}_{i}{q}_{j}}^{(l)} =\delta_{ij}z_{q\overline{q}}^{(l),\rm V} + z_{q\overline{q}}^{(l),\rm S}\, ,\nonumber\\
z_{qg}^{(l)} &= z_{gq}^{(l)} = z_{gg}^{(l)} = 0\,.
\end{align}
Here $i$ and $j$ denotes quarks flavors. The rest $z^{(l)}_{ab}$'s coefficients are given by: 
\begin{align}
z_{q_{i}q_{j}}^{(1)}(x) &= \delta_{ij} C_F \Bigg\{
       - 8 + 8 x\Bigg\}\, , ~~~ z_{{q}_{i}\overline{q}_{j}}^{(1)}(x) = 0 \,,\\    
z_{qq}^{(2),\rm V}(x) &=   C_F n_f T_F \Bigg\{
        \frac{16}{3}\Big( 1 - x \Big)\ln(x)  
       + \frac{80}{9} \Big( 1- x \Big)\Bigg\}
 -
   C_A C_F \Bigg\{
           \bigg(   \frac{80}{3} - \frac{8}{3} x \bigg)\ln(x)
       + 4   \Big(   1 -  x \Big)\ln^2(x)
\nonumber\\&       
       + \frac{592}{9} \Big(1-x\Big)  - 8 \zeta_2 \Big(1-x\Big)\Bigg\}
 +
   C_F^2 \Bigg\{
        16   \Big( 1- x \Big)\ln(x) \ln(1-x) 
       - 8   \Big(   2 +  x \Big)\ln(x)       
       - 16\Big(1-x\Big) \Bigg\}\,,\\
z_{q\overline{q}}^{(2),\rm V}(x) &= -\Bigg(C_F^2 - \frac{1} {2}C_A C_F\Bigg) \Bigg\{ 8 \Big(1+x\Big) \bigg(
        4 \text{Li}_{2}(-x)
       + 4 \ln(x) \ln(1+x)  
       - \ln^2(x)           
       - 3 \ln(x) 
       + 2 \zeta_2 \bigg)
       - 56 \Big( 1 - x \Big) \Bigg\}
\,,\\
z_{qq}^{(2),\rm S}(x) &= z_{q\overline{q}}^{(2),\rm S}(x) = C_FT_F \Bigg\{
        8 \Big( 3 -  x \Big) \ln(x)  
       +4 \Big( 2 +  x \Big) \ln^2(x)   
       + 16 \Big( 1 - x \Big) \Bigg\}\, .
\end{align}

\end{widetext}

\bibliographystyle{apsrev4-1}
\bibliography{main}

\end{document}